%% file: Shape_Dynamics.tex
\documentclass[12pt]{report}
\usepackage{amsthm,amssymb}
\usepackage[utf8]{inputenc}
\usepackage{amsmath}
\usepackage{xfrac}
\usepackage[backend=bibtex,style=ieee]{biblatex}
\usepackage{geometry}
\usepackage{authblk}
\usepackage[hidelinks]{hyperref}
\usepackage{ragged2e}
\usepackage{csquotes}
\usepackage{graphicx,scalerel}
\usepackage{float}
\usepackage{braket}
\usepackage{xcolor}
\usepackage{caption}
\usepackage{comment}
\usepackage{tikz}
\usepackage{cleveref}
\usepackage[stable]{footmisc}
\usepackage{enumitem}

\usetikzlibrary{mindmap,backgrounds,er,positioning}

\newtheorem{definition}{Definition}[chapter]
\newtheorem{theorem}{Theorem}[chapter]

\newcommand\sbullet[1][.5]{\mathbin{\vcenter{\hbox{\scalebox{#1}{$\bullet$}}}}}
\DeclareMathOperator{\Tr}{Tr}

\title{\LARGE \textbf{Shape Dynamics and The Universe: Foundations and Implications}\\ %\medskip {\Large Bachelor Thesis
}
\author{\Large {Pooya Farokhi\thanks{pooyafarokhi.p@gmail.com}}}
\affil{\vfill \normalsize Bachelor Thesis \\ \medskip \textit{Department of Physics, Sharif University of Technology, Tehran, Iran}}
\date{\normalsize Winter 2022}

\begin{document}
\maketitle

\pagenumbering{roman}

\chapter*{\Large Abstract}

Shape dynamics is an alternative background-independent approach to classical dynamics that implements Leibnizian philosophy and Mach's Principles. It is a formulation of the dynamics of the universe in terms of the intrinsic and relational degrees of freedom which are objectively observable and not properties defined with respect to an external frame of reference. Shape dynamics is not a very old field of study. Although it was already gradually coming alive out of Julian Barbour's early works on Mach's Principle back in the 1980s, it was invigorated by a series of rigorous works in the last decade.

This work is an exhaustive review of the historical and conceptual underpinning of the theory that extends to Leibniz and Newton's philosophy, and the currently established formulation of the theory, together with some of the major results of its cosmological applications. The structure of this work consists of two parts: In the first one we cover the foundations and the formulation of the theory (both the field and particle ontology), and in the second one we study its applications and reflect on the resolution of the problem of the arrow of time. We end our journey by contemplating some recent ideas and prospects for further developing the theory.

\chapter*{\Large Dedication}

I dedicate this work to all those who have pondered the mysteries of nature... since the time immemorial that man portrayed his inner wonder at the veiled secrets of nature on the walls of the caves hidden and protected at the heart of nature, until now that we have ink and paper, chalks and blackboards all over the world and do the same thing... extending this precious lineage.

\chapter*{\Large Acknowledgment}

There are a lot of people to whom I would like to express my gratitude.

I started my research independently and worked mostly on my own, but then I found my way to Julian Barbour's research group and since then, started writing this thesis. I can hardly find any words to convey my gratitude to Julian Barbour for many valuable and insightful discussions, his unconditional support and constant supervision and guidance. His unwavering dedication to understanding nature fundamentally and sincere curiosity are truly phenomenal. I was very lucky to conduct my research on shape dynamics with the help of the founder of it, which presented to me an opportunity to think deeply and critically and develop the courage to ask and ponder really fundamental questions, as well as a valuable source of constant inspiration for being a harmonious human being.

I would like to thank Adam Fountain and Kartik Tiwari for many interesting discussions. Also, thank you very much to Bahram Shakerin for his helpful remarks and heartening support.

During my undergraduate program, I greatly benefited from studying physics with some of my classmates, and the discussions I had with them played a major role in helping me learn to think more deeply and rigorously. I would like to thank Kurosh Allame, Mohammadamin Sadeghian, Mobin Moradi, and Hossein Mohammadi.

Finally, thank you to my family for providing for me and their support.

\tableofcontents
\listoffigures
\listoftables

\chapter*{Symbols}
\setlength{\tabcolsep}{0.075\linewidth}
\begin{tabular}{l l}
$Q^N$ & Configuration space \\
$Q^R$ & Relational configuration space\\
$CS^V$ & Conformal superspace plus volume\\
$Sup$ & Superspace\\
$Riem^3$ & Space of three-dimensional Riemannian metrics\\
$G$ & Lie group\\
$\mathfrak{g}$ & Lie algebra\\
$E(N)$ & Euclidean group (in N dimensions)\\
$\mathbb{R}^{+}$ & Scale transformations\\
$Sim(N)$ & Similarity group, i.e., $Sim(N) = \mathbb{R}^+ \ltimes E(N)$\\
$I,J,K,\cdots$ & Indices of particles\\
$a,b,c,\cdots$ & Spatial indices\\
$\mu,\nu,\rho,\cdots$ & Spatio-temporal indices\\
$q^a_I$ & generalized coordinates (particle mechanics)\\
$p^{J}_b$ & Conjugate momentum (particle mechanics)\\
$T_{\phi} q^a_I$ & Transformation of the $q_I$'s under a Lie group\\
$t_{\alpha b}^a$ & Elements of the generators of a Lie group\\
$D_{\phi} q^a_I$ & denotes $T^{-1}_{\phi} \frac{d}{d\lambda} T_{\phi} q^a_I$\\
$g_{ab}$ & 3D Riemannian metric\\
$p^{ab}$ & Momentum conjugate to metric\\
$K_{ab}$ & Extrinsic curvature\\
$\{\cdot,\cdot \cdot\}$ & Poisson bracket\\
$\{\cdot,\cdot \cdot\}_{DB}$ & Dirac bracket\\
$f(x)$ & Real-valued function defined on a manifold\\
$F[g]$ & Functional defined on the space of functions on a manifold\\
$f \cdot g$ & Smearing, i.e., $f \cdot g = \int d^3x \sqrt{g} f(x) g(x)$\\
$F[g,k,l,\cdots,x)$ & Functional $F$ depending on $x$ through the functions $g,k,l,\cdots$\\
\end{tabular}

\include{Introduction}

\part{The Foundations of Shape Dynamics}

\input{Chapter2}
\input{Chapter3}
\input{Chapter4}

\part{The Cosmological Application}

\input{Chapter5}
\input{Conclusion}

\appendix

\input{Appendices}

\printbibliography
\end{document}

%% file: Introduction.tex
\chapter{Introduction}
\pagenumbering{arabic}
\section{What is shape dynamics?}
Shape dynamics starts with an urge to find an answer to the question of how we can locate ourselves and other entities in the universe. In other words, what do we mean when we talk about location, position, motion, or more broadly, space or time? Shape dynamics strives to provide a meanigful, objective, and epistemologically sound answers to this question within a dynamical framework. As opposed to the Newtonian way of thinking about these questions, shape dynamics adopts a much simpler and intuitively more tangible point of view, from which we think about positions as `relative'. Ironically, shape dynamics presents a very childlike picture of the whole universe, in which everything is meaningfully defined in terms of how it is seen from each points of view. The whole reality is nothing but what is reflected in the views of all the entities within the universe. Following Leibniz, our anthem in shape dynamics is that `each part [of the universe] is a living mirror of the whole' \footnote{\emph{The Monadology} \cite{Leibniz.1714}, part 56.}, and shape dynamics is the physical theory that describes the dynamics of the whole universe in terms of the totality of those intrinsic pictures that each individual entity possesses of what the whole universe looks like, i.e., in terms of the relational properties.

And this childlike picture might ultimately turn out to be the key to unlock the deep puzzles of nature that have kept the adults in physics occupied for nearly a century. Shape dynamics provides a uniquely simple understanding of the arrow of time, and it seems to have the potential to tackle the most fundamental issues of cosmology and quantum gravity. At the moment, shape dynamics has been richly developed with both particle ontology and field ontology, and as an alternative theory of Newtonian mechanics and Einstein's relativity, it tells the story of classical physics in another way, much more illuminating and elegant, filled with many colorful insights that are normally hidden in the absolutist attitude towards physics which has worked quite satisfactorily in the laboratory, but not so much for unraveling the history of the whole universe.

To share with you more of the interesting story that follows in the next chapters, I clarify the relational properties and the underlying ontology of shape dynamics. All the objective information of the whole universe eventually boils down to the `angles' we observe in the universe. As I look around, I see dozens of different objects together forming complicated and intricate structures, all definable in terms of the pure angles I see as I look at them. The same thing follows for all the other entities of the universe, and these vast structures constitute what we consider to be the ontology of the whole universe: The totality of all the angles within the universe. Shape dynamics is a theory of the evolution of the shape of the universe, expressed through these pure angles.

In the standard Newtonian mechanics, the dynamical equations of motion are given with respect to a certain frame of reference (or as Newton himself originally said, motion is defined with respect to the absolute space). An intrinsic description of the evolution of the whole universe requires a different mathematical formulation. Julian Barbour and Bruno Bertotti found such a procedure in 1982. The key idea is to define an intrinsic `measure' on the relational shapes of the whole universe. Their method is to calculate the standard Euclidean distance between two specific shapes in the absolute Euclidean space, and then move the shapes relative to one another under any Euclidean transformation to minimize this distance. This minimized distance, the `best-matched' one, is identified as the intrinsic distance defined on the relational state space. We merely use Newton's absolute space to calculate all the possible distances between two shapes, but after finding out the minimum, we can simply cast it aside and work in the reduced relational configuration space. More explicitly, one might actually define the best-matched distance directly in terms of the relational degrees of freedom (the angles).

With a notion of relational distance at our disposal, we can take one step further and posit an action principle to find the dynamics of the whole universe given only the initial data: The solution curve is the one that extremizes the total path-dependent best-matched distance between an initial and final shape. This total distance is the sum of the distances between all successive shapes.

Time in this relational picture is nothing but the successive changes of the shapes which follow one another. Time has no fundamental role. Leibniz said that `time is the order of succession'. There is no gigantic divine metronome somewhere in the universe that synchronizes the universe. The evolving universe, along with its periodic subsystems, is its own clock and keeps everything synchronized. The idea that the flow of time is just the observable change of some system was more explicitly and clearly enunciated by Ernst Mach. It is not a challenge to sympathize with this idea. After all, we all use our wristwatches and wall clocks in our daily life. The way these clocks tell us the time is a clear order expressed in terms of the relative positions of the hands of the clocks, which can be read off. The relational status of the hands of these clocks and their change, encode what is interpreted as the passage of time. This idea is also true in a much broader sense about the whole universe. The whole universe, with its complicated structures and innumerable entities, records the passage of time.

In this sense, our relational theory of the whole universe does not implement a background notion of time. Mathematically, this means that the action of the theory is reparametrization invariant and the passage of time can be defined through gauge fixing.

More importantly, the solutions of the equations of shape dynamics tell us a very rich and profoundly simple story about the whole universe. To see that, we will define a mathematical quantity that measures the `amount' of structure formation in the whole universe in terms of the relational properties, called \emph{complexity}. What we ultimately aim for is to decree that \emph{complexity is the origin of the arrow of time}. The richness of the structure of the whole universe defines both a measure and an arrow of time. This will clearly show why the universe is as it is: The dynamics of the shape dynamics pushes any typical universe into increasing its complexity, and hence structure formation, as regions of higher complexity are the valleys of the `shape potential' which steers the dynamics in shape space. Hence, the history of the universe is one of more and more creation and formation of an increasing variety of structures. In this spirit, shape dynamics provides a sufficient reason for the way our universe works, the question that Leibniz was too concerned about.

\begin{figure}[h]
\centering
\includegraphics[scale=0.32]{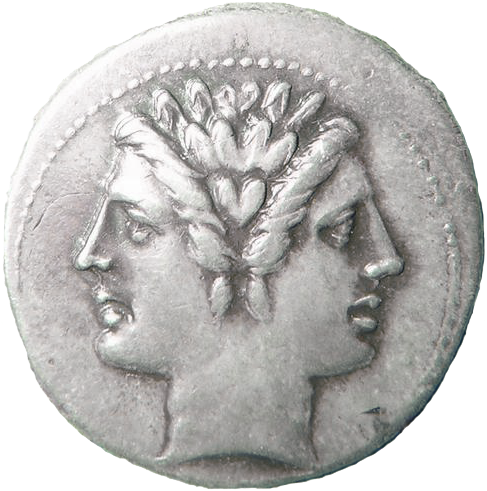}
\captionsetup{width=0.8\linewidth}
\caption[Janus depicted on a Roman coin]{\small The picture of a Roman coin with a depiction of the God `Janus'. Picture taken from Wikipedia}
\label{janus}
\end{figure}

The final plot is that our own universe we find ourselves in might be just one `branch' of a bi-universe. The point is that there is no reason to consider only an extendable solution curve in the configuration space, and we can in principle continue the curve representing the evolution of our universe in the other direction smoothly, and this takes us to another region of high complexity on the other side of shape space. The other universe also started at the same big bang that gave birth to our own universe, but its complexity (and hence time) flows in the other direction. The big bang, in this picture, resembles the Roman God `Janus', with its two protective faces looking at both of the universes in both directions, see Fig. \ref{janus}.

\section{The structure of this work}
My work has two major parts. The first part is on the foundations of shape dynamics and contains 3 chapters. I first start with a philosophical-historical account of the evolution of relational physics and the birth of shape dynamics in contemporary physics. I then move to the mathematical formulation of best-matching in the next chapter which is the procedure for purifying absolutist physical theories and distilling their inner relational core. Chapter 3 is the longest chapter of my work. I have very much cared about being rigorous and constructing a powerful mathematical skeleton for the theory that supports the heavy conceptual body of it that will form in parallel.

Chapter 4 is on the formulation of shape dynamics in the context of geometrodynamics, and this ultimately leads to an alternative geometrical theory of gravity which is equivalent to the standard theory of general relativity for globally hyperbolic and CMC foliable spacetimes. The role of these conditions will be clarified more properly. What is noteworthy at this stage, however, is the important role this alternative theory plays in laying bare the shiny Machian core of general relativity and sheds light on the decades-old problem of the status of Mach's Principle in Einstein's general relativity which left even Einstein, the prime figure behind Mach's Principle in a quandary.

The fifth chapter is on the cosmological applications of the theory and the resolution of the arrow of time. In this chapter, I give an account of the implications of the particle toy model as it is more maturely developed, and do not include the results of shape dynamics of geometry. There is still a long way to fully develop shape dynamics, and my focus in this work is to give a thorough review of what we are currently sure of. However, I will mention several ideas and prospects for research on shape dynamics in the last chapter.

This work might be useful to anyone, whether student or researcher, who is interested in learning the foundations of shape dynamics and being prepared to do research on the topic. This is meant to be a somewhat miniature textbook. No prior knowledge of shape dynamics and relational physics is assumed, but familiarity with classical dynamics, Lagrangian and Hamiltonian formulations, differential geometry and manifolds, and general relativity are needed. Knowing Lie groups, constrained Hamiltonian systems, and ADM formalism can be definitely beneficial and facilitate learning shape dynamics, but I have included a brief introduction to those topics in the appendices for possible readers who might need some help with them.

\section{A note on the symbols and notation}

Throughout this thesis, I have used many symbols and conventions for different purposes, many of which are not standard and common in physics but are relevant for the formulation of shape dynamics. Some of the symbols I use might be new and not conventionally used even in the literature on shape dynamics. I personally found these symbols a lot more succinct and hope that they facilitate following this work. The complete list of the symbols, along with their definition was presented on page vii before this chapter. Note that some of the symbols might not be familiar at this stage and will be defined properly in the body of the text. There are some comments on some conventions I have made in the work:

\begin{enumerate}
\item In all the calculations Einstein summation convention has been used for the spatial and spatio-temporal indices, but not for the particle indices, meaning that repeated spatial indices are summed over implicitly. For example, the expression

\[
q^a_I p^J_a q^c_J
\]

means

\[
\sum\limits_{a} q^a_I p^J_a q^c_J
\]

but there is no sum over $J$, and as another example we have

\[
T^{abc}_{l} g^{ls} p_{ar} v^r = \sum\limits_{a,l,r} T^{abc}_{l} g^{ls} p_{ar} v^r.
\]

\item The boundaries of the action integrals are often omitted for brevity. I hope this information can be understood and hence, it does not lead to confusion.

\item Functions or functional like $S[q]$ depend on \emph{all} of the configuration space variables and the individual indices are not explicitly written. Thus, $S[q]$ means $S[q^1_1,\cdots]$.

\item I use `$S$' to refer to the shape space of both particle ontology and field ontology. For particle model, it is simply the quotient of the Newtonian configuration space with respect to the similarity group, $S = Q^N/Sim(3)$. In geometrodynamics, it is the conformal superspace, i.e., the space of all Riemannian metrics defined up to a diffeomorphism and a conformal transformation.

\item I use the symbol `$\equiv$' to denote an identity \emph{based on definition}.
\end{enumerate}

%% file: Chapter2.tex
\chapter{Relationalism: A Philosophical-Historical Prelude}
\section{Newton's notion of absolute space and absolute time}
Back in 1687, Issac Newton formulated his novel theory. Aristotelian dynamics was already on the wane for centuries. Johannes Kepler had called for a new and different way of \emph{philosophizing}\footnote{
Following an important observation by Tycho Brahe of a comet, Johannes Kepler drew the far-reaching conclusion that crystalline spheres could not have existed, otherwise the comet would have gone through them. He said: \emph{``From henceforth the planets follow their paths through the ether like the birds in the air. We must therefore philosophize about these things differently.''}}
 , Galileo Galilei had started taking the first steps in this direction and formulated a new law of inertia. Issac Newton, a genius with a profound understanding of mathematics and philosophy as well as physical intuition, completed this feat. 

In doing so, Newton faced a problem in establishing a notion of \emph{equilocality}. In order to have a sensible  theory of dynamics, we must have a clear definition of `motion' and `rest'. The problem is to meaningfully talk of the same place at different times. How do we determine if a particle has remained in the `same' position if we look at it one hour later?

Newton, as sharp as he was, certainly anticipated this problem and addressed it in the Scholium. His solution was to reify space and time. In Newton's eyes, even if the universe is empty, still there exists the space of a translucent structure, and the flow of time as the tickings of an invisible clock resonate through nothingness. In Newton's words:

\begin{samepage}
\begin{displayquote}
\emph{
Absolute space, in its own nature, without relation to anything external, remains always
similar and immovable. Relative space is some movable dimension or measure of the absolute
spaces; which our senses determine by its position to bodies; and which is commonly taken for
immovable space...}
\bigskip

\emph{
Absolute, true, and mathematical time, of itself, and from its own nature, flows equably
without relation to anything external, and by another name is called duration: relative,
apparent, and common time, is some sensible and external (whether accurate or unequable)
measure of duration by the means of motion, which is commonly used instead of true time;
such as an hour, a day, a month, a year.}
\end{displayquote}
\end{samepage}

Following these principles, Newton proposed his three laws of motion. He tied the concepts of motion and rest to absolute space and absolute time and managed to consistently state his principles. For instance, Newton's first law which states that a body continues in its state of rest, or of uniform motion in a straight line, unless it is subject to a force, should be understood in the context in which state of rest or uniform motion is defined with respect to the absolute space.

Newton could see a difficulty in this approach. How do we determine the kinematical state of bodies in absolute space if we can only know the relative distances between bodies? Analogously, how can we hear the tickings of Newton's absolute time in this noisy universe full of mechanical clocks surrounding us? Newton's absolute space and absolute time are utterly beyond our observations. Ratios of the relative distances, angles between the objects, planetary motions, and the movements of the hands of the clocks are what we can directly observe.\footnote{
Actually, relative distances alone are also unobservable. In fact, \emph{ratios} of distances are the ultimate reality we observe. In other words, we can only see the \emph{angles} between objects. The relationality of size is a subtle issue that was not investigated by the earlier relationalists. We will come back to this issue for many times in this work and discuss that in detail.
}

In Newton's own words in the Scholium, this problem, known as the \emph{Scholium problem} is:

\begin{displayquote}
\emph{
It is indeed a matter of great difficulty to discover, and effectually to distinguish, the true motions
of particular bodies from the apparent; because the parts of that immovable space, in which those
motions are performed, do by no means come under the observation of our senses.}
\end{displayquote}

And he continues to comfort himself and also the worried reader\footnote{Newton concluded that identifying the true motions in absolute space is a fundamental problem. He promised to solve that, but he never mentioned it in Principia again! The Scholium problem was largely ignored for nearly two centuries.}:

\begin{samepage}
\begin{displayquote}
\emph{
Yet the thing is not altogether desperate; for we have some arguments to guide us, partly from
the apparent motions, which are the differences of the true motions; partly from the forces, which
are the causes and effects of the true motions.
}
\end{displayquote}
\end{samepage}

\begin{figure}[h]
\centering
\includegraphics[scale=0.4]{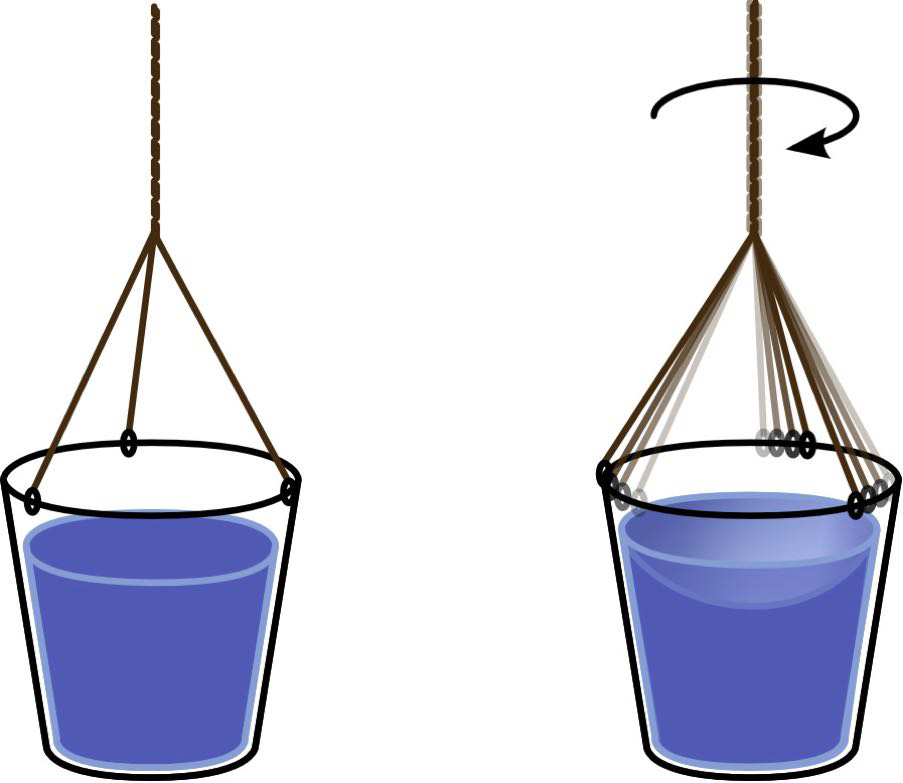}
\captionsetup{width=0.8\linewidth}
\caption[Newton's bucket thought experiment]{\small Newton's bucket. At first, the bucket is stationary (the left picture) and then both the bucket and the water inside are rotating (the right picture). In explaining the concave shape of the water in the right picture, one invokes something external to the bucket and the water inside, Newton's absolute space or inertial frames of reference. Picture taken from \cite{Mercati.2018}.}
\label{bucket}
\end{figure}

One such argument is the well-known Newton's bucket thought experiment illustrated in the Fig. \ref{bucket}.

Imagine a bucket half-filled with water hung by an elastic rope, capable of being wound by spinning the bucket. As the bucket is initially stationary, the surface of the water remains plain. Now as the rope starts untwisting itself, it sets the bucket spinning. As the bucket is spinning, the water at first remains static and its shape does not change. Finally, the water inside the bucket starts following the motion of the vessel and then its surface gradually becomes concave upwards. If we suddenly stop the rotation of the bucket the water continues its circular rotation and retains its concave shape. If we wait for long enough, the water gradually becomes stagnant. 

What this thought experiment shows is a phenomenon (change in the shape of the surface of the water) that needs explanation. From the experiment it follows that this phenomenon cannot depend upon the relative motion of the vessel and the water inside it: We see that the water does not immediately respond to the rotation of the vessel as the vessel is set in its motion and then later stopped, the water retains its previous shape. Thus what is the reason? What lies behind this observable phenomenon? According to Newton, the rotation of the water with respect to the absolute space is the reason. In this way, Newton hopes to convince us to endorse the reality of absolute time, even though it remains outside the domain of our direct empirical knowledge.

This sounds quite suspicious. Certainly, no wise man would doubt the reality of the change in the surface of the water. The problem is linking a `visible' phenomenon to something `invisible' and external. Should this phenomenon not be explained based on the relations between the observable entities within the universe? We saw that the relative state of the water and the sides of the bucket cannot be the cause, but this does not indicate that such a relational cause cannot exist at all. This will be addressed later in this and the next chapter. We may in fact say that this thesis explores this possibility and provides a review of what has been developed in that direction.

Newton, however, was satisfied and could carry on with developing his theory of motion withstood for more than two centuries. This conceptual flaw of his theory did not impede the ever-increasing practical usage of Newton's equations. After all the distant stars determined a proper reference system and the rotation of the earth with respect to them, a good measure of absolute time (called sidereal time). It is usually the case in the history of physics (and probably other fields) that when the generation of revolutionaries who establish a new paradigm grow older, the younger ones who come next are more practical and pragmatic and do not engage in these rather philosophical discussions.\footnote{
The same thing happened in the development of quantum mechanics to some extent. After the development of the Copenhagen interpretation, between the years 1930 and 1970, except for EPR paper and Schrodinger's reaction to it, as well as Everett's 1957 paper and Bohm's approach in 1952, until the full implications of Bell's theorems and Aspect's experiment, many younger physicists had no interest in the conceptual problems of the theory. However, unlike Newton's theory, the foundational problems of quantum mechanics are still unsolved.}

This is perhaps why the conceptual defect of Newton's theory went largely unnoticed for nearly two centuries until Ernst Mach revived this debate in the second half of the 19th century.

\section{Leibniz's relationalism}
Gottfried Wilhelm Leibniz, a prominent figure in the history of philosophy and mathematics, as well as the debates on Newtonian mechanics, was one of the first who criticized Newton's concepts of absolute space and time. In a heated exchange with Newton's representative, Samuel Clarke \cite{Leibniz.17156}, he mounted a condemnation of Newtonian philosophy. Leibniz advocated a relational understanding of space and time: Space is nothing but an order of co-existence of bodies, and time is a similar order of succession of events.

Leibniz was epistemologically a rationalist and believed that `reason' is our reliable main source of gaining knowledge. He then based his philosophy on his great \emph{Principle of Sufficient Reason}. Understanding this principle and Leibniz's philosophy, in general, helps us to lay a solid background of relationalism that proves to be of value in our future discussions. Leibniz's pivotal principle is:

\begin{samepage}
\begin{displayquote}
\textbf{The Principle of Sufficient Reason (PSR)}:
\emph{
...no fact can ever be true or existent, no statement correct, unless there is a sufficient reason why things are as they are and not otherwise - even if in most cases we can't know what the reason is.
}\cite{Leibniz.1714}
\end{displayquote}
\end{samepage}

In a nutshell, this principle encourages us to keep searching and do not easily content ourselves by taking things for granted. The term `sufficient reason' definitely needs more clarification. In the context of relationalism, a reason must be given in terms of the inner relations between the elements of the universe. In that respect, invoking external causes to explain the phenomena that we directly observe in the universe, violates this principle. In the example of Newton's bucket, we have an unsolved phenomenon, the concave shape of the water in the rotating bucket, and there must be a \emph{sufficient} reason behind that.

Newton's absolute space and absolute time are similarly in direct opposition to the above principle: Why is the universe located at a certain point in absolute space and not anywhere else? Why are the physical processes of the universe unfolding at `this' moment and not any moment sooner or later? Absolute space and time prompt us to ask these questions but render any possible solution unattainable. The Principle of Sufficient Reason points out this very flaw in the Newtonian understanding of the universe.

The Principle of Sufficient Reason is at the heart of relationalism from which various other principles stem. A direct consequence of the PSR was stated by Leibniz as the \emph{Principle of the Identity of Indiscernibles (PII)}. PII states that if two elements share the same set of features such that there are completely indistinguishable, they are in fact \emph{identical}: There are not two elements but only \emph{one}. Discernibility is to be stated in terms of all the \emph{relations} of the elements to other entities. What this principle suggests is that in the entire history of the universe, no two things have the same set of relations to the rest of the universe. In other words, every event, every structure in the vast expanse of the whole universe is unique. Entities of the universe might be similar, but never identical, as they are always distinguished based on their totality of relations to the whole.

Another major consequence of PII is that it undermines the concept of `symmetry'. Physical symmetries are widely used in theories, from Galilean symmetry in classical mechanics to Poincaré symmetry in relativity and supersymmetry in high energy physics. The key picture behind symmetries is to change the physical systems such that the dynamics of the system does not alter. An example is the translation of a closed system in space. However, it is very suspicious: we claim to change the physical state of the system while not changing anything observable! In the spirit of PII, how can we even say we have transformed anything if all properties of the system remain indistinguishable and unchanged? We will elaborate on this remark later and see that in relational physics the concept of physical symmetries of this type is simply meaningless. Physical symmetries of this type can only be `emergent' and defined with respect to the whole universe. The only symmetry we have is \emph{gauge symmetry} or rather, \emph{gauge freedom}. Some even use the term \emph{gauge redundancy}\footnote{
This expression can be problematic because of the negative connotation of ``redundancy''. Some argue that gauge is more than just a mathematical redundancy \cite{Rovelli.2014}. While I personally agree to that, I think that the concept of gauge eventually boils down to just a valuable mathematical language for describing nature. It is empty of any observational content or physical interpretation.
}. These symmetries are mathematical in nature and reflect our mere freedom in choosing physically equivalent, but mathematically distinct variables to describe nature. Gauge transformations are not tied to any physical interpretation like those we mentioned in the case of physical symmetries.

Despite his deeply suggestive principles, we can say that Leibniz lost the argument to Newton and Clarke. Mere philosophical reasoning could not be convincing while Newton had already developed a theory based on his standpoint. A relational physical theory requires a more sophisticated mathematical machinery that Leibniz could never have dreamed of. We will see that a relational theory of mechanics is possible using the language of gauge theories which is much more advanced than the mathematics available at that time. Although being a great mathematician, the co-founder of calculus could not cast his profound philosophical insights into mathematical expressions and build a theory upon them. Fortunately, that feat was finally completed a little more than two centuries later, the story of which is the main content of this thesis.

\section{Mach's critique}
Following Ernst Mach's demise in 1916 only a few months after the completion of the general theory of relativity in Berlin, in his obitary Einstein wrote:

\begin{samepage}
\begin{displayquote}
\emph{
It is a fact that Mach has had tremendous impact upon our generation of natural scientists, in particular with his historical-critical writings where he follows the evolution of individual sciences with so much love, where he probes practically the most remote brain cells of researchers who broke new paths in their fields.
}\cite{Einstein.1916}
\end{displayquote}
\end{samepage}

Ernst Mach was a 19th-century eminent experimental physicist as well as a key philosopher behind the second wave of positivism. He is considered to be one of the most influential critics of Newton's absolute space and time. In 1883 he wrote \emph{The Science of Mechanics}, a book on the history of mechanics which immensely influenced Einstein's pursuit of general relativity. This book is also an important critique of Newtonian space and time.

Starting with the problem of time, we find these words:

\begin{samepage}
\begin{displayquote}
\emph{
It is utterly beyond our power to measure the changes of things by time. Quite the contrary, time is an abstraction
at which we arrive by means of the changes of things; made because we are not restricted to
any one definite measure, all being interconnected.
}
\end{displayquote}
\end{samepage}

This is very suggestive. What do we really refer to when talking of time? After reflecting on the concept of time and freeing ourselves from the deep-rooted Newtonian absolute time, we find Mach's view quite tenable. It is always the \emph{change} of some thing that we identify as the passage of time, never the other way around. Look at the wristwatches we use all over the world. It is the \emph{change} in the position of the hands of the clock that tells the time. The days and nights are in turn the rotation of the earth around its axix and years and seasons are its rotation around the sun, all of which are essentially the \emph{change} in the earth's position.

If we lay bare the core of our usual way of formulating the laws of nature with respect to an external time, we will see that we are actually formulating the \emph{changes} of some physical system with respect to the \emph{changes} of some clock (which is also a physical system). Time, at least in the domain of classical physics\footnote{
The status of time as an independent concept is likely to be elevated in passing from classical physics to quantum mechanics. We will touch on this delicate matter (of which we have yet no proper understanding) in the second part of this thesis.
}, is a non-existent imaginary entity, resting on the changes in the real observable entities, or more appropriately, \emph{relations}.

Therefore, time should be understood as the total change in the whole universe. As Julian Barbour puts it, \emph{the universe is its own clock}. It is absolutely meaningless to say how fast the whole universe is evolving. The universe determines time. However, this question is meaningful if we talk of a certain subsystem in the whole universe. This is why we can tell the difference if we watch a movie in slow-motion: We compare the rate at which the changes unfold on the screen with the changes of other parts of the universe, i.e., our watch, our personal sense of the flow of time, etc.

It should be noted that there are two aspects to the concept of time: \emph{duration} and \emph{direction}. So far, our analysis has addressed the former. Duration is the measure of the passage of the time we attribute to physical processes. The latter is related to the direction along which the universe evolves and changes. In the Newtonian framework, absolute time determines both of these aspects. The second aspect is much more delicate and is tightly bound to the problem of time's arrow(s). The problem and its possible resolution in relational physics will be explained in the second part of this work.

Accompanying Leibniz and other advocates of relationalism, Mach was opposed to the concept of absolute space as an invisible container. Everything must be characterized in terms of the visible entities of the universe, in terms of the relations between parts of the whole universe. However, unlike Leibniz, Mach did not attempt to base his view on some rich metaphysical foundation. Instead, being an influential figure in the development of positivism, as well as a competent experimentalist, Mach proposed that we stay within the boundaries of observables and do not rush into going beyond what observations grant us in hope of finding a better explanation. Mach's anthem is that we have a phenomenon at our disposal, and as scientists, we have to proceed to explain that phenomenon in terms of the internal structures rather than external causes. In his words on the relational understanding of space:

\begin{samepage}
\begin{displayquote}
\emph{
When we say that a body $K$ alters its direction and velocity solely through the influence of another body $K'$, we have inserted a conception that is impossible to come at unless other bodies $A, B, C, \cdots$ are present with reference to which the motion of the body K has been estimated.
}
\end{displayquote}
\end{samepage}

What would Mach say as regards the bucket thought experiment by which Newton felt that he had cleared up all these suspicions? Mach remarked that Newton's mechanics has always been put to test by relying on fixed stars to represent absolute space and sidereal time as the measure of absolute time. Thus, why not consider the possibility that the concave shape of the water in Newton's bucket is the result of background stars, or in general, \emph{all the other masses} in the universe? This is indeed a radical proposal. In another well-known quote from Mach we read:

\begin{samepage}
\begin{displayquote}
\emph{
No one is competent to say how the experiment would turn out if the sides of the vessel increased
in thickness and mass till they were ultimately several leagues thick.
}
\end{displayquote}
\end{samepage}

What can we make of all these audacious, but not entirely clear remarks? The honest answer is that it is not very clear. Unfortunately, Mach never stated his ideas precisely enough and it did led to some confusions. Anyway, the correct and precise criterion was found about one century later by Julian Barbour which will be covered in Sec. \ref{contemporary}. What is worth noting, however, is that Mach did tentatively suggest that we include the \emph{whole universe} in our considerations. We will see that this insight is exactly what we have to do if we aspire to have a relational understanding of the universe: a relational theory must necessarily be a theory of the whole universe.

Einstein definitely played a significant role in popularizing Mach's ideas. Having read his works as a young man, Mach's critical thinking made a significant impact on Einstein's perspective. Even it was Einstein who dubbed the term \emph{Mach's Principle} in his 1918 paper, \textit{On the Foundations of the General Theory of Relativity} \cite{Einstein.1918}. We will come back to the story of Einstein and Mach's Principle later. It is fascinating and deserves some detailed remarks in its proper place.

\section{Lange and Tait's contributions}\label{Tait's problem}
Mach's critique captured Lang and Tait's attention to ponder the Scholium problem in a new light \cite{Barbour.2010}. Lange considered three freely-moving particles and took their successive positions to define a frame of reference based on the material particles alone. He called it an \emph{inertial system}. Ever since this term has become standard and replaced Newton's absolute space and time. But the point is that in the textbooks an inertial frame of reference is defined as a frame in which Newton's first law holds. But unfortunately, they do not get into the details of their construction based on only observational information (which are relational). After all, the core of the Scholium problem is exactly about determining the state of motion and rest of particles given only the observational data. Lange's work was an important step towards the resolution of the problem. We will see that the problem can be solved in this way, but only partially.

A cleaner procedure for the determination of inertial frames had been proposed by Peter Guthrie Tait in 1883 \cite{Tait.1883}. A generalization of his work in a more illuminating context can be found in Barbour's 2010 paper on Mach's Principle \cite{Barbour.2010}.

Tait assumed that size is given. Alhough we essentially follow his approach, we take size relational. Suppose a system of $N$ point particles. The observational information is encoded in `snapshots' of the configuration of particles at certain unspecified instants. In accord with relationalism, only relational data are physically meaningful, i.e, only ratios matter. These ratios are between separations of particles, or in other words, the angles between particles as seen from other ones. These angles define the `shapes'. Only shapes observed within the whole system can be physically allowed.

Now we do some counting. In an arbitrary Cartesian system, we can assign $3N$ coordinates to the $N$ particles in three-dimensional space. However, three of them determine the system's center of mass position which is unobservable from the relational point of view, 3 more are related to the total orientation of the system which is also unphysical. One last number determines the total scale of the system. As ratios are only relevant, this too must be omitted. In fact, translating, rotating, and rescaling the total system does not change the relational data. The $3+3+1$ data we arrived at are exactly linked to the action of these groups on the configuration space. Thus, we are eventually left with $3N - 3 - 3 - 1 = 3N - 7$ observable data.\footnote{One might say that for the $N$-body system there are $\frac{N(N-1)}{2}$ possible numbers for each pairs of the particles. This counting, however, neglects the certain constraints that the configuration of the particles must satisfy. The simplest one is the triangle inequality for three particles. If $N \geqslant 5$, then there are also equalities they have to satisfy which greatly reduces the number of independent numbers.} For instance, if there are three particles, the relational data can only be the two angles that define the triangular shape of the system which is indeed correct: $3\times 3 - 7 = 2$. If this three-body system is the entire universe, the size, place, and orientation of the triangle in the three-dimensional space are not observable.

Now Tait's problem is that \emph{how many snapshots} are needed to determine the inertial motion of the particles. More generally, we can ask the same questions for determining the motion of particles according to Newton's law for a certain interaction.

The answer is not that hard. Given the configuration of the particles, we can choose one of them (particle 1) to be perpetually at rest. This fixes the origin of the frame. We need a standard rod to define distance. We can take the separation between an arbitrary particle (particle 2) and particle 1 at that specific instant to be of distance 1 by definition.\footnote{We assume that the two particles never collide.} To define an orientation, we can take the direction of the second particle from the first one at that instant to be one of the axes ($x$), and the direction of the motion of the particle 2 the other one ($z$). With these two axes, an orthogonal coordinate system is defined. Finally, the motion of particle 2 can measure the time. Based on our construction, in this frame the coordinate of particle 1 is $(0,0,0)$, and that of particle 2 is $(1,0,t)$. Now all that is left is the information of the rest $N-2$ particles. To determine their trajectories completely, we need 6 numbers for their positions and velocities with respect to the frame of reference we built with the other two particles. The coordinates of the other particles are of the form $(x_I,y_I,z_I) + (u_I,v_I,w_I)t$. Thus $6(N-2) = 6N - 12$ data are needed.

It follows that given only two snapshots is not enough to determine the evolution of the system. For each snapshot provides $3N-7$ data. Also, we do not know `when' these two snapshots were taken. Thus, each of them contains $3N-8$ usable data and there are $6N-16$ data in two snapshots. We are short of 4 more numbers to determine the entire dynamics in the frame we constructed from the particles. Three of the lacking data are related to the angular momentum of the system, one of them is the value of the overall expansion and contraction of the system, called \emph{dilatational momentum}\footnote{The dilatational momentum in an inertial frame of reference is $D = \sum_{I = 1}^{N} \vec{r}_I.\vec{p}_I$. This quantity is the momentum conjugate to the total size of the system. We will define and clarify its role in detail in the next chapter.
}. These conserved quantities cannot be determined by the observable relational data alone. In the standard Newtonian mechanics, we determine them in an inertial frame of reference. Here, however, our frame is the $N$-particle system itself. Angular momentum measures the change in the total orientation. As total orientation is completely unobservable from the relational point of view, its change is also irrelevant. The same thing holds for the dilatational momentum.

This is very significant. If we insist on only using the available data in the entire system (which is in fact a reasonable expectation), two snapshots are not enough for solving the simplest inertial motion of the non-interacting particles. If the number of particles is large enough ($N \geqslant 5$), three snapshots will be enough.

We could have left the size of the system untouched (as Tait did). Assuming that an external rod exists, each snapshot contains $3N-6$ data (only the total orientation and the location of the system remain undetermined). We can still use two particles to construct a frame. This time, however, the distance between particles 1 and 2 is physically relevant and is to be determined. Thus, we would need $6(N-2) + 1 = 6N - 11$ data. Two snapshots contain $2(3N-6) - 2 = 6N - 14$. Now there is a 3-number shortfall which is contained in the angular momentum of the system.

Although we only considered the inertial motion, an analogous calculation can be carried out for a system of interacting particles. But it is more complicated as we cannot simply fix the motion of the particles to define an inertial frame of reference due to the particles' non-inertial motion.

The important lesson we must take from Tait and Lange's work is that Newtonian mechanics can be made relational, but at the cost of needing more than two sets of data. This is a matter of \emph{causality}, not \emph{epistemology}. Newton's absolute space and absolute time were indeed plagued with an epistemological defect. But what we learned from the above discussion is that the concept of an inertial frame of reference can be built on only the observational data and it can take the place of absolute space and time. What we lose in taking this step forward is the predictive power of the theory: We would be needing more initial information to solve the equations of motion.

\section{The contemporary relational physics}\label{contemporary}
Before we proceed to today's relational physics, there is an important link in the chain of the works done on relationalism and that is Henri Poincaré's contribution to the discussion. He was sympathetic to the critique of Newton's absolute space and time and pointed out the repugnancy of attributing states of motion to the whole universe in invisible space. He asked an illuminating question: What precise defect, if any, arises within Newtonian dynamics from its use of absolute space?

Based on our discussion in the previous section, the answer should be clear: Newton's absolute space gives the theory an artificial predictive power, without which more information would be needed for predicting the evolution of the system if we insist on only the physical information.

Poincaré too noted that the separations between the particles in a $N$-body system, $r_{ij}$ and their rates of change, $\dot{r}_{ij}$ are the physically meaningful data.\footnote{
In Poincaré's approach, an external clock and a measuring rod are taken for granted. But this does not affect the conclusion he arrived at. One could also take the time and size to be relational, as in our account of Tait's problem.
}
He then pointed out that those are not enough to determine the evolution uniquely. As we discussed, the information of the total angular momentum of the system is not encoded in the relational data. We did the counting in the previous section. We quickly recap our result by giving the simple example of a two-body Kepler system, like the motion of earth and sun. Let the initial change in their separation be zero, that is $\dot{r} = 0$. But this does not determine the evolution of the system uniquely. All possible motions can follow: the two particles collide, or they orbit around each other in a circular or elliptical motion. We could solve the equations of motion in this way, however, by knowing some of the second derivatives of the separations $\ddot{r}_{ij}$ as well. This unpredictability rests exclusively on the difference between relative quantities and those with respect to an inertial frame of reference. This is the defect of Newton's mechanics.

What is worth emphasizing is Poincaré's insight that helped to identify the precise manner in which Newton's theory fails to be relational. It is about the \emph{amount of observable information} needed for the evolution to be determined uniquely. 

Despite being on the right track to formulating relational physics, Poincaré was pessimistic about the prospect of finding a relational description of nature. The failure of Newtonian mechanics to be relational for subsystems led him to regretfully conclude that nature does not respect our philosophical aspirations.

But there is no reason for hopelessness. Mach already told us to look at the whole universe. Following this intuition, it is possible to find a relational theory of the whole universe, which is precisely what Barbour and Bertotti showed in their seminal paper \cite{Barbour.1982}. The method they discovered, now called \emph{best-matching}, is a mathematical procedure that helps to construct relational dynamics irrespective of the configuration space. Best-matching and relational particle mechanics will be covered in the next chapter extensively and in detail. After nearly three hundred years, finally a consistent relational framework was found that captured the core philosophy of Leibniz, the critique of Mach, and works of all the others who pondered relationalism. Relational field theory was also developed later in the early 2000s \cite{Barbour.2002}. We will cover them in this first part.

Finally, we state what we believe to be the precise statement of Mach's Principle as formulated by Julian Barbour. The starting point is to identify the true physical configuration space. It is done by finding the underlying symmetry group of the theory. For instance, for the $N$-particle system in three-dimensional Euclidean space, in the absence of external orientation, size, and spatial specification, the symmetry group is the similarity group $Sim(3)$ which is the semidirect product of the rotation group $SO(3)$, the translation group $T(3)$, and the rescaling group which is simply the set of the positive numbers that rescale the total size of the system, denoted by $\mathbb{R}^+$: $Sim(3) = \mathbb{R}^+ \ltimes E(3)$, where $E(3)$ is the Euclidean group: $E(3) = SO(3) \ltimes T(3)$. The symmetry group is mostly determined based on empirical considerations. The goal is to identify the directly observable quantities and satisfy the Principle of the Identity of Indiscernibles.

The next step is to quotient the initial configuration space in which we do the calculations by the symmetry group. The result is the relational configuration space. In the $N$-body system the configuration space can initially be taken to be $R^{3N}$: 3 coordinates for each of particles. But due to the redundancy of this space, orientation, size, and positions of the particles in the space must be eliminated by quotienting.

Finally, for the theory to have maximal predictive power, we demand that only an initial point and its change in the reduced (quotiented) configuration space suffice to determine the evolution of the system uniquely. Hence, we define the Mach-Poincaré Principle \footnote{As Poincaré's illuminating discussions are of significant relevance to our criterion of Machianity, we included his name in the principle, in accord with the terminology of \cite{Mercati.2018}.}:

\begin{samepage}
\begin{displayquote}
\textbf{Mach-Poincaré Principle}:
\itshape
Given the configuration space $Q$, and the underlying symmetry group $G$, we identify the quotient of the configuration space with respect to the group $G$ as the relational configuration space: $Q^R = \sfrac{Q}{G}$. Then the following are respectively the strong and weak form of Mach's Principle \normalfont \cite{Barbour.2010}\itshape:
\begin{enumerate}
\item Specification of an initial point $q \in Q^R$ together with a direction $\delta q$ in $Q^R$ at $q$ defines a unique curve in $Q^R$.

\item Specification of an initial point $q \in Q^R$ together with a tangent vector $\dot{q}$ in $Q^R$ at $q$ defines a unique curve in $Q^R$.
\end{enumerate}
\end{displayquote}
\end{samepage}
\normalfont

Some remarks should be made on the distinction between the strong and the weak form. The distinction is made based on Mach's philosophy of time as a measure of change. In this light, the strong form of the principle respects the relational understanding of time as well as the relational structure of the configuration space.\footnote{Some refer to Mach's temporal relationalism as \emph{Mach's second principle}, and his spatial relationalism as \emph{Mach's first principle}.} Explicitly, the set of admissible initial conditions in the first principle has one less datum: the directions in the configuration space provide us with one less piece of information. For instance, if the relational configuration space is represented by a two-dimensional sphere, as in the case of 3-body system, a direction at each point is determined only by one number. A tangent vector, or in other words, the rate of change of a point, is determined by two numbers. As the result, an initial tangent vector provides us with more information, but at the cost of positing an external clock with respect to which the magnitude of the tangent vector can be physically meaningful.

Although the concept of relational time is very precious to us from the relational point of view, nature is unfortunately not sympathetic to us. There are some difficulties in constructing a totally relational physics. We cannot have both a relational time and a relational size. One solution could be to eliminate size by using the change of which to define an external time. The downside is that only the weak form of the principle would be satisfied. We will discuss this problem more properly later in the second part.

\section{Why relationalism?}
Should we take relationalism seriously? If yes, why?

First of all, I believe one should not dismiss metaphysical thinking and trying too much to find objective reasons for proceeding in science can be detrimental to its progress.\footnote{We should criticize the ``Shut up and Calculate!'' attitude in this regard.} I do not mean to suggest that any personal preference and philosophical view can be blindly allowed to contribute to science. But if we study the history of science and the great men behind the great discoveries, it becomes quite hard to separate logical objective reasoning from personal views. Newton's advocacy of absolute space and absolute time, Nils Bohr's background and its influence on his approach to quantum mechanics, and Einstein's critique of quantum mechanics are some examples. Einstein's enthusiasm for Mach's ideas - which is in fact relevant to our discussion - is also a revealing example. After Einstein found the correct field equations in November 1915, the non-Machian nature of his theory based on his definition of Mach's principle bothered him so much that he set out to modify his theory to incorporate Mach's principle. He worked for some years on this program until he gradually became disappointed with it. Why would Einstein do that in the first place if it were not for his ingrained passion for Mach's ideas? Nevertheless, it is noteworthy that one must be very cautious in indulging in subjective thinking in science. After all, the outcome of science must be based on objective facts.

Thus, one support for relationalism has certainly been the personal `caring' about understanding the universe as a whole without reference to something external; a universe as a totality which includes its causes effects altogether.

It is not the only thing we can hope for. Shape dynamics as a successful classical relational theory, although a very young field, has already brought us many interesting facts and there are many hopes that it gives us some hints on the problem of quantum gravity. One notable success of the theory is the possible resolution of the old problem of the origin of the arrow of time which will be discussed in the second part of this work.

Philosophically, there are some reasons for taking relationalism as a guiding principle seriously. First of all, as relationalism describes nature in terms of the inner relations between its constituents, not against a background structure, it is more restrictive and hence more predictive. We will see that both in the case of particle dynamics and geometrodynamics, implementing Mach-Poincaré Principle imposes some restrictive constraints on the whole system. In the case of particle dynamics, these are the vanishing of the total angular momenta and linear momenta. Similarly, relational geometrodynamics demands the whole spacetime be globally hyperbolic, CMC foliable, and spatially closed which renders the theory more restrictive (thus more predictive) than the standard general relativity.

Predictive power is an important feature in science, the lack of which leads scientific pursuit astray. Some suggest it is exactly one of the reasons for the current stagnancy of the foundations of physics in the past decades \cite{Smolin.2006}.

Finally, I would like to elaborate on the strong bond between relationalism and the aspiration for a theory of the whole universe. We saw that how including the whole universe helped to state the Machian criterion more precisely. The point is that a relational theory must be a theory of the whole by nature: all relations play a dynamical role in such a theory and restricting the scope of our investigation to parts of the universe can only be approximate. This is very illuminating. In the standard Newtonian paradigm\footnote{Newtonian paradigm is different from Newtonian physics. Newtonian physics was overthrown by the advent of modern physics in the 20th-century. But Newtonian paradigm as a general way of doing science still lives on.}
the theory is for subsystems only and then we extrapolate it to larger and larger systems in our attempt to have a theory of the whole. This method - which is only possible because we posit an external background against which we study a certain system by neglecting the environment - has led to some difficulties in searching a fundamental theory of the whole \cite{Smolin.2015}. Relationalism is quite the opposite to that: we start with a theory of the whole and under certain conditions, we can study subsystems of the universe approximately. \emph{The whole universe is its own background}; relational physics describes the systems within the universe against this background.

It is also true the other way around. If a fundamental theory is to apply to the whole universe, it \emph{must} be background independent and relational, otherwise the theory would include some concomitant external causes and elements. This contradicts the fundamentality of such a theory since a fundamental theory must explain all causes by definition.

The distinction between physics of the whole universe and physics in laboratory cannot be overemphasized. Relationalism is the portal of a theory of the whole.

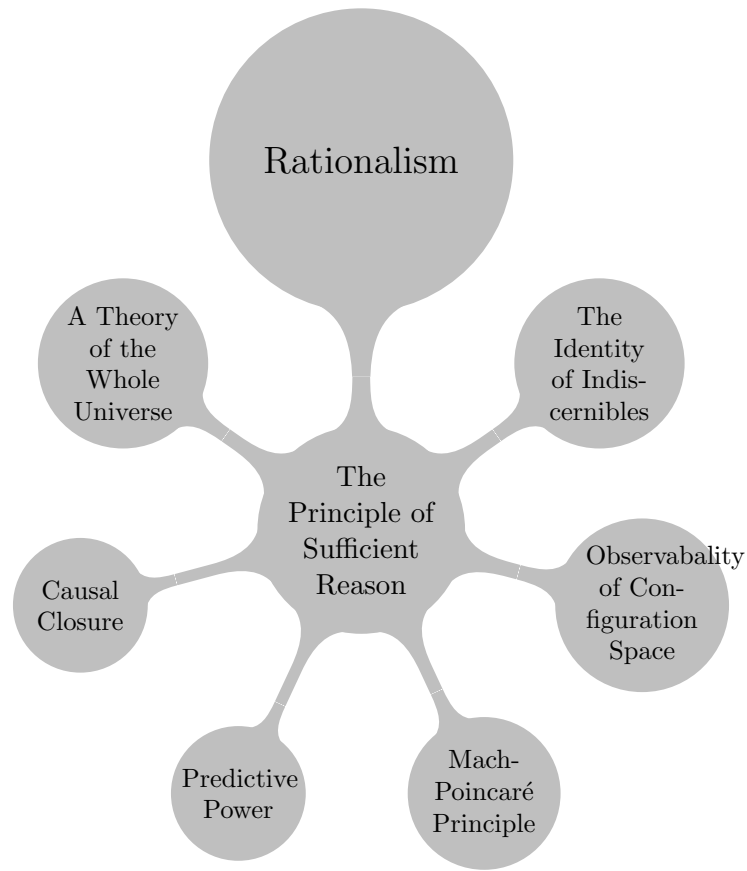
\begin{figure}[h]
\centering

\begin{tikzpicture}[mindmap,
  level 1 concept/.append style={level distance=140, sibling angle=0}
  ,level 2 concept/.append style={level distance=110, sibling angle=50}
]

  \begin{scope}[mindmap,every node/.style=concept, concept color=lightgray, text=black]
    \node [concept] {Rationalism} [clockwise from=-90]
      child {node [concept] (SR) {The Principle of Sufficient Reason} [clockwise from=35]
      child {node [concept] (PII) {The Identity of Indiscernibles}}
      child {node [concept] (O) {Observabality of Configuration Space}}
      child {node [concept] (MPP) {Mach-Poincaré Principle}}
      child {node [concept] (P) {Predictive Power}}
      child {node [concept] (E) {Causal Closure}}
      child {node [concept] (T) {A Theory of the Whole Universe}}
  };
  \end{scope}
  
\end{tikzpicture}

\captionsetup{width=0.8\linewidth}
\caption[Mind map of concepts in relationalism]{\small The mind map illustrating different concepts in relationalism and relational physics all stemming from the Principle of Sufficient Reason.}
\label{mindmap}
\end{figure}

All the results that follow from relationalism are shown in Fig. \ref{mindmap}. They all emanate from the cornerstone of relationalism, the Principle of Sufficient Reason.\footnote{Julian Barbour suggested to me to call this illustration `the map of God's mind'. We may say it is how Leibniz's God might have thought when creating the universe.}

We summarize the key features that make a compelling case for relationalism:

\begin{samepage}
\begin{enumerate}
\item Relational physics is more restrictive and hence more predictively powerful.
\item Relational physics describes the universe in terms of the directly observable real quantities. Moreover, the nexus of causal relations between the entities stays within the observable universe.
\item Relational physics is by construction applicable to the whole universe and makes it possible to have a fundamental theory of the whole.
\item Relational physics has thus far led us to many enthralling results which spur even more studies.
\end{enumerate}
\end{samepage}

%% file: Chapter3.tex
\chapter{Best-Matching and Relational Particle Mechanics}
\section{Building the intuition behind best-matching}\label{intuition}
In the last chapter, we tracked the story of relationalism and found our way to the Mach-Poincaré Principle. We must now proceed with constructing a proper mathematical framework for implementing that principle in various dynamical theories.

As we mentioned, Barbour and Bertotti came up with a novel approach in 1982 \cite{Barbour.1982}. Their method is now called best-matching. The idea behind that is rather simple. It is about finding the quantitative `difference' between certain configurations in terms of their innate properties without referring to anything external. We illustrate the procedure with a simple 3-particle model in two-dimensional space. We assume that the particles are distinguishable and label them with some distinct numbers. Consider two snapshots of a 3-particle system, represented by two different triangles in space. For the moment, we assume that an external size is given and that only the separations between the particles are observable. How do we compare these two configurations with one another without locating them in a fictional background space?

Given the two configurations located arbitrarily in the Euclidean space, a natural measure of the `distance' between them can be defined to be simply the sum of the Euclidean distances between each particle of the configurations (each vertex of the triangles in our illustration). What best-matching does is to use one of the triangles as the reference, and then transform the other one all over an imaginary space\footnote{It might seem that we are implicitly using the redundant Newtonian absolute space here. The point is that the best-matched difference that is established in this method is completely independent of the initial space and is only dependent on the relational properties of the two configurations. Mathematically, we use this background space to define the gauge variables. Eventually, all the physical evolutions and the ontological configurations reside in the reduced space we defined in Sec. \ref{contemporary}.} and rotate it under all arbitrary rotations until the Euclidean distance between the two configurations is minimized. This distance is the \emph{best-matched distance} and the resultant configuration in this procedure is called the \emph{best-matched configuration}.

Interestingly, this method solves the Scholium problem by establishing the concept of equilocality in an alternative way. The dynamical history of the whole universe is nothing but a succession of the relational configurations in the corresponding configuration space. The significance of best-matching is that it determines exactly how the configurations must follow one another. We no longer need to worry about the location of a certain system at different times and try to conceive that in absolute space (or an inertial frame of reference). Best-matching requires all the successive configurations to be the best-matched ones.\footnote{There are some subtleties here. We use the Euclidean distance between the particles in the procedure. However, in general, the dynamics of the configurations can be more complicated than that as there might be some interactions between particles. This does not affect our intuitive approach however, as we shall see below that interactions can be implemented through the `metric' we define on the configuration space for defining distance.\label{caveat!}}

Several remarks follow. First, we should note that this method is defined for the whole universe. The `snapshots' of the whole universe are to be compared to one another and then \emph{best-matched}. Second, best-matching kills two birds with one stone.\footnote{Actually, we may say it kills infinitely many birds!} All possible motions, from inertial ones to those under different interactions, are all different manifestations of the best-matching procedure carried out with different metrics (as hinted in the footnote \ref{caveat!}. And finally, the third remark is that best-matching imposes certain conditions on the evolution of the whole universe in configuration space (or equivalently on the phase space). For instance, best-matching with respect to translations results in the vanishing of the total linear momenta. Including the rotation group, in turn, enforces the vanishing of the angular momentum of the whole universe.

Mach said that it is meaningless to talk of a spinning universe. We are now in the position to endorse that: The total angular momentum of the whole universe is zero according to our relational approach based on best-matching. The same is true for the vanishing of the total linear momenta which renders the motion of the whole universe hollow, as it should be.

Leibniz's God is without a doubt more rational than to flaunt his mighty power to set the entire universe in motion. The rationality behind nature decrees that the momenta of the whole universe vanish! Relational physics is the rational naturalists' religion that rests on Leibniz's great metaphysical Principle of Sufficient Reason, and best-matching is its ritual. Now we proceed with learning best-matching in detail.

\section{Best-matching is minimizing the incongruence}
\subsection{Jacobi's action}\label{jac}
We now take the first steps to formulate best-matching mathematically. We first consider the simple $N$-body problem and then develop this technique more abstractly in a general case for any given configuration space with a certain symmetry group.

Following William Rowan Hamilton's great principle (1834), known as Hamilton's principle, we know that the Newtonian evolution of the $N$-body dynamical systems can be formulated as the following action:

\begin{equation}\label{standardaction}
S[q] = \int_{q^a_{Ii},t_i}^{q^a_{Ii},t_f} dt \, \left( T(\dot{q}^a_I,\cdots) - V(q^a_I,\cdots) \right) \equiv \int_{q^a_{Ii},t_i}^{q^a_{Ii},t_f} dt \, L.
\end{equation}

Here $t_i$ and $t_f$ refer to the initial and final Newtonian absolute time respectively, $V$ is the potential energy dependent on the particle coordinates, and $T$ is the Kinetic energy of the system, i.e., $T = \frac{1}{2} \sum\limits_{I=1}^{I=N} m_I \dot{q}_I^2$, and $L = T-V$ is the Lagrangian. The dots denote the derivatives with respect to the Newtonian time, and by $\dot{q}_I^2$ we mean $\sum\limits_a (\dot{q}_I^{a})^2$ by virtue of Einstein's summation convention for the spatial indices $a$.

There is a rather less known method, called \emph{Routhian reduction procedure}, by which we can transform the above action into a dynamically equivalent action with an explicit timeless character. See Appx. \ref{jacobi}. The result, known as \emph{Jacobi's action}, is of utmost importance in relational physics and is very suitable for formulating best-matching. Jacobi's action is\footnote{The boundaries of the integral - which are simply the initial and final configurations - are not shown for simplicity.}

\begin{equation}\label{jacobii}
S_J[q] = \int d\lambda \, \sqrt{(E-V(q))T({q'})},
\end{equation}

where $T = \frac{1}{2} \sum\limits_{I=1}^{I=N} m_I {q'}_I^2$, and $E$ is a constant. The role of $E$ is different here from its role in our derivation in Appx. \ref{jacobi}. In the latter derivation, it serves as a supplementing condition and is equal to the momentum conjugate to the Newtonian time. Here, there is no `time' at all, and our starting point is Jacobi's action, not the standard Newtonian action. $E$ is simply a constant that is to be determined experimentally. However, best-matching may restrict this constant term, as we will see. We say more about $E$ in the theory in Sec. \ref{best-matchingparametrizaion}. Jacobi's action can be expressed in the following suggestive way which manifests the timeless nature of the action more explicitly:

\begin{equation}
S_J[q] = \int \sqrt{(E-V(q))\sum\limits_I dq_I^2}.
\end{equation}

This action is the sacred mantra we keep chanting throughout this thesis. We will see how all our classical dynamical theories, from particle mechanics to general relativity, can be conceived and born from a simple Jacobi's action through best-matching with respect to the corresponding symmetry group. I doubt if Carl Gustav Jacob Jacobi ever thought of the profound significance behind this action and the remarkable role it would play in relational physics one day.

\subsection{Introducing the best-matching}\label{introbestmatching}

Let $q^a_I$ represents the coordinates of the $I$th particle in three-dimensional Euclidean space, where $a$ runs from 1 to 3 and $I$ takes the numbers $1,2,\cdots,N$. Under the translation group $T(3)$ acting on the whole system, the coordinates transform as $q^a_I \rightarrow q^a_I + \alpha^a$, where $\alpha^a$ is an arbitrary vector belonging to the three-dimensional space. The action of the rotation group is given by $3\times3$ rotation matrices $\Omega^a_b (\omega^c)$ parameterized by three variables denoting the axis and the angles of rotation: $q^a_I \rightarrow \Omega^a_b \, q^b_I$, where the summation over the repeated index $b$ is omitted according to our convention. Hence, under the Euclidean group\footnote{For the moment, we restrict our attention to the Euclidean group instead of the similarity group for the sake of simplicity. We will shortly develop the best-matching procedure more generally and the relationality of scale can be easily included.} the coordinates transform as

\begin{equation}\label{transform}
q^a_I \rightarrow \bar{q}^a_I = \Omega^a_b \, q^b_I + \alpha^a.
\end{equation}

Following our intuitive account of best-matching in Sec. \ref{intuition}, we `shift' the variables in Jacobi's action (\ref{jacobii}) and replace them with the transformed ones (\ref{transform}):

\begin{equation}\label{jacobiii}
S_J[q,\alpha,\omega] = \int d\lambda \, \sqrt{(E-V(\bar{q}))T({\bar{q}'})}.
\end{equation}

The important point is that the action is now a functional depending on `both' the trajectories in configuration space and the group parameters $\alpha^a(\lambda)$. Note that the action (\ref{jacobiii}) is not `derived' but rather `proposed' as a principle underpinning the relational physics. Its role in satisfying the Mach-Poincaré Principle will be shortly clarified.

First of all, we look at the potential term in the action (\ref{jacobiii}). In general, the potentials used in Newtonian physics are translationally and rotationally invariant, such as the Newtonian potential $V = - \sum\limits_{I < J} \frac{m_I m_J}{r_{IJ}}$, where $r_{IJ}$ is the separation between the particle $I$ and $J$.\footnote{We omitted the Newtonian gravitational constant for simplicity. Anyway, it is not important and it can be eliminated by simply redefining our unit of time.} As a result, we need to only calculate the kinetic term.

\begin{equation}
T(\bar{q}^{'2}_I) = \frac{1}{2} \sum\limits_{I} m_I \bar{q}^{'2}_I = \frac{1}{2} \sum\limits_{I} m_I \left( \Omega q'_I + \Omega' q_I + \alpha^{'}_I \right)^2,
\end{equation}

where by $\Omega'$, the derivative of the rotation matrix $\Omega \left(\alpha(\lambda),\beta(\lambda),\gamma(\lambda)\right)$ with respect to $\lambda$ is meant.

Because of the invariance of the Euclidean inner product, we can write the squared term in the above expression in the following way

\begin{equation}\label{turningtoliealgebra}
\left( \Omega q'_I + \Omega' q_I + \alpha' \right)^2 = \left( q'_I + \Omega^T \Omega' q_I + \Omega^T \alpha' \right)^2.
\end{equation}

The operator $\Omega^T \Omega'$ appearing in this term is related to `Lie algebra' of the rotation group.\footnote{For a brief account of Lie groups and Lie algebras see the Appx. \ref{lie}} Working with the matrix representation of the group elements, we can represent the group elements as the exponent of an element of the algebra and the numbers $\omega^a$ which can be identified as the components of that specific element:

\begin{equation}
\Omega = \exp{ \left(\omega^a O_a\right)}.
\end{equation}

Thus,

\begin{equation}
\Omega^T \Omega' = \frac{d}{d\lambda} \log{(\Omega)} = \frac{d}{d\lambda} \left(\omega^a O_a\right) = \omega^{'a} O_a
\end{equation}

Where $O_a$ is the matrix representation of the elements of the Lie algebra. Explicitly, $O_a$'s can be taken to be the matrices

\begin{equation}\label{explicitrotgen}
O_1 =
\begin{pmatrix}
0 & 0 & 0 \\
0 & 0 & 1\\
0 & -1 & 0
\end{pmatrix},\quad
O_2 =
\begin{pmatrix}
0 & 0 & -1 \\
0 & 0 & 0\\
1 & 0 & 0
\end{pmatrix},\quad
O_3 =
\begin{pmatrix}
0 & 1 & 0 \\
-1 & 0 & 0\\
0 & 0 & 0
\end{pmatrix}.
\end{equation}

If we first extremize\footnote{Actually, we are effectively `minimizing' the action with respect to the group parameters. It is easy to show that the transformation (\ref{transform}) increases the value of the action. After all, it takes some effort to set the entire universe in a certain translational or rotational motion!} the action (\ref{jacobii}) with respect to the parameters $\omega^{a}$ and $\alpha^a$, it will do exactly what we described in Sec. \ref{intuition}: it aligns the successive snapshots of the whole universe in a such a way that the difference between them (given by Jacobi's action) is minimized. However, because of the invariance of Euclidean inner product with respect to Euclidean group, and our assumption that the potential depends on only separations and thus also invariant under Euclidean group, only the derivatives of the group parameters appear in the action. As the result, we vary the action with respect to the parameters $\omega^{'a}$ and $\alpha^{'a}$ directly.\footnote{The sharp-witted reader can certainly spot a loophole here. Since when do we vary the action (or any other functional) with respect to only the derivative of a function appearing inside it? This is not what `variation' means in the calculus of variations. I ask the readers to be a bit patient and let me continue with this procedure to arrive at some illuminating results. I will address this issue at length in the following section.} Thus, we have the following `best-matched action'.

\begin{equation}\label{jacobiiii}
\begin{gathered}
S_J^{bm}[q] = \left.S_J[q,\omega',\alpha']\right|_{\omega^{'a} = \omega^{'a}_0, \alpha^{'a} = \alpha^{'a}_0} = \int d\lambda \, \sqrt{(E-V(\bar{q}))T(\bar{q}^{'2})},\\
\left.\frac{\delta S_J}{\delta \omega^{'a}}\right|_{\omega^{'a}_0} = \left.\frac{\delta S_J}{\delta \alpha^{'a}}\right|_{\alpha^{'a}_0} = 0.
\end{gathered}
\end{equation}

The parameters $\omega^a_0, \alpha^a_0$ are the ones that minimize the previous action (\ref{jacobiii}), and $\bar{q}^a_I(\lambda)$ is the `best-matched' trajectory: $\bar{q}^a_I(\lambda) = \Omega(\omega_0(\lambda)) q^a_I(\lambda) + \alpha^a_0(\lambda)$. Thus, the action (\ref{jacobiiii}) depends on the trajectories in the configuration space for any of which, we must first solve the two conditions in (\ref{jacobiiii}) and then insert the `refined' trajectory (the one transformed with the variables $\omega_0$ and $\alpha_0$) back into the action.

Best-matching introduces a gauge symmetry in the action. For any given trajectory, the best-matched action is `gauge invariant' and the gauge group is simply the Euclidean group with respect to which we carried out the best-matching. To see that, the reader can easily verify that

\begin{equation}
S_J[\tilde{q},\tilde{\omega},\tilde{\alpha}] = S_J[q,\omega,\alpha],
\end{equation}

where the transformed variables $\tilde{q},\tilde{\omega},\tilde{\alpha}$ are given by the transformation

\begin{equation}
\begin{gathered}
\tilde{q} = \Omega(\bar{\omega}) q + \bar{\alpha}\\
\Omega(\tilde{\omega}) = \Omega(\omega) \Omega^{T}(\bar{\omega})\\
\tilde{\alpha} = \alpha - \Omega(\omega) \Omega^{T}(\bar{\omega}) \bar{\alpha}.
\end{gathered}
\end{equation}

for arbitrary smooth functions $\bar{\omega}^a (\lambda)$ and $\bar{\alpha}^a (\lambda)$'s.

Hence, Jacobi's action (\ref{jacobiii}) is gauge-invariant under the above parameter-dependent (local) transformation.

\subsection{The best-matching constraints}
We now see what best-matching really implies. For this, we explicitly calculate the two conditions we derived in (\ref{jacobiiii}). For translations, we have

\begin{equation}
\frac{\delta S_J}{\delta \alpha^{'a}} = \sum\limits_{I} \Omega^T \frac{\delta S_J}{\delta q_I^{'a}} = \sum\limits_{I} \Omega^T p_a^I.
\end{equation}

Thus, best-matching with respect to translations enforces the vanishing of the total momentum:

\begin{equation}\label{totalmomentumI}
P^a \equiv \sum\limits_{I} p^I_a = 0.
\end{equation}

It means that the parameters $\alpha^a$'s must be chosen in such a way that the total momentum vanishes. Of course, this condition fixes only the derivative of $\alpha^a$'s and their initial value, i.e., their global value remains arbitrary. But the essential point is that they can be solved in terms of the $q^a_I$'s.

Regarding the second condition, in light of (\ref{turningtoliealgebra}) we have

\begin{equation}\label{bmrot}
\frac{\delta S_J}{\delta \omega^{'a}} = \sum\limits_{I} \frac{\delta S_J}{\delta q_I^{'b}} (O_a q_I)_b = \sum\limits_{I} (O_a)^b_c q_I^c p^I_b.
\end{equation}

Based on (\ref{explicitrotgen}), the matrix representations of the elements $O_a$s are all asymmetric and are indeed the Levi-Civita symbol. Thus, the condition (\ref{bmrot}) is equivalent to the vanishing of the total angular momentum.

\begin{equation}\label{totalangularmomentumI}
J^a = \sum\limits_{I} \epsilon^{abc} q^b_I p_c^I = 0.
\end{equation}

So it follows that best-matching with respect to rotations demands the total angular momentum of the whole universe vanishes. Again, we can use this condition to solve for the parameters $\omega^a$ (up to their initial value).

The above calculations have revealed what best-matching essentially does: it aligns the succession of configuration states of the whole universe by making the translational and rotational momenta vanish. More strikingly, the best-matching conditions we derived are the vanishing of the momenta that generate the very transformations used in the procedure. There is a bigger and more general picture behind this observation which will be developed in Sec. \ref{bmg}.

From these conditions, we see that relational particle physics is not so foreign to Newtonian mechanics. We already know that for rotationally and translationally invariant potentials the total momentum $P$ and total angular momentum $J$ are constant. Now best-matching requires them to vanish. Thus, we can say that relational physics is `part of' Newtonian mechanics, with only the difference that the value of the constants of motion, the total conjugate momenta, is set to zero. In other words, relational particle mechanics is equivalent to Newtonian mechanics, provided that the total momenta vanish.

What is striking is the relevance these constraints bear to Tait's problem. As we saw, to completely solve the Scholium problem, i.e., to determine the Newtonian evolution of a system based on the observable data encoded in the configuration of particles, we needed more than two snapshots. The constraints imposed on the possible solution curves in the configuration space exactly compensate for those missing data. Given that the total linear and angular momentum of the whole universe must vanish, the two snapshots representing the initial configuration and its change should be enough to determine the evolution of the system. This hint is in the direction of the Mach-Poincaré Principle as will be discussed in the following part.

\subsection{The solution to Newton's bucket problem}

As best-matching is the key to relational physics, we can see how it provides a better explanation of Newton's bucket thought experiment. We can think of the whole universe as being composed of the bucket, the water, and the distant stars. The best-matching condition corresponding to the rotational group is the vanishing of the total angular momentum. Thus, as the water is spinning, the distant stars must be rotating in the opposite direction to ensure that condition holds. However, due to their large mass and distance, the contribution of the motion of the stars to the total angular momentum is extremely more significant, and thus, their rotation is negligible. Even with a tiny amount of rotation which is unnoticeable, the total angular momentum can vanish in accord with the best-matching condition. This leads to the resolution of the problem: the rotation of the water is indeed meaningful with respect to distant stars. The distant stars create a background with respect to which the bucket and the water inside rotate, resulting in the change in the surface of the water.

There is no need to posit an absolute space or a background inertial frame of reference. Best-matching ensures that the whole universe provides the background.

It is tempting to think about the scenario in which there are no distant stars and there is only a bucket filled with water. Best-matching implies that there are only two acceptable possibilities: either the bucket and water are not rotating and the surface of the water is flat or the bucket and water are rotating in opposite directions and the surface of the water is concave. Unlike the results of Newtonian mechanics, there are no further possibilities. This highlights the restrictive character of relational physics and also emphasizes the role of the \emph{whole universe} in providing a background for motion.

\subsection{Mach-Poincaré Principle}

The content of the Mach-Poincaré Principle is to first quotient the redundant Newtonian configuration space with respect to the relevant symmetry group, and then construct a theory that uniquely evolves the whole universe given only the initial configuration and its change (tangent vector or the direction of change). We considered $N$ particles in the three-dimensional Euclidean space. Hence, the initial crude configuration space is $\mathbb{R}^{3N}$: one assigns three coordinates to each particle in the Euclidean space that plays the role of Newton's absolute space. Assuming that the size is given by some external measuring rod, the symmetry group of Euclidean space, i.e., the group of transformations that leave the distinguishable states of the whole universe untouched, is the Euclidean group. Thus, the real relational configuration space is $Q^R = \sfrac{\mathbb{R}^{3N}}{E(3)}$. Moreover, the action we arrived at after best-matching (\ref{jacobiiii}) is defined on this reduced configuration space. The distinct trajectories that give rise to different values for this action are the ones that are distinguishable in the reduced configuration space.

We can picture the way best-matching affects the configuration space in the following way. The Euclidean group slices the initial configuration space $\mathbb{R}^{3N}$ into orbits\footnote{For the definition of group orbits see Appx. \ref{lie}.}. We surely cannot visualize the $3N$-dimensional configuration space. But we can proceed with the simpler example of a two-dimensional space and the rotation group $SO(2)$ playing the role of the symmetry group. The group orbits are the centered circles around the origin. What the Mach-Poincaré Principle demands is a theory defined in the reduced configuration space that consists of only these circles, described by a positive number denoting the radius. Best-matching smoothly moves any trajectory in $\mathbb{R}^2$ along those circles so that the action is minimized. The curves which minimize the action `cut through' the orbits. Hence, the best-matched action is an action defined in terms of the trajectories in the quotiented configuration space.

The same scenario plays out in the general case of $N$ particles in the three-dimensional space. The orbits of the Euclidean group are the true degree of freedom on which the best-matched action is defined: The best-matched trajectories cut through the gauge orbits. The Euler-Lagrange equations for this action give the equations of motion.

Hence, we infer that \emph{spatial relationalism} is embodied in this method. We now turn our attention to \emph{temporal relationalism}.

With a glance at the action (\ref{jacobiii}), we notice its timeless character as we explained near the end of Sec. \ref{jac}. Moreover, we know that an action theory admits a unique solution given any initial data comprising of an initial point and an initial tangent vector. Because of the reparametrization invariance of the action (\ref{jacobiii}) only the initial `directions' are relevant. It is impossible to meaningfully talk about the tangent vectors as their magnitude can change under a simple change of the parameter:

\begin{equation}
q^{'a}_I = \frac{dq^a_I}{d\lambda} = \frac{d\lambda'}{d\lambda} \frac{dq^a_I}{d\lambda'} \equiv \zeta \tilde{q}^{'a}_I.
\end{equation}

Where $\tilde{q}^{a'}_I$ is another equally acceptable tangent vector.

There is a more illuminating way of seeing how Mach's relational understanding of time is embodied in the theory. The Euler-Lagrange equations for the action (\ref{jacobiii}) are

\begin{equation}\label{el}
\frac{d}{d\lambda} \left(\sqrt{\frac{(E-V)}{T}} m_I \bar{q}_I^{a'} \right) = - \sqrt{\frac{T}{(E-V)}} \frac{dV}{d\bar{q}^a_I}.
\end{equation}

Now we fix the parametrization of the equations of motion such that $E = V + T$. Thus, by definition

\begin{equation}
V(\bar{q}) + \frac{1}{2} \sum\limits_{I} m_I \dot{\bar{q}}^{2}_I = E.
\end{equation}

Where $\dot{\bar{q}}^a_I = \frac{d\bar{q}^a_I}{dt}$ and $t$ is the parameter that satisfies the above condition, called \emph{ephemeris time}. More explicitly,

\begin{equation}\label{eph1}
dt = \sqrt{\frac{\sum\limits_{I} m_I d\bar{q}^{a2}_I}{2\left(E-V(\bar{q})\right)}}.
\end{equation}

With resepect to this time variable, the equation (\ref{el}) turns into the familiar Newton's 2nd law:

\begin{equation}
m_I \ddot{\bar{q}}_I^{a} = - \frac{dV}{d\bar{q}^a_I}.
\end{equation}

Take another look at the definition (\ref{eph1}). It does exactly entail what Mach meant when he said `time is an abstraction we arrive at by means of changes'. The LHS of the equation is a measure of \emph{time}, while the RHS is solely dependent on the \emph{changes} of the best-matched variables, and (\ref{eph1}) defines the former in terms of the latter. This is a suitable choice of gauge which leads to formally simple equations of motion.

We learned the rudiments of best-matching and saw how it embodies the strong version of the Mach-Poincaré Principle. From the mathematical point of view, however, our account is not thorough. We now plunge into the mathematical details of this procedure and lay a solid foundation to dispel any confusion and unclear point we glossed over in this section.

\section{Best-matching: the general approach}\label{bmg}
To my knowledge, the best modern and thorough account of best-matching in a general theory with all the details is given by Sean Gryb in his doctoral thesis \cite{Gryb.2012}. We basically give a distillation of his work with some minor changes here.

Let $Q$ be the differentiable manifold representing the configuration space of a $N$-body system\footnote{Although we develop this general approach in particle mechanics, the method can be easily generalized to other configuration spaces. In particular, relational field theory and Einstein's general relativity will be developed in the next chapter with this method.} with the elements denoted by $q_I^a$ representing the $a$th coordinate of the $I$th particle. Let $G$ denote the underlying symmetry group. In accord with Sec. \ref{groupaction}, we represent the action of the symmetry group on $q_I^a$ by $T_{\phi} q^a_I$ where $\phi^\alpha$ denotes a specific element of $G$. Based on our discussion of the action of the Lie group on a manifold, we can write the transformation in the following way

\begin{equation}\label{actionalgebra}
T_{\phi}q_I^a \equiv \exp{\left( \phi^\alpha t_{\alpha b}^a \right)} \, q_I^b,
\end{equation}

where $t^a_{\alpha b}$ is the elements of the generator of the Lie group and is generally a smooth function on the configuration space manifold, depending on the points it acts on as explained in Sec. \ref{groupaction}.

We can define the trajectories in $Q$ and their transformation under $G$ analogously. For the trajectory $q_I^a(\lambda)$, we have

\begin{equation}\label{transformcurve}
q_I^a(\lambda) \rightarrow T_{\phi(\lambda)}q_I^a (\lambda)
\end{equation}

Note that $\lambda$ is a nominal variable used to represent the curve in $Q$. But the curve must be a smooth function of $\lambda$. The transformed curve on the RHS of (\ref{transformcurve}) must also be smooth which means that $\phi^\alpha(\lambda)$ must be smooth.

We assume that $Q$ is equipped with the Riemannian metric\footnote{A Riemannian metric is a symmetric type $(0,2)$ tensor field on the manifold, provided that $g(v,v) > 0$ for all non-zero tangent vectors $v$ and $g(0,0) = 0$. See \cite{Nakahara.2003}, Chapter 7.} $g$. Moreover, we assume that in our coordinatized representation, $g_{ I a \, J b} = \delta_{IJ} m_I h_{a b}$. We will construct an action in the next section in which $m_I$'s represent the \emph{masses} of the particles and are just a set of pure numbers, and $h$ the metrical structure of the \emph{space} underlying the configuration space of the whole system, depending on the configuration space variables, i.e., $h_{ab} = h_{ab}(q)$.

\subsection{Action principle and the best-matching}

To incorporate best-matching, we conceive of two snapshots representing two configurations at two infinitesimally separated moments. We can then define the quantity

\begin{equation}\label{distance}
\delta q^a_I(\lambda) = T_{\phi(\lambda+ d\lambda)}q_I^a (\lambda+ d\lambda) - T_{\phi(\lambda)}q_I^a (\lambda).
\end{equation}

along every smooth curve in $Q$. Then consider the quantity

\begin{equation}
{ds}^2 = \sum\limits_I m_I \, \delta q^a_I \, h_{a b} \, \delta q^b_I.
\end{equation}

$ds$ is the \emph{magnitude} of the distance (\ref{distance}) in $Q$ with respect to the metric $g$.

Hence, we define the following functional which assigns a value to each smooth curve in $Q$ that can be thought of as the length of the paths in the configuration space.

\begin{equation}\label{actionrep}
S[q,\phi] = \int ds = \int \sqrt{\sum\limits_{I} m_{I} \, \delta q^a_I(\lambda) h_{a b} (\bar{q}) \delta q^b_I(\lambda)}.
\end{equation}

By restoring the parameter, we have

\begin{equation}\label{action}
S[q,\phi] = \int d\lambda \, \sqrt{\sum\limits_{I} m_{I} \, T_{\phi(\lambda)}D_{\phi} q^a_I(\lambda) h_{ab}(\bar{q}) T_{\phi(\lambda)}D_{\phi} q^b_I(\lambda)},
\end{equation}

where

\begin{equation}\label{Dphi}
D_{\phi} q_I^a = T^{-1}_{\phi(\lambda)} \frac{\delta q_I^a}{\delta \lambda} =
T^{-1}_{\phi(\lambda)} \frac{d}{d\lambda} T_{\phi(\lambda)} q^a_I(\lambda).
\end{equation}

Moreover, we can write the action in terms of a Lagrangian defined smoothly on the tangent bundle of $Q$, $T^*Q$,

\begin{equation}\label{squareaction}
S[q,\phi] \equiv \int d\lambda \, L(q,\dot{q},\phi), \quad L(q,\phi) = \sqrt{\sum\limits_{I} m_{I} \, T_{\phi}D_{\phi} q^a_I h_{a b}(\bar{q}) T_{\phi}D_{\phi} q^b_I}.
\end{equation}

By direct calculation we have

\begin{equation}\label{Differentialq}
D_{\phi} q_I^a(\lambda) = \dot{q}_I^a + T^{-1}_{\phi(\lambda)} \frac{d}{d\lambda} \left( T_{\phi(\lambda)} \right) q_I^a (\lambda),
\end{equation}

where $\dot{q}_I^a = \frac{d}{d\lambda} q_I^a (\lambda)$. Fortunately, the second term in (\ref{Differentialq}) looks familiar. It is related to the generators of the action of $G$ in light of (\ref{actionalgebra}):

\begin{equation}
T^{-1}_{\phi(\lambda)} \frac{d}{d\lambda} \left( T_{\phi(\lambda)} \right) q_I^a (\lambda) \equiv \dot{\phi}^\alpha t^a_{\alpha b}(\phi) q^b_I,
\end{equation}

where we have introduced $t^a_{\alpha b}(\phi)$ as the generators evaluated at the group element represented by $\phi$ (see (\ref{Bdefintarbit})). Thus,

\begin{equation}\label{bestmatchingderivative}
D_{\phi} q_I^a(\lambda) = \dot{q}^a_I + \dot{\phi}^\alpha t^a_{\alpha b} (\phi) q^b_I.
\end{equation} 

The action we postulated in (\ref{action}) is reminiscent of Jacobi's action. Our general approach in this section illuminated why Jacobi's action is so important in relational physics. However, our starting point here is the action (\ref{action}) and note that there is no supplementing energy condition. Moreover, the action (\ref{action}) is reparametrization invariant as is explicit in (\ref{actionrep}), that is, it is invariant under any transformation of the form $\lambda \rightarrow f(\lambda)$ provided that $f$ is a strictly monotonic smooth function. This symmetry and the best-matching condition ensure that the strong version of the Mach-Poincaré Principle holds as we will show below.

The action (\ref{action}) can also describe interacting theories. The interaction terms can be included in the spatial metric $h_{ab}$. It does not change the nature of the equations of motion. The simplest choice could be to consider a conformally flat metric with the conformal factor $-V$ where $V$ is a function on the configuration space. This is enough to reproduce Newtonian particle dynamics if $V$ is taken to be the potential for the system. So $h_{ab} = -V \delta_{ab}$ where $\delta_{ab}$ is the Kronecker delta and is the flat metric in Cartesian coordinates. Thus,

\begin{equation}\label{jacobiconformal}
L_{J}(q,\dot{q},\phi) = \sqrt{-V(T_{\phi(\lambda)}q_I^a(\lambda)) \sum\limits_{I} m_I \, T_{\phi(\lambda)}D_{\phi} q^a_I(\lambda) \, \delta_{a b} \, T_{\phi(\lambda)}D_{\phi} q^b_I(\lambda)}.
\end{equation}

This is now precisely the Lagrangian appearing in Jacobi's action (\ref{jacobiii}) for $E=0$ if $G$ is taken to be the Euclidean group.

We now turn our attention to the variation of the action (\ref{action}), implement the best-matching procedure to clarify our variation method in (\ref{introbestmatching}).

The variation with respect to the $\phi^\alpha$ must be carried out differently as they are not physical variables. Normally, in the calculus of variations, the endpoints of the fields are fixed while the curve varies between those two points. However, we demand that the value of the variables $\phi^\alpha$ at the endpoints of \emph{any} infinitesimal interval along the curve must remain arbitrary. This means that we no longer can use the vanishing of $\delta \phi^\alpha$ at the endpoints.

To mathemtically realize this condition, we first vary the action (\ref{action}) with respect to $\phi^\alpha$:

\begin{equation}\label{variationphi}
\delta_{\phi^\alpha} S = \int d\lambda \, \left[ \frac{\partial L}{\partial \phi^\alpha} - \frac{d}{d\lambda} \left( \frac{\partial L}{\partial \dot{\phi}^\alpha} \right) \right] \delta \phi^\alpha + \delta \phi^\alpha \left. \frac{\partial L}{\partial \dot{\phi}^\alpha} \right|_{\lambda_i}^{\lambda_f}.
\end{equation}

Where $\lambda_i,\lambda_f$ denote the endpoint values of $\lambda$. If $\delta \phi^\alpha$ is arbitrary at both of the end points, besides the Euler-Lagrange equations, the vanishing of $\delta_{\phi^\alpha} S$ implies the vanishing of $\frac{\partial L}{\partial \dot{\phi}^\alpha}$ at the endpoints.

Following this intuition, we demand that this condition holds regardless of the endpoints of a certain trajectory. Thus, we impose this `additional' condition for all values of $\lambda$ and arrive at the \emph{best-matching condition}

\begin{equation}\label{bestmatchingcondition}
\frac{\partial L}{\partial \dot{\phi}^\alpha} = 0.
\end{equation}

It then follows that the best-matching variation of $\phi^\alpha$ is essentially the usual variation of $\phi^\alpha$ which leads to the Euler-Lagrange equations

\begin{equation}\label{eulerlagrangefields}
\frac{\partial L}{\partial \phi^\alpha} - \frac{d}{d\lambda} \left( \frac{\partial L}{\partial \dot{\phi}^\alpha} \right) = 0,
\end{equation}

with the additional condition (\ref{bestmatchingcondition}).

The next step is to use the equations (\ref{eulerlagrangefields}) and (\ref{bestmatchingcondition}) to solve for $\phi^\alpha(\lambda)$ in terms of the $q^a_I$'s. By inserting them in the action (\ref{action}) we arrive at the \emph{best-matched action}, the variation of which with respect to the configuration space variables gives us the equations of motion.

Now we use the explicit form of the generators we found in table \ref{generators} and find the best-matching condition (\ref{bestmatchingcondition}) for the similarity group.

First, we note that because of the form of (\ref{action}) and (\ref{bestmatchingderivative})

\begin{equation}
\frac{\partial L}{\partial \dot{\phi}^\alpha} = \sum\limits_I \frac{\partial L}{\partial (D_\phi q^a_I)} \frac{\partial (D_\phi q^a_I)}{\partial \dot{\phi}^\alpha} = \sum\limits_I \frac{\partial L}{\partial \dot{q}^a_I} t^a_{\alpha b} (\phi) q^b_I.
\end{equation}

Thus, as the momentum conjugate to $q^a_I$ is $p_a^I = \frac{\partial L}{\partial \dot{q}^a_I}$, we have

\begin{equation}
\frac{\partial L}{\partial \dot{\phi}^\alpha} = \sum\limits_I p^I_a t^a_{\alpha b}  (\phi) q^b_I.
\end{equation}

Hence the best-matching conditions corresponding to the groups whose generators are listed in the table \ref{generators} for $\phi = 0$ (evaluated at the identity element) are respectively

\begin{gather}
\sum\limits_I p^I_a = 0 \label{totalmomentum}\\
\sum\limits_{I} \epsilon^{abc} q^b_I p^I_c = 0\label{totalangularmomentum}\\
\sum\limits_I q^a_I p^I_a = 0.\label{dilatationalmomentum}
\end{gather}

The first two conditions are familiar and remind us of (\ref{totalmomentumI}) and (\ref{totalangularmomentumI}). Now we see that removing absolute size enforces another condition. We call it the dilatational momentum, i.e., the momentum conjugate to size:

\begin{equation}\label{dilatationalmomentum}
D \equiv \sum\limits_I q^a_I p^I_a.
\end{equation}

Best-matching with respect to the scale transformation requires the vanishing of the dilatational momentum. It is a highly restrictive condition as the dilatational momentum is not even conserved in many models, including Newtonian gravity, let alone setting it equal to zero as a constraint. Actually, we will see that for a scale invariant theory to be consistent, the potential must be homogeneous of degree -2. This ensures the constancy of the dilatational momentum and the best-matching condition sets it to zero in such a case.

\subsection{Mach-Poincaré Principle revisited}\label{revisitingmachpoincare}
The action (\ref{action}) is invariant under the double gauge transformation

\begin{equation}\label{banal}
\begin{gathered}
q_I^a(\lambda) \rightarrow T_{\omega(\lambda)}q_I^a (\lambda),\\
\phi^\alpha(\lambda) \rightarrow \phi^\alpha(\lambda) -\zeta^\alpha(\lambda),
\end{gathered}
\end{equation}

for any $\omega$, where $\zeta^\alpha$ satisfies the condition $T_{\phi-\zeta} T_\omega = T_\phi$. This double gauge symmetry is named \emph{banal transformation} in \cite{Barbour.2003}.

Thus, the action (\ref{action}) is practically defined on the orbits of $Sim(3)$ in $Q$. More precisely, after choosing the initial conditions in $\sfrac{Q}{Sim(3)}$, in light of our discussion that the best-matched curves in the configuration space cut through the orbits, there will be only one curve in the quotiented space extremizing the action and given the initial condition in the reduced space. This is essentially the first main content of the Mach-Poincaré Principle.

As regards the second principle, the reparametrization invariance of the action (\ref{action}) ensures that the directions are only relevant. This action is a geodesic principle and only requires an initial point in $Q$ and a direction in the corresponding tangent space as the initial conditions, for the magnitude of the initial vector is redundant due to the arbitrariness of the parameter.

More strikingly, we can choose the ephemeris time as the parameter with respect to which the equations of motion take the familiar form of Newton's equations (in the case that interaction is also included in the metric, i.e., $h_{ab} = -V \delta_{ab}$. The variation of the action (\ref{action}) with respect to the best-matched coordinates, $\bar{q}^a_I = T_{\bar{\phi}} q^a_I$, where ${\bar{\phi}}^\alpha$ is the solution of the best-matching conditions, leads to the equation of motion

\begin{equation}
\frac{d}{d\lambda} \left( \frac{-V(\bar{q})}{L(q,\dot{q},\bar{\phi})} m_I \dot{\bar{q}}^b_I (\lambda) \delta_{ab} \right)
=
- \frac{T(\dot{\bar{q}})}{L(q,\dot{q},\bar{\phi})} \frac{\partial V(\bar{q})}{\partial \bar{q}^a_I}.
\end{equation}

With the choice of the parameter such that $\sqrt{-\frac{T}{V}}=1$, we have

\begin{equation}
m_I \ddot{\bar{q}}^a_I
= - \frac{\partial V(\bar{q})}{\partial \bar{q}^a_I}.
\end{equation}

This specific time parameter is the ephemeris time (\ref{eph}) and its definition is based on the change of the variables of the entire island universe:

\begin{equation}\label{eph}
dt = \sqrt{\frac{\sum\limits_{I} m_I d\bar{q}^{2}_I}{-2 V(\bar{q})}}.
\end{equation}

We recapitulate the best-matching procedure in the Lagrangian formulation:

\begin{enumerate}
\item First, we begin with a well-defined configuration space $Q$ endowed with a differentiable manifold and a metric $g$.

\item We identify the symmetry group $G$ on empirical grounds, consider its action on $Q$, and make the substitution $q_I^a \rightarrow T_{\phi^\alpha} q_I^a$.

\item We posit a geodesic principle in the configuration space of the form

\[
S = \int d\lambda \, \sqrt{ g(T_{\phi}D_{\phi} q,T_{\phi^\alpha}D_{\phi} q)}.
\]

\item Finally, we perform the best-matching variation of $\phi^\alpha$ by imposing the Euler-Lagrange equations and the additional condition

\[
\frac{\partial L}{\partial \dot{\phi}^\alpha} = 0.
\]

\item We solve the above equations, find the best-matched coordinates, vary the aforementioned action with respect to them, and arrive at the relational equations of motion.
\end{enumerate}

Depending on the nature of the metric, there are in principle two different types of best-matching. One is \emph{equivariant best-matching} which is the case that the metric $g$ remains invariant under the group transformation $T_{\phi}$, i.e., this transformation is an isometry of $g$. More precisely, it means that $ T^{-1*}_{\phi} g = g$ where $T^{-1*}_{\phi}$ is the pullback corresponding to the diffeomorphism $T^{-1}_{\phi}$ on $Q$. The second type is \emph{non-equivariant best-matching}, corresponding to the case that the group transformation is not an isometry of the configuration space metric.

As these two types can differ in many respects, we study them in the following sections separately.

\section{Best-matching and reparametrization invariance}\label{best-matchingparametrizaion}
In our general approach, we posited an action defined in terms of the trajectories in the configuration space equipped with a metric:

\[
S[q] = \int d\lambda \, \sqrt{\sum\limits_{I} m_{I} \, h_{ab} \dot{q}^a_I \dot{q}^b_I}.
\]

We saw that with a particular choice of a conformally flat metric the action becomes Jacobi's action (\ref{jacobii}). In fact, what is striking is that one can look at the reparametrization invariance of Jacobi's action in light of best-matching with respect to \emph{time}. To see that, let us assume $h_{ab} = -V \delta_{ab}$. It is easy to check that the action

\begin{equation}\label{parametrizedaction}
S[q,N] = \int d\lambda \, \left( \frac{1}{N} \sum\limits_{I} m_I \dot{q}_I \dot{q}_I - N V \right)
\end{equation}

also reproduces the same equations of motion and is equivalent to the above action provided that we identify $N$ as

\begin{equation}
N = \sqrt{\frac{\sum\limits_{I} m_{I} \, \dot{q}_I \dot{q}_J}{V}}.
\end{equation}

In general, $N$ is an additional variable in the action (\ref{parametrizedaction}) and should be varied independently. If we extermize (\ref{parametrizedaction}) with respect to $N$, we arrive at the above condition.

Action (\ref{parametrizedaction}) resembles the reparametrization invariant form of the Newtonian action we have introduced in Appx. \ref{jacobi}:

\begin{equation}
S[q,t] = \int d\lambda \, \left( \frac{1}{\dot{t}} \sum\limits_{I} m_{I} \dot{q}_I \dot{q}_J - \dot{t} V \right)
\end{equation}

We should just take $N = \dot{t}$ to make them identical. For this reason, we might say that the action (\ref{parametrizedaction}) performs best-matching with respect to time: We have an initial action written in terms of an external time $t$, we give it an arbitrary parametrization, i.e., $ t \rightarrow t(\lambda)$, we eventually vary the action with respect to $t$ and impose the best-matching condition

\begin{equation}\label{timebestmatchingcondition}
\frac{\partial L}{\partial \dot{t}} = 0.
\end{equation}

This condition ensures that we could have made the identification $N = \dot{t} = \frac{dt}{d\lambda}$ and vary the action with respect to $N$, in accord with what we did with the action (\ref{parametrizedaction}).

Overall, what it all means is that mathematically, Jacobi's action is effectively the standard Newtonian action best-matched with respect to the time that is used in its formulation. It should be emphasized that this is only a mathematical analogy, and not as part of the procedure developed in relational physics. Time is essentially different from the configuration space variables, and using the term `best-matching' for reparametrization invariant theories, as we did, has merely a heuristic meaning.

One last note should be made regarding the status of $E$ which we identified as the energy in Jacobi's action. In general, one can conceive of a conformally flat metric with an additional constant added to the potential term, $h_{ab} = (E-V) \delta_{ab}$. In this approach, $E$ is a universal constant existing originally in the action. Also, as we discuss in Appx. \ref{jacobi}, in the Routhian reduction procedure for time-independent Newtonian potentials, we can set the momentum conjugate to time equal to a constant value, e.g., $-E$. This again adds a constant term to the potential term in Jacobi's action. Here, unlike the above case, $E$ is a constant of motion that is only meaningful in the case that Newtonian external time is posited. In relational physics, the momentum conjugate to time is essentially set to zero and is a constraint by virtue of the best-matching condition (\ref{timebestmatchingcondition}). The point is that both of the above approaches result in the same action and in both cases, $E$ can be seen as the energy of the system. In Newtonian physics, $E = -p_t$ is the energy by definition, while in relational physics, if we introduce the ephemeris time we have

\begin{equation}
E = T + V, \quad T = \sum\limits_{I} \frac{1}{2} m_{I} \frac{dq_I}{dt} \frac{dq_J}{dt},
\end{equation}

where $t$ is the ephemeris time again by definition.

The above discussion shows that at least in classical physics, as far as the evolution of the relational degrees of freedom of the system is concerned, these two ways of defining $E$ result in the same thing. However, in the spirit of Leibniz's Principle of Sufficient Reason, we may not want to introduce an arbitrary universal constant in the action and set $E$ equal to zero. It should be noted that in quantum mechanics these two ways do not remain equivalent, the discussion of which goes beyond the scope of this work. See Section 3.5 of \cite{Gryb.2012}.

\section{Equivariant best-matching}
The focus of this chapter is the equivariant actions. We recap the Lagrangian formulation of equivariant actions, perform the best-matching, and finally analyze the implications of equivariance best-matching for interaction potentials. Then we will develop the Hamiltonian formulation.

The Hamiltonian formulation is a lot more powerful than the Lagrangian formulation in many respects. The most significant aspect is the elegantly clear perspective it provides on the concept of gauge transformations as we will see.

\subsection{Lagrangian formulation}\label{2lagrangeequivariancebm}
Our starting point is the action (\ref{action}) we found in the last section

\[
S[q,\phi] = \int d\lambda \, \sqrt{\sum\limits_{I} m_{I} \, T_{\phi(\lambda)}D_{\phi} q^a_I(\lambda) h_{a b} T_{\phi(\lambda)}D_{\phi} q^b_I(\lambda)}.
\]

Based on definition, the equivariance condition means that $T^{-1*}_{\phi} h = h$, implying that

\begin{equation}\label{equivariance}
\bar{h}_{ab}(\bar{q}) \equiv (T_{\phi}^{-1})_a^c  (T_{\phi}^{-1})_b^d  \, h_{cd}(q) = h_{ab}(\bar{q}),
\end{equation}

where $(T^{-1}_{\phi})_a^c$ denotes the matrix elements of the inverse of the diffeomorphism induced by the action of the symmetry group $G$ on $Q$, i.e.,

\begin{equation}\label{Tab}
(T_\phi)^a_b \equiv \partial_{q^b_I} T_{\phi} q^a_I = \frac{\partial \bar{q}^a_I}{\partial q^b_I},
\end{equation}

where $\phi(q)$ is a smooth function defined on $Q$. Thus, the equivariance condition leads to the action

\begin{equation}\label{equiaction}
S[q,\phi] = \int d\lambda \, \sqrt{\sum\limits_{I} m_{I} \, D_{\phi} q^a_I(\lambda) h_{a b}(q) D_{\phi} q^b_J(\lambda)}.
\end{equation}

The equations of motion can be determined by performing the best-matching variation with respect to $\phi^\alpha$'s and then $q^a_I$'s. For equivatiant theories, what best-matching variation essentially does is that it elevates the role of the parameters $\alpha^\alpha$'s as a \emph{cyclic} variable to a \emph{Lagrange multiplier}. In an equivariant action of the form (\ref{equiaction}), $\phi^\alpha$'s drops out of the theory completely. Formally,

\begin{equation}
\frac{\partial L}{\partial \phi^\alpha} = 0.
\end{equation}

This means that the variables $\phi^\alpha$'s are cyclic and the momentum conjugate to them is a constant of motion. Now the best-matching condition (\ref{bestmatchingcondition}) is effectively equivalent to varying the action (\ref{equiaction}) with respect to the variable $N^\alpha = \dot{\phi}^\alpha$. In light of this consideration, we can say that $\dot{\phi}^\alpha$'s appearing in the term $D_\alpha q^a_I$ in (\ref{equiaction}) act as Lagrange multipliers, enforcing the best-matching conditions.

We know that every isometry like (\ref{equivariance}) leads to the existence of some global Killing vectors in $Q$, for the Lie derivative of the metric in the direction of the flow of the symmetry group vanishes, i.e., $\mathfrak{L}_v h = 0$, where $v$ is the vector field associated with the infinitesimal group transformation. As to the first degree we have $q^a_I \rightarrow {\phi}^\alpha t^a_{\alpha b} q^b_I$, $v^a = \phi^\alpha t^a_{\alpha b} q^b_I$.

Consequently, the equivariance condition explicitly implies\footnote{We drop the $\phi$ dependence of the generators in this section for brevity.}

\begin{equation}\label{killing}
\mathfrak{L}_v h_{ab} = \sum \limits_{I} \phi^\alpha \left(t^c_{\alpha d} q^d_I \partial_{q^c_I} h_{ab} + h_{cb} \partial_{q^a_I} (t^c_{\alpha d} q^d_I) + h_{ac} \partial_{q^b_I} (t^c_{\alpha d} q^d_I)\right) = 0,
\end{equation}

for all $\phi^\alpha$'s. When exponentiated, the above condition yields the global isometry (\ref{equivariance}). In fact, (\ref{killing}) is the infinitesimal form of the aforementioned condition.

Now we turn our attention to the implications for particle dynamics of the equivariance condition with respect to the similarity group.

As we explained in the previous section, one can choose a conformally flat metric and arrive at the standard Newtonian mechanics given by Jacobi's action (\ref{jacobiconformal}). For brevity, we drop the constant term $E$ and take the factor to be proportional to only the potential. With this choice, $h_{ab} = -V \delta_{ab}$. In light of the above discussion, the equivariance condition (\ref{killing}) with respect to translations, rotations, and scale transformation implies the conditions

\begin{gather}
\sum\limits_I \partial_{q^a_I} V(q) = 0 \label{potsym1},\\
\sum\limits_{I} \epsilon^{a r t} q^{r}_I \partial_{q^t_I} V(q) = 0 \label{potsym2},\\
\sum\limits_{I} q_I^a \partial_{q^a_I} V(q) + 2N V(q) = 0 \label{potsym3}.
\end{gather}

for $\phi = 0$ respectively.

%The first two conditions are familiar and also striking! It means that the total potential of the island universe (the whole system) must be translationally and rotationally invariant. Of course we always `assume' that this holds in Newtonian mechanics. But that is exactly the point! We `assume' that. There is nothing in the Newtonian framework that forbids other interactions. However, we actually `derived' these conditions in relational physics as the consistency

The significance of the first two conditions is quite clear. It means that the total potential of the island universe (the whole system) must be translationally and rotationally invariant, as they should be if we want a relational theory of the whole. The third condition, however, is new. Using Euler's theorem, it means that the potential is homogeneous of degree -2 in the variables, i.e., $V(q) \propto \sfrac{1}{q^2}$. Unfortunately, the Newtonian gravitational potential is of degree -1 and does not satisfy this condition. This is an issue if we want to include the standard gravitational interaction in a relational theory of the whole universe. This shakes our hope for an empirically valid consistent scale-free theory. We will come back to this issue in the second part of this work. For now, we just stay focused on developing the mathematics of best-matching.

\subsection{From the Lagrangian to the Hamiltonian formulation}

To make the transition to the Hamiltonian formulation, our starting point in the canonical formulation of the equivariant action (\ref{equiaction}). We can then find the momentum conjugate to the variables $q^a_I$'s and $\phi^\alpha$'s:

\begin{equation}\label{conjugatemomenta}
\begin{gathered}
p_a^I \equiv \frac{\partial L}{\partial \dot{q}^a_I} = \frac{m_I h_{ab} D_\phi q^b_I}{\sqrt{\sum\limits_{I} m_{I} \, D_{\phi} q^a_I(\lambda) h_{a b} D_{\phi} q^b_J(\lambda)}},\\
\pi_\alpha \equiv \frac{\partial L}{\partial \dot{\phi}^\alpha} = \frac{ \sum\limits_{I} \, m_{I} h_{ab} D_\phi q^b_J \, t^a_{\alpha c}(\phi) q^c_I}{\sqrt{\sum\limits_{I} m_{I} \, D_{\phi} q^a_I(\lambda) h_{a b} D_{\phi} q^b_J(\lambda)}}.
\end{gathered}
\end{equation}

Then the phase space $\Gamma$ can be coordinatized by the variables $(q^a_I,p^a_I,\phi^\alpha,\pi_\alpha)$. Also, as the fundamental and essential structure of phase space, we posit the Poisson brackets

\begin{equation}\label{poissonbrackets}
\{q^a_I,p_b^J\} = \delta^a_b \delta^J_I, \qquad \{\phi^\alpha,\pi_\beta\} = \delta^\alpha_\beta.
\end{equation}

From the definitions of the conjugate momenta in (\ref{conjugatemomenta}), it can be easily verified that the constraints

\begin{gather}
\mathcal{H} = \sum\limits_{I} \, \frac{h^{ab} p^I_a p^I_b}{m_I} - 1 \approx 0,\label{hamconstraintbm}\\
\mathcal{H}_\alpha = \pi_\alpha - \sum\limits_I p_a^I t^a_{\alpha b} q^b_I \approx 0 \label{vecconstraintbm}.
\end{gather}

immediately follow, where $h^{ab}$ is the inverse of $h_{ab}$.

The first constraint is the Hamiltonian constraint as the result of the reparametrization invariance. This quadratic constraint arises from the fact that only the \emph{directions} of the variables $q^a_I$ matter. From the definition of $p^I_a$ in (\ref{conjugatemomenta}) we see that it is a \emph{unit} vector on phase space. These are all consistent and ultimately come down to temporal relationalism and Mach's second principle. The vector constraints (\ref{vecconstraintbm}) are related to the symmetries of the configurations and Mach's first principle.

Thus, $\mathcal{H}$ reflects the physically insignificant magnitude of $\dot{q}^a_I$, meaning that the evolution of the whole universe can be parametrized arbitrarily. $\mathcal{H}_\alpha$'s reflect the gauge invariance of the action with respect to the symmetry of the configuration space expressed by $G$.

Hence, as the Hamiltonian corresponding to the Lagrangian (\ref{equiaction}) vanished, we follow Dirac's approach (see Appx. \ref{constrained}), and write total Hamiltonian as

\begin{equation}\label{totalhamiltonian}
H_T = N \mathcal{H} + N^\alpha \mathcal{H}_\alpha,
\end{equation}

with $N$ and $N^\alpha$ as some arbitrary coefficients. Following the standard terminology used in ADM formalism, we call them the \emph{lapse} function and the \emph{shift} vector respectively.

It is very important to calculate the Poisson brackets between the constraints and see that they are all first-class:\footnote{One has to use (\ref{Bpartialtstructure}) for calculating the Poisson brackets between the vector constraints. Thanks to Sean Gryb for the discussion and the clarification of this point.}  \footnote{In working our the second bracket note that $\partial_{q^c_J} t^a_{\alpha b} q^b_I = \delta^J_I ...$, as is clear from the fact that $G$ acts non-trivially only on the \emph{spatial} part of $Q$ and treats all particle on the same footing.}

\begin{equation}\label{bracketsconstraints}
\begin{gathered}
\{\mathcal{H}_\alpha,\mathcal{H}_\beta\} = 0 ,\\
\{\mathcal{H},\mathcal{H}_\alpha\} = \left(\sum\limits_{I} m^{I} p_a^I p_b^J \right) \left(\sum\limits_{K} t^c_{\alpha d} q^d_K \partial_{q^c_K} h^{ab} - h^{cb} \partial_{q^c_I} t^a_{\alpha d} q^d_I - h^{ac} \partial_{q^c_I} t^b_{\alpha d} q^d_I\right).
\end{gathered}
\end{equation}

The first relation merely reflects the structure of the Lie algebra and clearly shows that $\mathcal{H}_\alpha$'s are first-class with respect to one another. The second relation is also zero in light of the equivariance condition (\ref{killing}).\footnote{Note that contrary to (\ref{killing}), in (\ref{bracketsconstraints}) the inverse of the spatial metric appears, resulting in a relative minus sign between the first term and the other two. It can be easily verified by multiplying (\ref{killing}) by $h^{ae} h^{bf}$ and using Leibniz rule.}

As we know from Dirac's formulation of constrained Hamiltonian systems, first-class constraints generally generate gauge transformations in the phase space. The status of the single scalar constraint $\mathcal{H}$ is clear: In light of \cite{Barbour.2008} and our account of it in Sec. \ref{Creparinvariant}, it generates \emph{true} evolution but with an \emph{arbitrary} parametrization. The parametrization, expressed through $N$ in (\ref{totalhamiltonian}), is the gauge. The vector constraints $\mathcal{H}_\alpha$'s generate the transformations

\begin{equation}
\begin{gathered}
\delta q^a_I \equiv N^\alpha \{q^a_I,\mathcal{H}_\alpha\} = - N^\alpha t^a_{\alpha b} q^b_I,\\
\delta p_a^I \equiv N^\alpha \{p_a^I,\mathcal{H}_\alpha\} = N^\alpha p^I_b \partial_{q^a_I} t^b_{\alpha c} q^c_I.
\end{gathered}
\end{equation}

When exponentiated, these infinitesimal transformations yield the banal transformation (\ref{banal}) we noted in the Lagrangian formulation. This ensures that Mach's first principle is satisfied in the theory.

\subsection{Best-matching conditions in the Hamiltonian formulation}

Apart from the standard variation with respect to $q^a_I$'s and the equations of motion, we posited the additional condition (\ref{bestmatchingcondition}) by which the best-matching variation was defined. So far, in our analysis of the canonical formulation of best-matching, we have not addressed this condition. It can be done analogously to the Lagrangian formulation through the phase space action.

The phase space action which gives the correct equations of motion of the phase space variables is

\begin{equation}\label{phasebestmatchingaction}
S_{PS}[q,p,\phi,\pi,N,N^\alpha] = \int d\lambda \, \left( \sum\limits_I p^I_a \dot{q}^a_I + \pi_\alpha \dot{\phi}^\alpha - H_T(q,p,\phi,\pi,N,N^\alpha) \right).
\end{equation}

As we discussed in the previous section, the best-matching variation means to vary the fields $\phi^\alpha$'s freely on the endpoints of \emph{any} interval along the trajectory. In (\ref{phasebestmatchingaction}) only the in the variation of $\dot{\phi}$ and $\pi_\alpha$ this method becomes relevant, since only these two are the parameters associated with the action of the symmetry group. Following the condition of the vanishing of the variation with respect to the $\pi_\alpha$, we find out that the values of the field at the endpoints do not appear and there would be no best-matching condition for them.

However, the variation of (\ref{phasebestmatchingaction}) with respect to $\phi$ is

\begin{equation}
\delta_{\phi} S = -\int d\lambda \, \left(\frac{\partial H_T}{\partial \phi^\alpha} + \dot{\pi}_\alpha \right) \delta \phi^\alpha + \left. \pi_\alpha \delta \phi^\alpha \right|_{\lambda_i}^{\lambda_f}.
\end{equation}

The vanishing of the first term is simply the Euler-Lagrange equation for the $\phi^\alpha$ field. However, $\delta \phi^\alpha$ is not equal to zero at the endpoints and if we demand that the above variation vanishes irrespective of the endpoints along any trajectory, we arrive at the \emph{canonical best-matching condition}

\begin{equation}\label{canbestmatchingconditon}
\pi_\alpha \approx 0.
\end{equation}

It is the counterpart of the best-matching condition in the Hamiltonian formulation. This condition is new and actually changes the Poisson structure. Using Dirac's terminology, it is a constraint to be enforced weakly, same as $\mathcal{H},\mathcal{H}_\alpha$. This constraint is first-class in the equivariant best-matching.

This condition leads to a remarkable conclusion in its connection with the vector constraints $\mathcal{H}_\alpha$. The two of them lead to the first-class constraint

\begin{equation}\label{firstclassconstraint}
C_\alpha \equiv \sum\limits_{I} p^I_a t^a_{\alpha b}q^b_I \approx 0
\end{equation}

This constraint is also first-class as it is the offspring of two first-class constraints. The meaning of this constraint is evident: It generates a gauge transformation induced by the symmetry group $G$ on the $(q^a_I,p^I_a)$ variables alone. The best-matching condition breaks the banal symmetry (\ref{banal}) in two parts, throws the transformation of the unphysical field $\phi$ away, and keeps the other part which corresponds to the transformation of the configurations.

\subsection{The equations of motion}
We can now work out the equations of motion.

In light of the best-matching condition and that the constraint (\ref{firstclassconstraint}) is first-class, Dirac's conjecture is applicable here and we define the extended Hamiltonian as (\ref{Cextendedhamil}):

\begin{equation}\label{totalhamiltonianequivariant}
H_E = N\mathcal{H} + N^\alpha C_\alpha + \zeta^\alpha \pi_\alpha.
\end{equation}

The equations of motion are

\begin{equation}\label{equmotions}
\begin{gathered}
\dot{q}^a_I = 2N \frac{h^{ab}}{m_I} p_b^I + N^\alpha t^a_{\alpha b} q^b_I,\\
\dot{p}^I_a = - N \sum\limits_{J} \frac{1}{m_J} \partial_{q^a_I} h^{bc} p_b^J p_c^J - N^\alpha p^I_b \partial_{q^a_I} t^b_{\alpha c} q^c_I,\\
\dot{\phi}^\alpha = \zeta^\alpha,\\
\dot{\pi}_\alpha = 0.
\end{gathered}
\end{equation}

The last equation is obvious. As we noted, the constraints are first-class and the consistency conditions ensure that the constraint $\pi_\alpha$ remains zero. The third equation merely reflects the arbitrariness of $\phi_\alpha$ in light of the best-matching condition. The first two equations govern the evolution of the configurations. We find the equation of motion of $q^a_I$'s alone and make sure that it is in accord with our results in the Lagrangian formulation. First, we rewrite the equations of motion for $q^a_I$ and $p^I_a$ more concisely in the form\footnote{Strictly speaking, the equations (\ref{equmotionsT}) with $-N^\alpha$ follow from (\ref{equmotions}). But this is not important due to arbitrariness of $N^\alpha$.}

\begin{equation}\label{equmotionsT}
\begin{gathered}
T_{N}^{-1} \frac{d}{d\lambda} T_{N} q^a_I = -2 N \frac{h^{ab}}{m_I} p_b^I,\\
T_{N} \frac{d}{d\lambda} T_{N}^{-1} p^I_a = N \sum\limits_{J} \frac{1}{m_J} \partial_{q^a_I} h^{bc} p_b^J p_c^J.
\end{gathered}
\end{equation}

We can inverse the first equation in (\ref{equmotionsT}) and find the conjugate momentum:

\begin{equation}\label{conjugatemomentabtransformed}
\bar{p}^I_a = -\frac{1}{2N} m_I h_{ab}(\bar{q}) \dot{\bar{q}}^b_I,
\end{equation}

where $\bar{q}^a_I = T_{N} q^a_I$ and ${\bar{p}}^I_a = T^{-1}_{N} p^I_a = (T^{-1}_{N})_a^b p^I_b$, and we also used (\ref{equivariance}).

And the second equation can be rewritten as

\begin{equation}
\dot{\bar{p}}^I_a = N \sum\limits_{J} \frac{1}{m_J} p^J_b p^J_c \partial_{\bar{q}^a_I} h^{bc}(\bar{q}).
\end{equation}

Thus, the equations of motion for $\dot{\bar{q}}^a_I$ is

\begin{equation}
\frac{1}{N} \frac{d}{d\lambda} \left(\frac{1}{2N} m_I h_{ab}(\bar{q}) \dot{\bar{q}}^b_I \right) = -\sum\limits_{J} \frac{1}{m_J} p^J_b p^J_c \partial_{\bar{q}^a_I} h^{bc}(\bar{q}).
\end{equation}

It becomes more familiar if we take $h_{ab} = -2V \delta_{ab}$. Moreover, the Hamiltonian constraint $\mathcal{H} = 0$ given in (\ref{hamconstraintbm}) gives

\begin{equation}\label{energyzerobm}
\sum\limits_I \frac{ p^I_a p^I_a}{2m_I} + V = 0.
\end{equation}

Hence,

\begin{equation}
-\frac{V}{N} \frac{d}{d\lambda} \left(-\frac{V}{N} m_I \delta_{ab} \dot{\bar{q}}^b_I \right)
=
- \partial_{\bar{q}^a_I} V.
\end{equation}

By using the reparametrization freedom and identifying

\begin{equation}\label{lapsecondition}
-\frac{N}{V} d\lambda \equiv dt,
\end{equation}

we fix $N$, and arrive at Newton's 2nd law:

\begin{equation}
m_I \frac{d^2}{dt^2} \bar{q}^a_I = -\partial_{\bar{q}^a_I} V(\bar{q}).
\end{equation}

By using (\ref{conjugatemomentabtransformed}) and (\ref{energyzerobm}), it is straightforward to see that

\begin{equation}
\frac{N^2}{V^2} = -\frac{T}{V}.
\end{equation}

Therefore, this reparametrization fixing condition is in accord with the result we arrived at Sec. \ref{revisitingmachpoincare} and (\ref{lapsecondition}) actually defines the ephemeris time (\ref{eph}).

Hence, compatibly with the results we derived in the Lagrangian formulation, from the point of view of relational particle physics, Newtonian mechanics is born when a particular conformally flat metric, and a specific time gauge are chosen. We must not forget the constraints $\mathcal{H}_\alpha$. From this perspective, we could say relational physics is more fundamental.
%\subsection{Solving the constraints in terms of the variables}

\section{Non-equivariant best-matching}\label{2nonequivariancebm}

In the last section, we explored the best-matching technique for equivariant metrics, i.e., those which are invariant under the action of the symmetry group:

\begin{equation}
T^\ast_\phi h = h.
\end{equation}

We saw that best-matching elevates the global symmetry of such theories to a gauge symmetry in a larger configuration space. This mathematical `transfiguration' turned the standard Newtonian mechanics into a Machian theory. In this chapter, we consider non-equivariant metrics which are by definition, not invariant under the action of the group. They do not even respect the symmetry induced by the group as a global symmetry. If we insist on performing best-matching, we will have the best-matching constraint to be satisfied, same as before. However, the conservation of such constraint is not even assured in the first place, let alone its vanishing. This is the issue: best-matching with respect to a group that is not a symmetry group of the action results in an inconsistency.

This can be quite disheartening, especially in light of its connection with our own universe. So far, our attempts in this first part of the thesis have been in the direction of developing the theory mathematically and we keep the physical application of relational physics to our universe for the second part. But the point is that the Newtonian $\sfrac{1}{r}$ potential - which is strongly supported by observational evidence - renders the configuration space metric non-invariant under scale transformation. However, from the relational point of view, we resolutely need scale invariance if we aspire to have a relational theory of the whole, i.e., to satisfy the Mach-Poincaré principle for the group $Sim(3)$. What can we do?

I think it is safe to say that a completely scale-free theory compatible with observations is nearly unobtainable if not entirely impossible. We will discuss this issue in more detail later in part II. For now, the matter is fortunately not that hopeless. A \emph{partially} scale-free theory can be constructed - even if the underlying metric is not equivariant - which is the main focus of this section. The physical intuition of this approach was first realized by Julian Barbour, et al. in \cite{Anderson.2005} in its application to general relativity and was later developed more mathematically by Tim Koslowski, Sean Gryb, and Henrique Gomes at Perimeter Institute \cite{Gomes.2011}.

To give an intuitive picture of how it can be done, suppose $N$ particles in the three-dimensional Euclidean space. Scaling means changing the positions of the particles by a factor. If best-matching with respect to this transformation does not work, why do we not weaken the symmetry? We can conceive of a transformation that changes the `local' scale while being subject to a global restriction. As we will see in the next chapter, In geometrodynamics, this translates into working with \emph{volume preserving} conformal transformations. We can then proceed with implementing the best-matching and Mach-Poincaré principle for that weaker symmetry.

There is a price to pay in the course of this procedure. The metric remains non-equivariant under this weaker symmetry. In order to resolve the inconsistency that might follow from enforcing the best-matching condition, we strengthen the reparametrization invariance of the theory, making it possible to fit the best-matching constraint into the theory consistently as we have now more freedom in fixing the parametrization. There is a deeper picture behind this procedure, called \emph{symmetry trading}, that essentially captures the mathematical essence of what we are doing here: We trade a larger reparametrization invariance for a symmetry in configuration space we would like to have.

In other words, we must `bribe' a non-equivariant theory into accepting a more humble symmetry as its gauge freedom. If we accept that compromise, we can have a partially scale-free theory.

We now proceed to formulate this intuition mathematically. We use a particle toy model in this section to sharpen our technique, leaving the more physically profound case of general relativity for the next chapter.

\subsection{The Lagrangian point of view}

The procedure we outlined above can be developed more powerfully and clearly in the Hamiltonian formulation. it is nevertheless insightful to see how it generally works in the Lagrangian picture.

The toy model we work with comprises $N$ independent harmonic oscillators.\footnote{There is nothing special about harmonic oscillators in this approach. It is just for making the calculations simpler.} In light of (\ref{parametrizedaction}), the reparametrization invariant action is

\begin{equation}\label{toymodelaction}
S[q,N] = \int d\lambda \, \sum\limits_{I=1}^{N} \left( \frac{\dot{q}_I^2}{2N_I} + N_I \frac{k}{2} q_I^2 \right),
\end{equation}

where the masses of the particles are set to one for convenience. The metric that is used for the Kinetic term is the Euclidean metric. We dealt with a general conformally flat metric on configuration space before and for now, we focus on a simpler toy model. The quadratic potential and its negative sign are also for simplicity purposes. As we will see, what is most important is that the $N$ systems must be independent and in other words, `local'. The reader might spot something conspicuously non-relational about the action (\ref{toymodelaction}). The concept of an independent system does have a non-relational aura indeed. Nonetheless, this is a good toy model for studying best-matching and scale transformations. Note that this approach serves solely a heuristic role as a toy model for working with general relativity later. The $N_I$'s are meant to model the local nature of time in general relativity. Here, the concept of locality is modeled by working with the individual particles. We are effectively using a theory with a finite number of degrees of freedom as a `naive' model of field theories (general relativity in particular) with continuous degrees of freedom.

Now we define the symmetry to best-match. We are looking for a scale transformation of the form

\begin{equation}
q^a_I \rightarrow e^{\phi_I} q^a_I,
\end{equation}

where $\phi_I$'s are arbitrary conformal factors for each $I$. However, this is not going to work as it is equivalent to scale transformation of the $N$ independent systems separately and in light of our above discussion this leads to an inconsistency. We define the \emph{mean} operator:

\begin{equation}
\langle \cdot \rangle \equiv \frac{1}{N} \sum\limits_{I} \cdot_I,
\end{equation}

and restrict the scale transformation parameters to obey the condition

\begin{equation}
\langle \hat{\phi} \rangle = 0.
\end{equation}

We can satisfy this identity by writing $\hat{\phi}$ as

\begin{equation}
\hat{\phi}_I =\phi_I - \langle \phi \rangle.
\end{equation}

This explicitly reduces the number of independent $\phi_I$'s to $N-1$. We can still work with the parameters $\phi_I$'s but use the transformation $e^{\hat{\phi}_I}$. This weakens the scale transformation as promised.

We now turn our attention to best-matching with respect to the parameters $\hat{\phi}_I$. The action (\ref{toymodelaction}) changes to

\begin{equation}\label{toymodelactionII}
S[q,N,\phi] = \int d\lambda \, \sum\limits_{I=1}^{N} \left( \frac{1}{2N_I} \left( \frac{d}{d\lambda} \left( e^{\hat{\phi}_I} q_I \right) \right)^2 + N_I \frac{k}{2} e^{2\hat{\phi}_I}q_I^2 \right),
\end{equation}

Using the notation we introduced in (\ref{Dphi}), we have

\begin{equation}
\dot{\bar{q}}_I^a \equiv \frac{d}{d\lambda} \left( e^{\hat{\phi}_I} q_I^a \right) = e^{\hat{\phi}_I} D_{\hat{\phi}} q^a_I = e^{\hat{\phi}_I} \left( \dot{q}^a_I + \dot{\hat{\phi}}_I q^a_I \right)
\end{equation}

If the action (\ref{toymodelactionII}) is to be best-matched, we have to impose the best-matching condition

\begin{equation}\label{toymodelcondition}
\pi_\phi^I \equiv \frac{\partial L}{\partial \dot{\phi}_I} = \frac{ \dot{\bar{q}}_I \cdot \bar{q}_I}{N_I} - \langle \frac{\dot{\bar{q}} \cdot \bar{q}}{N} \rangle = 0
\end{equation}

as can be simply checked.

The remarkable aspect of (\ref{toymodelcondition}) is that it does not impose the vanishing of the total dilatational momentum. It only requires them to be the same:

\begin{equation}
D_I(\lambda) = D_J(\lambda) = C(\lambda)
\end{equation}

at each instant for all $I,J$, where $D_I$ is the dilatational momentum of the system $I$ and $C$ is a function of $\lambda$, independent of the configuration space variables.

The vanishing of the dilatational momentum is a restrictive condition that is not satisfied in a non-equivariant theory, whereas the above condition can in principle be satisfied by fixing the lapse functions $N_I$. (\ref{toymodelcondition}) enforces only $N-1$ independent constraints. Thus, $N-1$ lapses can be fixed, and one global lapse remains that evolves the whole system. More significantly, the solutions of (\ref{toymodelcondition}) are in general \emph{non-local}, meaning that the $N_I$ satisfying (\ref{toymodelcondition}) would be generally dependent on the variables $q_J$.

This is what happens in the Lagrangian formulation. $N$ independently evolving systems `link up' together and evolve in a universal manner to keep up a local scale symmetry. Of course, this is only an intuitive picture of what happens in our abstract mathematical description of the system. Neither of the $N_I$ is physical or related to something inherently observable.

We leave out the complete calculations of $N_I$ here. More rigorous calculations and details will be given in the Hamiltonian formulation. This section is to only give an overview of how to deal with non-equivariant actions.

\subsection{The Hamiltonian formulation and symmetry trading algorithm}
The Hamiltonian formulation provides a remarkably deeper perspective into what happens in a non-equivariant model by using Dirac's approach and the language of constraints. Our starting point is to perform the Legendre transform of the Lagrangian in (\ref{toymodelaction}). The momentum conjugate to $q^a_I$ is

\begin{equation}
p^I_a = \frac{\dot{q}^a_I}{N_I}.
\end{equation}

Because of the local reparametrization invariance of each system, there are $N$ constraints associated with the variation of lapse functions,

\begin{equation}
\frac{\partial L}{\partial N_I} = -\frac{\dot{q}_I^2}{2N_I^2} + \frac{k}{2} q_I^2 = 0.
\end{equation}

Thus, we identify the Hamiltonian constraints\footnote{We have deviated from our notation convention here and used the subscript $I$ for the square of $p^I_a$ to make the expression neater.}

\begin{equation}
\chi_I = \frac{1}{2} \left( p_I^2 - k q_I^2\right) \approx 0.
\end{equation}

The Hamiltonian constraints are all first-class. The total Hamiltonian is\footnote{Strictly speaking, we should have also considered the variables $N_I$ and the momenta conjugate to them (which vanish as some constraints) in the Hamiltonian formulation. This only complicates the procedure and does not change the final Hamiltonian, as those momenta simply generate gauge transformations, rendering the $N_I$'s arbitrary. This is already respected in the Hamiltonian we arrived at.}

\begin{equation}\label{totalhamiltonianI}
H = \sum\limits_{I=1}^N N^I \chi_I.
\end{equation}

This theory is defined on the original phase space $\Gamma(q,p)$ labeled by $(q^a_I,p^J_b)$. To implement canonical best-matching, we enlarge the phase space by adding the conformal factors $\phi_I$'s and their conjugate momenta $\pi_\phi^I$'s.

From (\ref{toymodelactionII}), the conjugate momenta are

\begin{equation}
p^I_a = \frac{1}{N_I} e^{2\hat{\phi}_I} D_\phi q^a_I.
\end{equation}

Moreover, a straightforward calculation shows that the momentum conjugate to $\phi_I$ is

\begin{equation}
\pi_\phi^I = q_I \cdot p^I - \langle q \cdot p \rangle.
\end{equation}

The Hamiltonian constraints associated with the theory (\ref{toymodelactionII}) are

\begin{equation}
\chi_I = \frac{1}{2} \left( e^{-2\hat{\phi}_I} p_I^2 - e^{2\hat{\phi}_I} k q_I^2 \right)
\end{equation}

Thus, we have a theory in the extended phase space $\Gamma_E (q,p,\phi,\pi)$. In this theory, the following constraints have to be satisfied:

\begin{equation}\label{explicitconstraints}
\begin{gathered}
\chi_I = \frac{1}{2} \left( e^{-2\hat{\phi}_I} p_I^2 - e^{2\hat{\phi}_I} k q_I^2 \right) \approx 0 \\
C_I = \pi_\phi^I - \left( q_I \cdot p^I - \langle q \cdot p \rangle \right) \approx 0\\
\end{gathered}
\end{equation}

This is reminiscent of the constraints in the equivariant case (\ref{hamconstraintbm}). The total Hamiltonian is

\begin{equation}\label{totalhamiltonianII}
H = \sum\limits_{I=1}^N \left( N^I \chi_I + \theta^I C_I \right).
\end{equation}

It can be checked that the constraints $\chi_I$'s and $C_I$'s are first-class. It should not be a surprise. We have not done anything suspicious beyond the `tame' extended action (\ref{toymodelactionII}). As $C_I$'s are first-class, they generate gauge transformations. The infinitesimal gauge transformations are easy to calculate:

\begin{equation}\label{infguagetrans}
\begin{aligned}
&\delta_{\theta} q_I^a = - \hat{\theta}_I q_I^a &\qquad & \delta_{\theta} p^I_a = \hat{\theta}_I p^I_a\\
&\delta_{\theta} \phi_I = \theta_I & & \delta_{\theta} \pi^I_{\phi} = 0,
\end{aligned}
\end{equation}

where $\delta_\theta \cdot = \{ \cdot, \sum\limits_I \theta^I C_I \} $. We can easily fix this gauge freedom by imposing the condition $\phi_I = 0$ and go back to the reduced phase space $\Gamma(q,p)$ with the Hamiltonian (\ref{totalhamiltonianI}). However, we bravely tread forward and perform the best-matching in the extended phase space. The best-matching condition implies

\begin{equation}
\pi_\phi^I \approx 0.
\end{equation}

The point is that this constraint is not in general first-class with respect to $\chi_I$'s and hence, it does not propagate. This is what we did not face in the equivariant best-matching and is now solely the result of non-equivariance. This is the main issue, the solution of which is the focus of the next section.

We should note that the constraints $\pi_\phi^I$'s are redundant and this reflects the lack of complete independence of $\phi_I$'s from one another. It is easy to see

\begin{equation}
\langle C \rangle = \langle \pi_\phi \rangle \approx 0.
\end{equation}

are satisfied in the original theory regardless of best-matching. In fact, the constraint $\langle \pi_\phi \rangle$ acts trivially on phase space. Following (\ref{infguagetrans}), one can show that the change of $q^a_I,p^I_a,\hat{\phi}_I,\pi_\phi^I$ under the transformation generated by $\langle \pi_\phi \rangle$ vanish.

This is actually a blessing! Had we insisted on the full scale transformation with respect to $\phi_I$, we would have had a problem: The theory would be inconsistent and lead to `frozen' dynamics as we will shortly see.

For now, the lesson we learn from the above consideration is that only $N-1$ of the $C_I$'s are some non-trivial constraints to be satisfied. It is reasonable to simply ignore $\langle C \rangle$.

\subsection{Symmetry trading}
In the spirit of Dirac's approach, all constraints on phase space must propagate according to the Hamiltonian evolution. This is the consistency condition. The best-matching condition we imposed is first-class with respect to $C_I$'s as can be easily checked. It is intuitively clear why this should be the case: The $C_I$'s are manifestly $\phi_I$ independent and $\pi_\phi^I$ essentially generates translations in $\phi_I$. Hence, our focus should be the Hamiltonian constraints. We have the following condition:

\begin{equation}\label{propogatingnonequi}
\sum\limits_{I = 1}^N N^I \{ \chi_I , \pi_\phi^J \} = \left[
\langle N( \bar{p}^2 + k \bar{q}^2) \rangle - N^J (\bar{p}_J^2 + k \bar{q}^2_J)
\right] = 0.
\end{equation}

We must solve this condition for $N^I$. Although the expression in the bracket is generally non-zero, we can be sure that a solution exists. The reason is that the rank of the matrix $\{\chi_I , \pi_\phi^J\}$ is not $N$. We must not forget about the $\langle \pi_\phi \rangle$ and that it acts trivially on $\chi_I$. Thus, there is in principle a one-dimensional Kernel and a unique lapse function $N^I_0$ that satisfies the above constraint.\footnote{There is a mathematical subtlety here. The `matrix' appearing in the LHS of (\ref{propogatingnonequi}) is defined over the phase space and its rank is in principle non-constant and it might be $N$. However, for all of the physically interesting states, our argument is valid and we can find a non-trivial lapse function as the solution.} Therefore, the constraint

\begin{equation}
\chi_{f.c.} \equiv \sum\limits_I N^I_0 \chi_I
\end{equation}

is first class and propagates the best-matching constraint, leading to a consistent partially scale-free theory. It is now clear why we had to weaken the symmetry group in the context of the Hamiltonian formulation. If $\phi_I$ had not been restricted, the rank of the aforementioned matrix would have been $N$, leading to the trivial solution $N^I = 0$ for (\ref{propogatingnonequi}). This is the `frozen' dynamics we mentioned before. By restricting the symmetry group, we made room for the existence of a non-trivial solution to evolve the system.

It is worthwhile to note that $\alpha N^I_0$ for any $\alpha$ also satisfies (\ref{propogatingnonequi}). Thus, one global freedom is still left for the lapse function which reflects Mach's temporal relationalism. The local direction of time is only fixed by (\ref{propogatingnonequi}). Time as an external entity has not come into existence. Time is still a measure of change, but now it is the global change of the $N$ independent systems linked together.

We can redefine the remaining second-class Hamiltonian cosntraints by using some arbitrary numbers $\alpha^I_i$ where $i=1,2,\cdots,N-1$:

\begin{equation}
\chi_i \equiv \sum\limits_I \alpha^I_i \chi_I.
\end{equation}

$\alpha^I_i$ are chosen in such a way that the $N-1$ constraints $\chi_i$ are linearly independent, provided that the complete set $\{\chi_{f.c.},\chi_i\}$ includes all of the original $N$ linearly independent Hamiltonian constraints.

The constraints $C_I$ are still first-class. All $\pi_\phi^I$ are also second-class, apart from one corresponding to $\langle \pi_\phi \rangle$. In the same way, we introduce some numbers to define $N-1$ linearly independent constraints $\pi_\phi^i$. Thus, the final theory we arrived at, defined on $\Gamma_E(q,p,\phi,\pi_\phi)$, is characterized by the first-class constraints

\begin{equation}
\chi_{f.c.} \approx 0, \qquad C_I \approx 0,
\end{equation}

along with the second-class constraints

\begin{equation}
\chi_i \approx 0, \qquad \pi_\phi^i \approx 0.
\end{equation}

The total Hamiltonian is

\begin{equation}\label{thedualhamiltonian}
H_T = \chi_{f.c.} + \sum\limits_{I=1}^N \theta^I C_I.
\end{equation}

We can now use the method we explained in Sec. \ref{Celimseconclasssec} to eliminate the second-class constraints of the theory by using the Dirac bracket and imposing the second-class constraints strongly. First of all, note that

\begin{equation}\label{eliminatedsecondclasstoymodel}
\bar{H}_T = H_T|_{\chi_i = 0, \pi_\alpha^i = 0} = \chi_{f.c.}(q,p,\phi_0(q,p)) + \theta^I \bar{C}_I,
\end{equation}

where

\begin{equation}
\bar{C}_I = \sum\limits_{I=1}^N \theta^I \left( p^I_a q^a_I - \langle p \cdot q \rangle \right),
\end{equation}

and $\phi_0^i (q,p)$'s are the solutions of $\chi_i = 0$'s. It then follows that if we have a function $F$ of the variables $(q,p)$, its evolution is given by

\begin{equation}
\dot{F} \approx \{F,\bar{H}_T\}_{DB} \approx \{F,\bar{H}_T\},
\end{equation}

as the Dirac bracket and the usual Poisson bracket give the same equations of motion here.\footnote{It is always the case if the second-class constraints are functions of an equal number of some of the configuration space variables and their conjugate momenta. Thanks to Mohammad Khorrami for that observation.} We now find the explicit form of the first-class Hamiltonian and the equations of motion.

\subsection{Solving the constraints and the explicit form of the equations of motion}

The toy model we used in developing non-equivariant best-matching is fortunately simple enough to solve the constraints and derive the equations of motion explicitly. This will also enable us to construct a \emph{dictionary} between the best-matched theory and the original theory. We saw that in the non-equivariant case, best-matching can be consistently performed by trading part of the gauge symmetry of the original theory. Thus, at the level of the equations of motion for the physical variables, the two theories must match.

We have the constraints $\chi_i \approx 0$'s to be satisfied. Using the explicit expression (\ref{explicitconstraints}) we can solve the first set for $\hat{\phi}_0^i$:

\begin{equation}
e^{2\hat{\phi_{0}^{i}}} = \sqrt{\frac{p^2_i}{kq_i^2}}.
\end{equation}

These $N-1$ solutions together with the condition $\sum\limits_I \phi_{0}^{I} = 0$ determine all the $\phi_{0}^{I}$'s. We now proceed to work out the first-class Hamiltonian. Solving the lapse-fixing equation (\ref{propogatingnonequi}) can be quite complicated. But it is easy to use $\phi_{0}^{i}$'s we found and find the general form of $\chi_{f.c.}$:

\begin{equation}
\begin{aligned}
\chi_{f.c.} = \sum\limits_{I} N_0^I \chi_I &= \sum\limits_{I} N_0^I \left( e^{-2\hat{\phi}_{0}^{I}} p_I^2 - k e^{2\hat{\phi}_{0}^{I}} q_I^2 \right)\\
&= \sum\limits_{I} N_0^I \left( \exp{\left(2\sum\limits_{J \neq I} \hat{\phi}_{0}^{J}\right)} p_I^2 - \exp{\left( -2 \sum\limits_{J \neq I} \hat{\phi}_{0}^{J}\right)} k q_I^2 \right)\\
&\propto \prod\limits_{I=1}^N p_I^2 - k^N \prod\limits_{I=1}^N q_I^2.
\end{aligned}
\end{equation}

Thus, in general, we can write the first-class Hamiltonian as

\begin{equation}
\chi_{f.c.} = N \left(\prod\limits_{I=1}^N p_I^2 - k^N \prod\limits_{I=1}^N q_I^2 \right)
\end{equation}

for the arbitrary lapse function $N$ which reflects the reparametrization invariance of the theory.

Two points follow. First of all, note that the explicit expression for $\chi_{f.c.}$ is $\hat{\phi}$ invariant as expected. The whole procedure we followed was to ensure that the best-matching constraint, $\pi_\phi^I \approx 0$ propagates and the invariance of $\chi_{f.c.}$ under translations in $\hat{\phi}$ ensures that.

The next point is that we could have started with a certain initial condition in phase space that satisfies the second-class constraints. This essentially means that $\hat{\phi}_{0}^{I} = 0$ in this approach. Consistent propagation of the constraints then ensures that $\hat{\phi}_{0}^{I}$ remains zero under the evolution of the system. This method is completely equivalent to the above one and is more suitable when an explicit solution for $\hat{\phi}_{0}^{I}$ is difficult to find. The drawback, however, is that in this method the lapse fixing equation must be explicitly solved. Otherwise, the complete expression for $\chi_{f.c.}$ cannot be found. In the case of a system with three particles, this method can be followed quite easily. See \cite{Gryb.2012}, Sec. 4.2.4 for the calculations.

\subsection*{The dictionary}
We can now compare the two theories we have. One is the original theory with local reparametrization invariant. The other one, the dual theory, is a theory with local conformal symmetry but a fixed local parametrization. The two theories are essentially defined on the same configuration space with the following Hamiltonians:

\begin{equation}\label{theoriganddualtheory}
\begin{gathered}
H_0 = \sum\limits_{I=1}^N N^I \chi_I, \\
H_{dual} = N \left(\prod\limits_{I=1}^N p_I^2 - k^N \prod\limits_{I=1}^N q_I^2 \right) + \sum\limits_{I=1}^N \theta^I \left( p^I_a q^a_I - \langle p \cdot q \rangle \right).
\end{gathered}
\end{equation}

We see more explicitly how the symmetry of the original theory has been traded: Those $N^I$'s generate local reparametrization transformations for each of the variables, $N-1$ of which have been traded for the independent $\theta^I$'s that generate local scale transformations. The non-local nature of the dual Hamiltonian is also note-worthy.

We complete the construction of the dictionary between the two theories by finding the equations of motion for both of them. Those of the original theory are

\begin{equation}\label{equationorig}
\begin{gathered}
\dot{q}^a_I = 2N^I p_a^I,\\
\dot{p}^I_a = 2N^I k q^a_I,\\
\end{gathered}
\end{equation}

and those of the latter are

\begin{equation}\label{equationnonequi}
\begin{gathered}
\dot{q}^a_I = \left(2N \prod_{J\neq I} p_J^2 \right) p_a^I + \frac{N-1}{N} \theta^I q^a_I,\\
\dot{p}^I_a = \left(2N \prod_{J\neq I} k q_J^2 \right) k q^a_I + \frac{1-N}{N} \theta^I p^I_a.\\
\end{gathered}
\end{equation}

Now we see that the non-local dependence of the evolution of the phase space variables is the result of the gauge freedom and is not physically relevant. Finally, we see that the equations (\ref{equationnonequi}) turn into (\ref{equationorig}) with the gauge choice

\begin{equation}
\begin{gathered}
N^I = N \prod_{J\neq I} p_J^2,\\
\theta^I = 0.
\end{gathered}
\end{equation}

By virtue of the constraints $\chi_I \approx 0$, we have $N^I \approx N \prod\limits_{J\neq I} k q_J^2$ as well and with this gauge fixing, the two sets of equations of motion would be identical.

This establishes a complete connection between the two theories. Thus, non-equivariant actions can in principle be best-matched (at least for the simple model we considered) and the resultant theory is equivalent to the original one in a specific gauge.

\section{Canonical best-matching: An independent approach}\label{2canonicalbmindependent}

Both in equivariant and non-equivariant best-matching, we started with a geodesic principle on configuration space and then made the transition to the Hamiltonian formulation. However, although less intuitive, one can literally start independently in the canonical formulation and implement best-matching. Configuration space is a more tangible structure as it by definition consists of the ontological states that can be realized in nature. In this particular sense, phase space might be to some extent less comprehensible but it is mathematically more powerful. The importance of considering canonical best-matching more deeply and independently was mostly invigorated by \cite{Gomes.2011,Gomes.2012b} which is the subject of the next chapter.

We recapitulate the canonical best-matching in this section by constructing an algorithmic procedure for both equivariant and non-equivariant metrics. For simplicity, we set the masses of all particles to one. We can easily include the masses by tensor producing the metric by the mass tensor as we did before.

\subsection*{Equivariant canonical best-matching}

\begin{enumerate}
\item We posit the phase space $\Gamma(q,p)$ coordinatized by the conjugate variables $q^a_I$ and $p^I_a$. $\Gamma$ is equipped with the Poisson brackets

\begin{samepage}
\begin{equation}
\{q^a_I,p^J_b\} = \delta_I^J \delta^a_b.
\end{equation}
\end{samepage}

To satisfy temporal relationalism (Mach's second principle), we need a geodesic theory in the configuration space. In phase space, this translates into the Hamiltonian constraint $\mathcal{H}$ that renders the magnitude of the momenta irrelevant. We can define the Hamiltonian constraint by using a metric defined for the variables $q$:

\begin{equation}
\mathcal{H} = \sum\limits_{I} g^{ab}(q) p_a^I p^I_b \approx 0.
\end{equation}

\item We trivially extend the phase space to a larger one by adding the conjugate variables $(\phi^\alpha,\pi_\alpha)$ where $\alpha$ runs from 1 to the dimension of the symmetry group $G$. The larger phase space is denoted by $\Gamma_E (q,p,\phi,\pi) = \Gamma(q,p) \times T^\ast (\phi)$ and is equipped with the new Poisson bracket

\begin{equation}
\{\phi^\alpha,\pi_\beta\} = \delta^\alpha_\beta.
\end{equation}

To ensure that $\phi^\alpha$ is non-physical and arbitrary, we impose the primary constraint

\begin{equation}
\pi_\alpha \approx 0.
\end{equation}

Thus, we have a Hamiltonian theory on the extended phase space with the total Hamiltonian

\begin{equation}
H_T = N \mathcal{H} + \zeta^\alpha \pi_\alpha.
\end{equation}

\item We have to apply the group of transformations - which necessarily leaves the metric $g$ invariant - to the phase space variables. This can be done by performing a canonical transformation $(q,\phi,p,\pi) \rightarrow (\bar{q},\bar{\phi},\bar{p},\bar{\pi})$ generated by the generating function

\begin{equation}
F(q,\phi,\bar{p},\bar{\pi}) = \sum\limits_{I} \bar{p}^I_a T_\phi q^a_I + {\phi}^{\alpha} \bar{\pi}_{\alpha},
\end{equation}

Under this canonical transformation, we have

\begin{equation}\label{2equitransformcan}
\begin{aligned}
\bar{q}^a_I &= \frac{\partial F}{\partial \bar{p}_a^I} = T_\phi q^a_I,\\
\bar{\phi}^\alpha &= \frac{\partial F}{\partial \bar{\pi}_\alpha} = \phi^\alpha, \\
p_a^I &= \frac{\partial F}{\partial q^a_I} = T^b_a(\phi) \bar{p}_b^I, \\
\pi_\alpha &= \frac{\partial F}{\partial \phi^\alpha} = \bar{\pi}_\alpha + \sum\limits_{I} p_a^I t_{\alpha b}^a (\phi) q^b_I. \\
\end{aligned}
\end{equation}

This canonical transformation captures the action of the symmetry group $G$ on the phase space. The variables $q$'s have transformed exactly in the way we expected. The transformed momenta are also compatible with what we defined in (\ref{conjugatemomentabtransformed}) for working out the equations of motion.

\item The next step is implementing Mach's first principle and projecting the dynamics to the relational space. Mathematically, this is done by imposing the best-matching constraint

\begin{equation}
\pi_\alpha \approx 0
\end{equation}

after performing the above canonical transformation.

Note that the Hamiltonian remains invariant under this canonical transformation as

\begin{equation}
\begin{aligned}
\mathcal{H}(q,p) = \sum\limits_{I} g^{ab}(q) p_a^I p^I_b &= \sum\limits_{I} \bar{g}^{ab}(\bar{q}) \bar{p}_a^I \bar{p}_b^I\\
&= \sum\limits_{I} g^{ab}(\bar{q}) \bar{p}_a^I \bar{p}_b^I\\
&= \mathcal{H}(\bar{q},\bar{p}),
\end{aligned}
\end{equation}

where we have used the transformations (\ref{2equitransformcan}) and the equivariance condition. This ensures that the best-matching condition propogates and the theory is consistent, i.e., $\{\pi_\alpha, \mathcal{H}\} = 0$. Thus, we have the first-class Hamiltonian constraint and two first-class constraints: $\bar{\pi}_\alpha \approx 0, \pi_\alpha \approx 0$.

\item Finally, we include the first-class constraints in the Hamiltonian with some arbitrary coefficients and arrive at the Hamiltonian (\ref{totalhamiltonianequivariant}).

\end{enumerate}

\subsection*{Non-equivariant canonical best-matching}

Our canonical analysis of non-equivariant best-matching was also highly under the influence of the action principle on the configuration space. However, non-equivariant best-matching can also be independently developed. One of the main advantages of this approach is that it highlights the concept of `linking theory' and its role in general. We present a somewhat algorithmic account of the non-equivariant best-matching procedure for scale transformations.

\begin{enumerate}
\item Similarly to equivariant best-matching, we start with a phase space $\Gamma(q,p)$ equipped with the corresponding Poisson structure. We posit $N$ Hamiltonian constraints $\chi_I$ on the phase space. We then extend the phase space of the theory by introducing the non-physical variable $\phi^I$ along with the constraint $\pi_I \approx 0$. So we have a theory on the extended phase space $\Gamma_E = \Gamma \times T^\ast (\phi) $ with the following first-class constraints:

\begin{equation}
\chi_I \approx 0, \qquad \pi_I \approx 0.
\end{equation}

We call this extended theory a \emph{linking theory}. The reason is that this broad theory links two theories with smaller gauge groups: one with local reparametrization invariance expressed through $\chi_I$'s, and another one with local scale invariance but with a fixed local direction of time.

\item We introduce the symmetry by performing the canonical transformation generated by the generator

\begin{equation}
F(q,\phi,\bar{p},\bar{\pi}) = \sum\limits_{I} \left( q^a_I e^{\phi^\alpha_I} \bar{p}_a^I + {\phi}^I \bar{\pi}_I \right),
\end{equation}

Under this transformation, we have the following trivial constraint:

\begin{equation}
T_\phi \pi_I \equiv \bar{\pi}_I = \pi_I - \left(q^a_I p^I_a - \langle q \cdot p \rangle \right) \equiv C_I \approx 0,
\end{equation}

and the constraints

\begin{equation}
T_\phi \chi_I \approx 0.
\end{equation}

\item We impose the best-matching condition

\begin{equation}
\pi_I \approx 0.
\end{equation}

In non-equivariant best-matching, the constraint $\pi_I$ is generally not first-class with respect to the transformed constraints $T_\phi \chi_I$'s which might potentially lead to an inconsistency. However, provided that the symmetry group is not so large, under certain conditions there might exist a lapse function such that

\begin{equation}
\sum\limits_{I} N_0^I \{T_\phi \chi_I,\pi_J\} \approx 0.
\end{equation}

If such $N_0$ exists, then $\pi_I$'s can be treated as special gauge fixing conditions for the local reparametrization invariance generated by $\chi_I$'s. We call this step `symmetry trading'. In this case, $N_0 \sbullet T_\phi \equiv \sum\limits_{I} N_0^I T_\phi \chi_I$ is the first-class Hamiltonian constraint. $\pi_I$'s and the rest of the Hamiltonian constraints, represented by $\chi_i$'s, are the second-class constraints.

\item Elimination of the second-class constraints by defining Dirac bracket and imposing them strongly. Hence, we find $\phi^I_0$ by solving $T_\phi \chi_i = 0$ and arrive at the dual Hamiltonian 

\begin{equation}
H_T = N_0 \sbullet T_{\phi_0} + \sum\limits_{I=1}^N \theta^I \left( p^I_a q^a_I - \langle p \cdot q \rangle \right),
\end{equation}

in accord with (\ref{eliminatedsecondclasstoymodel}).

\item Finally, one can construct the explicit dictionary which relates the two physically equivalent theories with different underlying gauge groups. It must be checked that there exists a particular gauge in both theories that lead to equivalent equations of motion. This completes the construction of the dual theory as a consistent theory with the desired gauge group.

\end{enumerate}

Thus, the linking theory is the progenitor of two theories with two different gauge groups. Both of them are the result of a particular gauge fixing of the linking theory. In the non-equivariant theory we studied, the local reparametrization invariant theory is born by imposing the gauge condition $\phi^I \approx 0$ in the linking theory, while its local scale invariant twin (the dual theory) is the result of the gauge fixing condition $\pi_I \approx 0$. In the case of particle dynamics, both of these theories are physically equivalent in the sense that for any solution in one of them there exists a corresponding one in the other. It should be noted that both of these theories are also gauge theories. The gauge freedom of both of them can be fixed in a particular way and one arrives at one single `isolated' theory with physically realizable solutions: the dictionary.

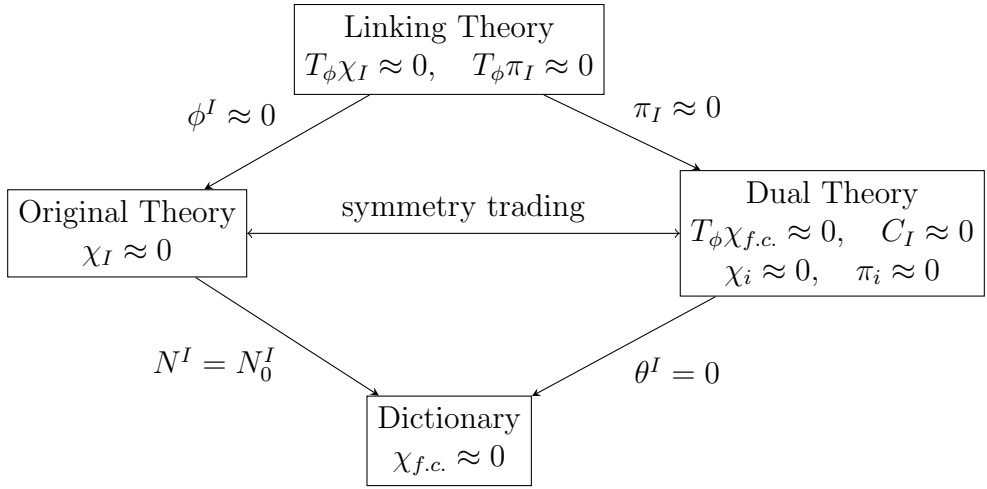
\begin{figure}[h]
\centering

\begin{tikzpicture}[every text node part/.style={align=center}]
  % Create an entity with ID node1, label "Fancy Node 1".
  % Default for children (ie. attributes) is to be a tree "growing up"
  % and having a distance of 3cm.
  %
  % 2 of these attributes do so, the 3rd's positioning is overridden.

  % Now place a relation (ID=rel1)
  \node [entity] (n1) {Linking Theory\\ $T_\phi \chi_I \approx 0,\quad T_\phi \pi_I \approx 0$}  [grow=up];
  % Now the 2nd entity (ID=rel2)
  \node [entity] (n2) [below right = of n1]	{Dual Theory\\ $T_\phi \chi_{f.c.} \approx 0,\quad C_I \approx 0$ \\ $\chi_i \approx 0, \quad \pi_i \approx 0$};
  \node [entity] (n3) [left = 5.75cm of n2]	{Original Theory\\ $\chi_I \approx 0$};
  \node [entity] (n4) [below = 4cm of  n1]	{Dictionary\\ $\chi_{f.c.} \approx 0$}; 
  % Draw an edge between rel1 and node1; rel1 and node2
  \path[-stealth] (n1) edge node [above  right] {$\pi_I \approx 0$} (n2)
                edge node [above left] {$\phi^I \approx 0$} (n3);
  \path[-stealth] (n2) edge node [below right] {$\theta^I = 0$} (n4);

  \path[-stealth] (n3) edge node [below left] {$N^I = N^I_0$} (n4);

  \path[<->] (n2) edge node [above] {symmetry trading} (n3);
\end{tikzpicture}

\captionsetup{width=0.8\linewidth}
\caption[Illustration of the symmetry trading procedure]{\small The diagram illustrating different gauge fixing of the linking theory which gives rise to two theories. The dictionary can be constructed by a final gauge fixing of both of the theories.}
\label{linkingtheory}
\end{figure}

Fig. \ref{linkingtheory} illustrates the structure of non-equivariant canonical best-matching and symmetry trading more clearly.

%% file: Chapter4.tex
\chapter{Shape Dynamics of Geometry and Fields}

\section{The historical evolution of shape dynamics}

\subsection{Einstein and Mach's Principle}

The story of Mach's Principle began with Einstein who picked up the idea from an early age and kept considering it as an important guiding principle until several years after the devlopment of his great theory of general relativity. It was actually Einstein who dubbed the term `Mach's Principle', a phrase which has stood with us for more than a century. Einstein's advocacy of Mach's Principle, however, was not that plain and simple. He repeatedly changed his attitude towards the principle, used different formulations over the course of his work on what later became general relativity, and finally dismissed the idea altogether and considered it obsolete. The lack of clarity of Einstein's formulations and understanding of the principle, along with Mach's own vague statements did inflict some lasting negative connotations even on the phrase itself.\footnote{I personally think this is another strong reason to refer to this principle as the `Mach-Poincaré Principle'!}

Many historians have analyzed the historical evolution and the role of Mach's ideas in Einstein's works and have come up with many interesting insights.

To see clearly why Einstein could not fit his understanding of Mach's Principle into the framework of his theory, which led to even more misunderstanding, we should start with the development of special relativity and Hermann Minkowski's contribution to it shortly afterwards, and pinpoint the source of the problem. 

Einstein's 1905 discovery of special relativity was the result of his long-standing discontent with Newton's absolute notion of space and time, and in particular, the disturbing discrepancy between Newtonian mechanics and electromagnetism. He solved the problem by putting the constancy of speed of light which is at the heart of Maxwell's equations first, and modify the theory of mechanics. His physical intuition was sound and valid. Shortly after that in 1908, Hermann Minkowski, Einstein's former teacher at Zurich polytechnic\footnote{Interestingly, Minkowski once called Einstein `a lazy dog' for not attending his lectures as a student!} put the theory in a concise mathematical formulation by constructing a mathematical structure, now known as the `Minkowski spacetime' \cite{Minkowski.1923}. 

The key thing about Minkowski spacetime is the unification of time and space into a single structure called spacetime. Time is considered as another physical dimension, same as space, and this way one formulates the dynamical laws of physics in a `timeless' manner. Einstein at first dismissed this idea, calling it a `a pointless mathematical complication', but soon he changed his mind and used Minkowski's mathematical spacetime as his starting point to unify gravity with his special relativity.

Einstein frequently talked about his three guiding principles in developing his theory: the relativity principle or the principle of general covariance, the equivalence principle, and what later became known as Mach's principle. I will focus on the status of Mach's Principle and how it shaped Einstein's thinking over time here.\footnote{There are very valuable papers in \cite{Barbour.1995} which came out of a conference on Mach's Principle in July 1993. I mostly rely on Hoefer's contribution \cite{Hoefer.1995} for my historical account here.} 

Einstein's earliest expression of what he later called Mach's Principle was the idea that the whole inertia of any material point should be an effect of the presence of all other masses, depending on a kind of interaction with them. At the time, Einstein had not started to use tensor calculus and geometry to develop his theory, and he still had the intuition behind Newtonian particle ontology in mind. Later in the period in which he was working on the first formulation of his theory in light of his equivalence principle, he equated general covariance, Mach's Principle, and the equivalence principle. 

As regards the connection between the equivalence principle and Mach's Principle, Einstein argued that by virtue of the equilvalence principle, one cannot differentiate a centrifugal force from a gravitational field in the case of a rotating object (like Newton's bucket). In this way, one may regard a rotating system as at rest and the centrifugal field as a gravitational field. He also wanted his theory to be generally covariant. This, according to Einstein, would ensure that Mach's Principle holds. He regarded the principle of general covariance as the generalization of his previous relativity principle and that would make the apparently priviledging role of uniform transformations disappear, and the lack of existence of any priviledged reference frame would incorporate and vindicate Mach's critique of Newton's absolute motion. 

Furthermore, in the 1913-1915 period that Einstein was vigorously working on his theory, he also articulated Mach's ideas as the demand that the metric field should be fully determined by the material distribution, in other words, in the absence of any matter there should not be any solution for the metric. Minkowski spacetime definitely posed a serious challenge for this way of understanding Mach's Principle, as it is a vacuum solution.

Einstein published his first formulation of his theory, the \emph{Entwurf theory} in the summer of 1913, and maintained that the theory is Machian. This is very strange! Entwurf theory lacked general covariance, to which Einstein referred as an `ugly dark spot' in his letter to Lorentz in August 1913. But Einstein had equated Mach's ideas with the covariance principle! What is more striking, is that the Minkowski spacetime was a valid vacuum solution of the Entwurf equations, and this is in direct contradiction with his understanding of Mach's Principle.

The problem of the general covariance of the Entwurf theory was resolved when Einstein finally found general relativity in November 1915. But the problem with the Minkowski spacetime solution remained. Einstein tried to solve this problem by imposing some suitable boundary conditions to rule out `non-Machian solutions' such as the empty Minkowski spacetime. He had begun working on this conditions by May 1916 as a letter to Besso indicates, but as Hoefer suggests, he might have had the idea much further back in the period in which he was working on the Entwurf theory. 

Until December 1916, Einstein kept working on `Machian boundary conditions', but gradually became disappointed with this approach. By that time, he had already focused on the cosmological solutions of his theory and tried to implement Mach's Principle by considering a topologically closed universe.

In the 1917-1918 period, Einstein formulated Mach's Principle as the requirement for a closed and finite universe with no boundary, and that there should be no matter-free and singularity-free solutions. He kept emphasizing that the $g_{\mu \nu}$ field should be determined by the matter alone both in his letter to de Sitter of March 24, 1917, and more explicitly in his 1918 review paper \cite{Einstein.1918b} in which he dubbed the phrase `Mach's Principle'.

However, de Sitter found a matter-free solution to the new field equations with the cosmological constant in early 1917. Einstein struggled to show that either this solution had a singularity and was physically unacceptable, or not matter-free at all. He failed. Eventually, he gave up his attempts in June 1918 and this marks the end of Einstein's strong advocacy of Mach's Principle. 

After 1918, Einstein kept referring to some Machian ideas in his writings occasionally. He mostly cited certain Machian phenomena, such as the frame dragging effect, which were implemented in his theory, but not the full principle itself in a rigorous and mathematical manner. 

Ultimately, a letter to Felix Pirani in 1954 indicates Einstein's complete repudiation of the principle:

\begin{samepage}
\begin{displayquote}
\emph{
One should not talk at all any longer of Mach’s principle, in my opinion. It arose at a time when one thought that `ponderable bodies' were the only physical reality and that in a theory all elements that are fully determined by them should be conscientiously avoided. I am quite aware of the fact that for a long time, I, too, was influenced by this fixed idea.}
\end{displayquote}
\end{samepage}

But why did Einstein have such a complicated attitude towards Mach's Principle? 

It seems that there are two main reasons behind it: one is the spacetime approach which was initiated by Minkowski, the other is Einstein's confusion and misunderstanding of the underlying ontology of his own theory. 

Space and time are inherently different entities, especially from the relational point of view. They should not be treated on the same footing. Space emerges from the relational states of the universe, and time emerges from the change of those relations. From this perspective, it must be the configuration space of the relational states that comes first. Unifying space and time goes against the relational understanding of either one of them, and obscures the essence of what we formulated as the Mach-Poincaré Principle. We will see how clearly relational ideas can be implemented in a dynamical theory of geometry and fields, and how spacetime can emerge from such a Machian approach. 

Moreover, in judging the value and usefulness of Mach's Principle near the end of his life, Einstein confused the matter of ontology with the true criterion of Machianity. The resolution of this conundrum has been the privotal matter in the contemporary relational physics project. 

To make the matter more clear, consider Einstein's field equations:

\begin{equation}
R_{\mu \nu} - \frac{1}{2} g_{\mu \nu} R = k T_{\mu \nu}.
\end{equation}

This equation is not for determining the metric $g_{\mu \nu}$ from the matter fields on the RHS. Actually, both the LHS and the RHS do in principle contain derivatives of the metric and the matter fields with respect to time and space in parallel. Einstein's field equations govern the evolution of the matter and the \emph{metric}. 

In light of this note and the above realization that the spacetime approach is the culprit behind the obscure status of Machianity of GR, we take a step forward to identify an alternative way of understanding the Machian core of Einstein's theory. Fortunately, we already have a formal definition of Mach's Principle. We should now discuss its application to general relativity.

\subsection{Mach-Poincaré Principle and general relativity}\label{3Mach-poincaregr}

We should begin this section with an important question: What is the configuration space of GR?

John Wheeler asked this question. The dynamical core of Einstein's theory is veiled behind the spacetime structure. If we think about this question, Einstein's theory describes the evolution of the geometry and the matter fields in a spatial extent through time. Hence, we can start with a Riemannian manifold, representing our universe, and introduce the metrical structure and the matter fields. Now we see that geometry has its own ontological status in the theory. Focusing on only geometry, we can identify the configuration space as `the space of all possible Riemannian metrics on a certain manifold with a given topology', denoted by $Riem^3$. 

Now we can apply the Mach-Poincaré Principle to such a configuration space. The ingredient we need is the symmetry group of this configuration space. The observable role of the metric is that we can define `length' with it. Now let us assume we have identified the points of a manifold, and have set an arbitrary coordinate system and assigned some coordinates to each point, e.g $x^a (P)$. What we can do is to measure the \emph{angles} between lines in different directions each point, i.e., quantities like

\begin{equation}\label{3ratiosgeo}
\frac{g(X,Y)}{\sqrt{g(X,X)}\sqrt{g(Y,Y)}}
\end{equation}

at each point of the space. We can therefore measure the angles between the lines emerging along the coordinate basis vectors $\partial_a$ at each point and this gives us certain ratios between the components of the metric. In the three-dimensional space, there are 2 measurable numbers at each point. So two of the total 6 components of a three dimensional Riemannian metric actually represent the observable content of the metric. This is in accord with the field-theoretic approach to gravity as a spin-2 field. The massless spin-2 field has 2 polarizations. 

The other 4 components of the metric are redundant and are the \emph{handles} of gauge transformation. Three of them correspond to diffeomorphisms. The coordinate system we considered was arbitrary and one could write the components of the metric in any other smooth coordinate system. The transformed metric under a diffeomorphism $\psi$ from the manifold to the manifold, $\psi^\ast g_{ij}$ is as good as $g_{ij}$. This introduces a local three-parameter group of transformations, the diffeomorphism group. Morover, in considering only the ratios (\ref{3ratiosgeo}) we removed the local scale of the metric. This corresponds to the action of the conformal group. The conformal group of transformations comprise of any smooth, positive definite function $\phi(x)$ which transforms the metric as $g_{ij} \rightarrow \phi g_{ij}$. This transformation also leaves the observable content of the metric invariant. 

Hence, the symmetry group should be $Conf \ltimes Diff$, the group of all conformal and diffeomorphism transformations on a certain manifold with a given topology. 

The quotient of $Riem^3$ by diffeomorphisms was called the \emph{superspace} by Wheeler, denoted by $Sup$. We can construct the \emph{conformal superspace} (the shape space) CS, by quotienting superspace by the conformal transformations. 

Einstein's theory satisfies the criterion of Machianity in superspace. The BSW action

\begin{equation}\label{3BSWaction}
S_{BSW}[g,N^a] = \int dt d^3x \, \sqrt{g}  \sqrt{R \, G^{abcd} \left( \dot{g}_{ab} - 2\nabla_{a} N_b \right) \left( \dot{g}_{cd} - 2\nabla_{c} N_d \right) },
\end{equation}

derived in Appx. \ref{CADMformalism}, is equivalent to Einstein-Hilbert action for globally hyperbolic spacetimes and as we soon will see more clearly, implements best-matching with respect to diffeomorphisms through the shift vector $N_a$. Einstein was confused about the Machian nature of his theory, but his theory was certainly not! Without even having the formulation of the Mach-Poincaré Principle in mind, he created a theory that not only satisfies that principle to some extent, but also embodies the best-matching procedure in a natural way! 

We are now in the position to dispell some of the conceptual difficulties Einstein faced in implementing Mach's Principle. First of all, by virtue of describing the universe as an evolution of geometry unfolding in the configuration space, we do not need to look for Machian ideas in spacetime. Machianity translates into the requirement for the dynamics of the universe to take place in the relational configuration space. Moreover, this prompts us to look at certain seemingly non-Machian solutions, e.g., the Minkowski spacetime, in a new light. If we remember that the theory is a dynamical one of both the geometry and the matter fields, the Minkowski spacetime should not be considered as an absolute and rigid structure. It is just one particular `dynamical' solution to the equations. As Julian Barbour put it more clearly in \cite{Barbour.1995b},\footnote{Note that at the time of writing that paper the role of scale and conformal transformations was not clearly understood. Therefore, the claim made in that paper must be moderated, and as we will see, general relativity is equivalent to an \emph{almost} Machian theory.}

\begin{samepage}
\begin{displayquote}
\emph{
Any solution of pure geometrodynamics, i.e., any Ricci-flat spacetime, is not to be analyzed as a matter-free structure in which test particles have inertia or as a structure that has a disconcerting resemblance to Newton's absolute space and time but as a dynamical history of 3-geometries. The Machian requirements apply to the structure of such a dynamical history, not the behavior of test particles within it... The real difference between the Newtonian absolute structures and all Einsteinian spacetimes is that the former are a rigid kinematic framework for dynamics, whereas the latter arise as a result of Machian dynamical evolution in a situation in which there are no preexisting frameworks at all.
}
\end{displayquote}
\end{samepage}

\subsection{Towards a theory of geometrodynamics on conformal superspace}

Just a few years after the Tubingen conference on Mach's Principle, Julian Barbour with the help of his collaborators set out to implement the relationality of size and moved towards a completely relational description of the universe both in the particle model and geometrodynamics. This task was not inherently unattainable from the technical point of view, as in 1982 Barbour and Bertotti already found a correct mathematical procedure to construct a relational dynamical theory with respect to \emph{any} Lie group. The matter was actually more conceptually challenging.

The first attempt to include the conformal transformations in geometrodynamics was Barbour and Ó Murchadha's joint paper \cite{Barbour.1999}. In that paper, they tried to develop a scale-invariant theory that satisfies the strong version of the Mach-Poincaré Principle in the conformal superspace. In a nutshell, what they did was to include the conformal transformations

\begin{equation}
g_{ab} \rightarrow e^{4 \phi} g_{ab},
\end{equation}

for an arbitrary smooth function $\phi$ defined on a closed Riemannian manifold in an action for geometrodynamics and perform the best-matching variation.

But one cannot use the standard BSW action (\ref{3BSWaction}) for that. The reason is that the BSW action is not invariant under a global rescaling of the metric in the first place, and performing best-matching with respect to the local form of this transformation would lead to inconsistency.\footnote{In the canonical formulation, this means that the best-matching constraint would not propagate to close the constraint algebra.} They proposed the following simple modification of a BSW-type action:

\begin{equation}\label{3conformalizedBSW}
S[g] = \int dt \, \frac{\int d^3x \, \sqrt{g} \sqrt{R \, G^{abcd} \left( \dot{g}_{ab}\right) \left(\dot{g}_{cd}\right)}}{V^{2/3}},
\end{equation}

where $V = \int d^3x \, \sqrt{g}$ is the volume and we ignored the shift vector $N^a$ for brevity.

The action (\ref{3conformalizedBSW}) is invariant under the global scaling $g_{ab} \rightarrow \alpha \, g_{ab}$ ($R$ transforms to $\alpha R$). But to implement best-matching with respect to the full local scale transformation, we add the $\phi(x)$ field to the action and perform the transformation $g_{ab} \rightarrow e^{4\phi} g_{ab}$ and write

\begin{equation}\label{3conformalizedBSWbm}
S[g,\phi] = \int dt \, \frac{\int d^3x \, \sqrt{g} e^{4\phi} \sqrt{R - \frac{8 \nabla^2 e^\phi}{e^\phi}} \sqrt{\bar{T}}}{\left(\int d^3x \, e^{6\phi} \sqrt{g}\right)^{2/3}},
\end{equation}

where the transformed kinetic term is

\begin{equation}
\bar{T} = G^{abcd} \left( \dot{g}_{ab} + 4 \dot{\phi} g_{ab} \right)
\left( \dot{g}_{cd} + 4 \dot{\phi} g_{cd} \right).
\end{equation}

Performing the best-matching variation leads to the condition

\begin{equation}\label{3zeromccondition}
p \equiv \Tr p^{ab} = g_{ab} p^{ab} = 0,
\end{equation}

in other words, the trace of the extrinsic curvature must vanish identically at each instant. This condition also fixes the lapse function. The Hamiltonian constraint of this theory has only a solution if and only if the initial metric belongs to the Yamabe-non-negative class (see \cite{Mercati.2018}, Chapter 8). This condition is overly restrictive.

Hence, a completely scale-free theory of geometry is mathematically possible, but is highly restrictive. Moreover, one has to modify the standard BSW action and use a globally scale-invariant action such as (\ref{3conformalizedBSWbm}). It could be very hard to reconcile the `global' coupling that the volume term in the denominator induces with the verified cosmological predictions of GR.

Scale-invariant particle mechanics was also developed in \cite{Barbour.2003}. But it too had some deep issues with observations. In particular, such a toy particle model cannot explain the structure formation that has happened in our universe and explains only a short range of the redshift we observe.

In short, it is fair to say that the project of constructing a completely scale-invariant relational theory, whether with particle ontology or field ontology, is defeated. We will come back to the issue with scale-invariant particle mechanics in the next chapter.

During this time, the full potential of the BSW action and the various matter fields one could introduce in the theory was also studied and developed rigorously in \cite{Barbour.2002}.

Julian Barbour pondered the issue of scale invariance for several years and finally came up with an idea to use the slightly weaker group of `volume-preserving conformal transformations' (VPCTs) instead \cite{Anderson.2005}. This idea worked! A straightforward way to mathematically realize this idea is to use the transformation

\begin{equation}
g_{ab} \rightarrow e^{4\hat{\phi}} g_{ab},
\end{equation}

where $\hat{\phi}$ denotes the VPCT, defined in terms of an unrestricted smooth function $\phi$ as

\begin{equation}
\hat{\phi}(x) \equiv \phi(x) - \frac{1}{6} \log \langle e^{6\phi} \rangle_g,
\end{equation}

where $\langle f \rangle_g = \frac{\int d^3x\, \sqrt{g} f}{V}$. One can show that $\hat{\phi}$ does indeed leave the total volume invariant.

The authors used the BSW action in \cite{Anderson.2005} and performed the best-matching with respect to VPCTs. That paper is not totally flawless, and the best-matching was not performed clearly and rigorously, leading to some confusion. However, the results they obtained withstood the lack of rigor. They correctly found that using VPCTs leads to the CMC foliation (constant-mean-extrinsic curvature) condition

\begin{equation}\label{3cmcconditionconstant}
\frac{p}{\sqrt{g}} = const.
\end{equation}

That is, the trace of the extrinsic curvature must be spatially constant, not identically zero as is required by (\ref{3zeromccondition}) in the previous model.

The resulting lapse fixing equation can also be solved for any initial metric, as opposed to the more restrictive case of the full scale-invariant theory. More strikingly, as discovered by James York quite by chance, the transformed initial Hamiltonian constraint, known as the Lichnerowicz-York equation, can also be solved uniquely for the conformal factor $\phi$ for any 3D Riemannian metric in this approach \cite{OMurchadha.1973}.\footnote{Following Lichnerowicz's method for solving the Hamiltonian constraint of the standard GR in the initial-value formulation, James York tried to remedy the shortcoming of the previous method by adding one extra number to the momenta to give it a non-vanishing trace, known as the \emph{York time}, $\tau = \frac{2}{3} \frac{p}{\sqrt{g}}$, which he found by trial and error. See \cite{Mercati.2018}, Chapter 8 for an account of his approach.} The remarkable thing is that Julian Barbour discovered the physical principle that explains the success of York's method for solving the initial-value problem: It is the implementation of VPCTs for constructing a relational theory. Once again, Einstein's equations seem much more Machian than what their father wished them to be!

The next step in the course of developing shape dynamics was the clear identification of the required initial conditions in light of the previous works on VPCTs and the initial-value problem \cite{Barbour.2010b}. Barbour and Ó Murchadha articulated all the previous findings and clearly identified the minimum initial data needed to solve Einstein's equations. Their finding showed that the dynamical core of GR satisfies the `weak' form of the Mach-Poincaré Principle.\footnote{Historically, it is noteworthy that the distinction between the strong and weak forms of the principle was made in \cite{Barbour.2010} around the same time, in connection with the aforementioned paper.}

Assume we start with a metric in $CS$, representing the shape of the universe, and its initial `change'. The metric in $CS$ contains 2 degrees of freedom per point, and its change contains 2 local degrees of freedom minus one global number which represents its `rate of change' with respect to some imaginary external clock. This is what we should `have' in principle. Now we work out what we `need'.

Einstein's equations of motion are invariant under 3D spatial diffeomorphism, and hence, we have the diffeomorphism constraints as the result of that gauge invariance which require the momentum to be transverse. So 3 degrees of freedom of the metric are irrelevant as they can be fixed by the gauge fixation, and also the transverse-lessness of the momentum removes 3 numbers per point. Now if we assume that we can employ York's method and work in CMC foliation, we need the initial value of $p/\sqrt{g}$ which is just one single pure number and is related to York time. Moreover, we have the Hamiltonian constraint to satisfy. Following York's method, we can start with a metric and apply a VPCT to it and solve for the conformal factor. Thus, in principle, the Hamiltonian constraint removes one local degree of freedom of the metric minus one global number. Moreover, the equations of motion are reparametrization invariant and this renders one more number unnecessary. Ultimately, we need $2+2$ local degrees of freedom minus one global number, one global number for the total volume, and one for the York time. This would completely solve the initial-value problem of GR and one could evolve the metric and arrive at a `spacetime' by putting the spatial geometries at all instants, the lapse function (which is already determined as the result of the CMC condition), and the shift vector together.

But we saw that the initial point and a direction in $CS$ give us only $2+2$ local degree of freedom (minus one number) for the initial shape and the change of the shape of the whole universe. Therefore, we end up needing $2$ pure numbers for the initial volume and the York time.\footnote{As we will see, York time is actually the momentum conjugate to the total volume and this way of counting fits together coherently.}

But the solution to this dilemma, as was found in \cite{Barbour.2010b} is that the equations of motion have a hidden symmetry corresponding to a simultaneous global scaling of the volume and the York time. This kind of global symmetry is called \emph{dynamical similarity} and maps solutions of a theory to solutions, parametrized by a set of numbers. A more familiar form of a dynamical similarity is the one corresponding to Newton's second law with a gravitational potential. One can see that under the transformation

\begin{equation}
q_I^a \rightarrow \alpha q_I^a, \quad t \rightarrow \alpha^{3/2} t
\end{equation}

the law

\begin{equation}
\frac{d^2 q_I^a}{dt^2} = - \sum\limits_{J\neq I} \partial_a \frac{m_J}{|q_I-q_J|}
\end{equation}

is invariant and the observable content of the evolution of the system would remain untouched. In fact, this transformation encapsulates the global scale invariance of Newtonian mechanics and what we are here suggesting is an analogous dynamical similarity corresponding to the global scale in GR.

Hence, one of the two global numbers we need for solving the equations of GR in CMC foliation is redundant and represents the irrelevant total scale of the evolution of the system. The punchline is that even if we fix the global scale of the theory, we would still need the initial value of York time and the strong version of the Mach-Poincaré Principle cannot be satisfied. This can be salvaged by introducing an external clock that measures the York time and the value of York time can be read off from the magnitude of the velocity of the change of the metric in $CS$ (which is unobservable in $CS$). This satisfies the weak form.

I will derive and explain this conclusion more rigorously in the next section, as I am more focused on the historical evolution of the theory here and want to put decades of work that has been done on the subject in a historical perspective.

The next leap in the formulation of shape dynamics in light of Julian Barbour's idea to use VPCTs - which not only solved the defects of the previous formulations but also revealed the inner working of the theory - was taken by Henrique Gomes, Sean Gryb, and Tim Koslowski in 2010, \cite{Gomes.2011}. They skillfully used the full power of Dirac's formulation of constrained systems and developed the theory rigorously. We already gave an account of how their approach works using a particle toy model in Chapter 3, Sec. \ref{2nonequivariancebm}. They showed that shape dynamics is equivalent to the ADM formulation of GR in the CMC condition.

Shortly afterward, Gomes and Koslowski discovered the Linking theory, the progenitor of GR and SD \cite{Gomes.2012b}, and also studied the coupling of matter to shape dynamics \cite{Gomes.2012}. See \cite{Gomes.2013} for a review of the basic features of shape dynamics and the clarification of some misunderstandings.

In the last decade, the cosmological applications of shape dynamics, both the particle model and the field model, have been explored and it has led to several breakthroughs, the most notable of which is the solution to the problem of the arrow of time \cite{Barbour.2013}. These findings are the subject of the next chapter of this work. For now, we should turn our attention to the final formulation of shape dynamics.

\section{The formulation of shape dynamics}
As we defined, shape dynamics is a theory of geometry and matter that satisfies the Mach-Poincaré Principle by embodying the conformal transformations and diffeomorphisms as the gauge symmetry. We have to first specify the configuration space. Working with pure geometry, we posit a closed (compact and without boundary) spatial manifold $\Sigma$ and identify the configuration space as $Riem^3$. As we argued in Sec. \ref{3Mach-poincaregr}, diffeomorphisms and conformal transformations do not alter the observable content of the geometry, hence, we consider the quotiented space $CS$ as the shape space of geometrodynamics. In light of our historical account, it soon became known that one extra ingredient (the total size) must be included as well. Thus, we work with the conformal superspace \emph{plus} size (CS+V), denoted by $CS^V$. We define a metric on $Riem^3$ in the form of the BSW-type action:

\begin{equation}
\langle \delta g^1 , \delta g^2 \rangle = \int_{\Sigma} d^3x \sqrt{g} \,\sqrt{V(g,\partial g, \cdots) G^{abcd} \delta g_{ab}^1 (x) \delta g_{cd}^2 (x)},
\end{equation}

with the potential $V$ and the supermetric

\begin{equation}
G^{abcd} = g^{ac} g^{bd} - \alpha g^{ab} g^{cd},
\end{equation}

for the kinetic term with an unspecified parameter $\alpha$. One could take the square-root \emph{outside} the integral and define another metric. This alternative model would be highly non-local, coupling all the distant points in $\Sigma$.\footnote{See Chapter 5 in \cite{Gryb.2012} for further details.}

Using the above metric, we can define the action

\begin{equation}\label{3actionriem3prim}
S[g] = \int\limits_{\Sigma} d^3x \, \sqrt{gV G^{abcd} \dot{g}_{ab} \dot{g}_{cd}}
\end{equation}

on $Riem^3$. The conjugate momenta are

\begin{equation}
p^{ab} (x) \equiv \frac{\delta \mathcal{L}}{\delta g_{ab}} = \sqrt{\frac{gV}{T}} G^{abcd} \dot{g}_{cd} (x)
\end{equation}

which immediately leads to a primary constraint corresponding to the reparametrization invariance of (\ref{3actionriem3prim}):

\begin{equation}
H[g,x) = \frac{G_{abcd} p^{ab} p^{cd}}{\sqrt{g}} - \sqrt{g} V.
\end{equation}

We call this the Hamiltonian constraint. Now, if we take one further step and implement best-matching with respect to diffeomorphisms, we see that diffeomorphisms leave the kinetic term invariant, provided that the potential $V$ is scalar. Hence, this procedure would be an equivariant best-matching. Under infinitesimal diffeomorphisms $x^a \rightarrow x^a + \xi^a$ , the metric transforms as

\begin{equation}
\begin{gathered}
g_{ab}(x) \rightarrow \bar{g}_{ab}(\bar{x}) = g_{ab} (x) - (\partial_a \xi^c) g_{bc} - (\partial_b \xi^c) g_{ac}\\
\implies \bar{g}_{ab}(x) = g_{ab} (x) - \mathfrak{L}_\xi g_{ab} (x) = g_{ab}(x) - 2\nabla_{(a} \xi_{b)}.
\end{gathered}
\end{equation}

Thus, under best-matching, the $\dot{g}_{ab}$ term in (\ref{3actionriem3prim}) transforms as\footnote{In light of our discussion near the end of Sec. \ref{groupaction}, we should note that for simplicity, we are here using the generators of the diffeomorphisms evaluated at the identity element for lifting the metric. It does not change the essence of equivariant best-matching with respect to diffeomorphisms.}

\begin{equation}\label{3transformdiffeoderivative}
\dot{g}_{ab} \rightarrow \dot{g}_{ab} - 2 \nabla_{(a} \dot{\xi}_{b)}.
\end{equation}

As the best-matching variation requires that we vary the action with respect to $\dot{\xi}$ directly, putting (\ref{3transformdiffeoderivative}) in (\ref{3actionriem3prim}) effectively leads to the BSW action (\ref{3BSWaction}), provided that $V=R[g]$ and $\alpha = 1$ to ensure that the theory would be consistent and the best-matching constraint propagates. Best-matching with respect to diffeomorphisms leads to the familiar momentum constraints

\begin{equation}
H^a = -2 \nabla_b p^{ab}
\end{equation}

we have derived in Appx. \ref{CADMformalism}.

As we intuitively indicated, to construct an almost scale-free consistent theory, we have to perform best-matching with respect to VPCTs. The action (\ref{3actionriem3prim}) is not invariant under the global form of these transformations, meaning that we have to perform non-equivariant best-matching. Fortunately, we already have a strong local gauge invariance by virtue of the Hamiltonian constraint, and we see the possibility of symmetry trading on the horizon. Thus, we construct canonical formulation of shape dynamics following our non-equivariance best-matching procedure outlined in Sec. \ref{2canonicalbmindependent}.

\subsection{Enlarging the phase space}

The phase space $\Gamma$ of ADM formulation of general relativity on $\Sigma$ is coordinatized by 3-dimensional Riemannian metrics $g_{ab}$ and its conjugate momenum density $p^{ab}$ of density weight 1.\footnote{It means that $p^{ab}/\sqrt{g}$ transforms as a 2-tensor.} $\Gamma(g_{ab},p^{ab})$ is equipped with the fundamental Poisson bracket

\begin{equation}
\{ A^{ab} \sbullet g_{ab} , B_{cd} \sbullet p^{cd} \} = A^{ab} \sbullet B_{ab} = \int_\Sigma d^3x \sqrt{g} \, A^{ab}(x) B_{ab}(x),
\end{equation}

for any smooth square-integrable scalars $A$ and $B$. The dynamics on $\Gamma$ is given by the first-class constraints

\begin{equation}
\begin{gathered}
H = \frac{1}{\sqrt{g}} \left( p^{ab} p_{ab} - \frac{1}{2} p^2 \right) - \sqrt{g} R, \\
H^a = - 2 \nabla_b p^{ab},
\end{gathered}
\end{equation}

where $p = g_{ab} p^{ab}$. We have used the DeWitt supermetric (\ref{CDeWittsupermetric}).

We extend the phase space to $\Gamma_E (g_{ab},p^{ab},\phi,\pi_\phi)$ by adding the scalar field $\phi$ and its conjugate momentum $\pi_\phi$ defined on $\Sigma$ and impose the first-class constraint

\begin{equation}
\pi_\phi \approx 0.
\end{equation}

to ensure that $\phi$ is redundant and a gauge variable.

\subsection{Introducing the VPCTs}

We now perform a canonical transformation that implements VPCT's on the extended phase space. The total volume is by definition

\begin{equation}
V_g \equiv \int_\Sigma d^3x \, \sqrt{g}.
\end{equation}

We define the restricted field $\hat{\phi}$ to apply a VPCT, as opposed to the unrestricted original $\phi$ in the following way:

\begin{equation}
\hat{\phi} = \phi - \frac{1}{6} \log \langle e^{6\phi} \rangle_g,
\end{equation}

where $\langle f \rangle_g = \frac{1}{V_g} \int_\Sigma d^3x \, \sqrt{g} f$ for any smooth function $f$ on $\Sigma$. One can then see that the conformal transformation $e^{4\hat{\phi}}$ leaves $V_g$ invariant.

The functional

\begin{equation}
F[g_{ab},P^{ab},\phi,\Pi_\Phi] = \int_\Sigma d^3x \, \left( g_{ab} e^{4\hat{\phi}} P^{ab} + \phi \Pi_\Phi \right)
\end{equation}

generates the canonical transformation corresponding to the VPCT. More explicitly,

\begin{equation}
\begin{aligned}
&G_{ab} \equiv \frac{\delta F}{\delta P^{ab}} = e^{4\hat{\phi}} g_{ab},\qquad
&&p^{ab} \equiv \frac{\delta F}{\delta g_{ab}} = e^{4\hat{\phi}} P^{ab} - \frac{1}{3} \left(e^{6\hat{\phi}} - 1 \right) \langle e^{4\hat{\phi}} g_{cd} P^{cd} \rangle_g \sqrt{g} g^{ab},\\
&\Phi \equiv \frac{\delta F}{\delta \Pi_\Phi} = \phi,\qquad
&&\pi_\phi \equiv \frac{\delta F}{\delta \phi} = \Pi_\Phi + 4 \left( e^{4\hat{\phi}} g_{ab} P^{ab} - \langle e^{4\hat{\phi}} g_{cd} P^{cd} \rangle_g \sqrt{g} e^{6\hat{\phi}} \right).
\end{aligned}
\end{equation}

These relations imply the transformations

\begin{equation}
\begin{aligned}
&G_{ab} = e^{4\hat{\phi}} g_{ab}, \qquad &&P^{ab} = e^{-4\hat{\phi}} \left( p^{ab} + \frac{1}{3} \left(e^{6\hat{\phi}} - 1 \right) \langle p \rangle_g \sqrt{g} g^{ab} \right),\\
&\Phi = \phi, \qquad &&\Pi_\Phi = \pi_\phi - 4 \left( p - \langle p \rangle_g \sqrt{g} \right).
\end{aligned}
\end{equation}

A long and careful calculation shows that the transformed constraints are weakly equivalent to

\begin{equation}\label{3linkingtheoryconstraints}
\begin{aligned}
&T_{\phi} H = \frac{e^{-6\hat{\phi}}}{\sqrt{g}} \left( p^{ab} p_{ab} - \frac{p^2}{2} + \frac{1}{3} \sqrt{g} (1-e^{6\hat{\phi}}) \langle p \rangle_g p - \frac{1}{6} g (1-e^{6\hat{\phi}})^2 \langle p \rangle_g^2 \right) - \sqrt{g} \left( R e^{2\hat{\phi}} - 8 e^{\hat{\phi}} \bigtriangleup e^{\hat{\phi}} \right) ,\\
& T_{\phi} H^a \approx -2 \nabla_b p^{ab} + \pi_\phi \nabla^a \phi,\\
& C \equiv T_{\phi} \pi_\phi = \pi_\phi - 4 \left( p - \langle p \rangle_g \sqrt{g} \right),
\end{aligned}
\end{equation}

where $\bigtriangleup$ denotes the invariant Laplacian, i.e., $\bigtriangleup = \nabla_a \nabla^a$. These constraints are of course first-class, as the cannonical transformation we performed does not change the Poisson brackets. The theory on $\Gamma_E (g_{ab},p^{ab},\phi,\pi_\phi)$, characterized by the constraints (\ref{3linkingtheoryconstraints}), is called the \emph{Linking theory} with the total Hamiltonian

\begin{equation}\label{3totalHamillinking}
H_{T} = N \sbullet T_{\phi} H + N_a \sbullet T_{\phi} H^a + \rho \sbullet T_{\phi} \pi_\phi.
\end{equation}

Note that $\int_{\Sigma} d^3x \, T_\phi \pi_\phi$ acts trivially on the phase space, just as in the toy particle model. Its Poisson bracket with $g_{ab}$,$p^{ab}$, $\hat{\phi}$, and $\pi_\phi$ vanish. This is the direct consequence of weakening the symmetry group and reducing it to only VPCTs, and is the key behind the success of shape dynamics as a consistent theory.

\subsection{Shape dynamics is gauge-fixing of the Linking theory}

The gauge-fixing $\phi \approx 0 $ fixes the $\rho = 0$ in (\ref{3totalHamillinking}) and gives us the standard ADM formalism of GR. However, if we impose the best-matching condition

\begin{equation}
\pi_\phi \approx 0,
\end{equation}

it would give us shape dynamics as an alternative gauge-fixing of the Linking theory. $\pi_\phi$ is first-class with respect to $C$ and $T_{\phi} H^a$, as

\begin{equation}
\{ \pi_\phi (x) , N^a \sbullet T_{\phi} H^a \} \approx \sqrt{g} \, \nabla_a \left(\frac{N^a \pi_\phi}{\sqrt{g}}\right) (x) \approx 0.
\end{equation}

But it has a non-vanishing Poisson bracket with $T_{\phi} H$ and except for a special choice of $N$, it does not propagate under the Hamiltonian evolution. Thus, we calculate

\begin{equation}
\{ N \sbullet T_{\phi} H, \pi_\phi (x)\} = S_N (x) - \langle S_N \rangle_g \sqrt{g} e^{6\hat{\phi}(x)},
\end{equation}

where

\begin{equation}
S_N(x) = \frac{\delta (N \sbullet H)}{\delta \phi(x)} \approx \sqrt{g} e^{\hat{\phi}} \left(56 N \bigtriangleup	(e^{\hat{\phi}}) + 8 \bigtriangleup (e^{\hat{\phi}} N) - 2N \left(4R e^{\hat{\phi}} + e^{5\hat{\phi}} \langle p \rangle_g^2 \right) \right).
\end{equation}

The \emph{lapse-fixing equation}

\begin{equation}\label{3lapsefixingeq}
S_N (x) - \langle S_N \rangle_g \sqrt{g} e^{6\hat{\phi}(x)} = 0
\end{equation}

has a one-parameter family of solutions for $N_0$, all related by a constant rescaling. It is not a surprise, as, for a given solution $N_0$, one can simply see that $a N_0$ is also a solution to (\ref{3lapsefixingeq}). Physically, this ensures that we have a reparametrization invariant dynamics. The `local' reparametrization invariance of the ADM dynamics is traded for spatial local conformal invariance, but one global degree of freedom remains and evolves the geometry. Had we not restricted the gauge transformations to VPCTs, only $N = 0$ would be the solution, leading to `frozen dynamics'. The solution to (\ref{3lapsefixingeq}) is a functional which depends on the transformed $g_{ab}$ and $p^{ab}$, thus, we write $N_0 = N_0 [g_{ab},p^{ab},\phi,x)$. The family of solutions of the lapse-fixing equation is known to be unique \cite{OMurchadha.1973}.

Ultimately, we have the global first-class Hamiltonian

\begin{equation}
T_{\phi} H_{f.c.} \equiv N_0 \sbullet T_{\phi} H = \int_{\Sigma} d^3x \, N_0[g_{ab},p^{ab},\phi,x) T_{\phi} H[g_{ab},p^{ab},x),
\end{equation}

along with the first-class constraints

\begin{equation}
T_{\phi} H^a \approx 0, \qquad C \approx 0,
\end{equation}

and the second-class constraints

\begin{equation}
\tilde{T}_{\phi} H \approx 0, \qquad \pi _\phi \approx 0,
\end{equation}

on the extended phase space $\Gamma_E$. $\tilde{T}_{\phi} H = T_{\phi} H - T_{\phi} H_{f.c.}$, denotes the second-class part of the Hamiltonian constraint. The constraints $\pi_\phi \approx 0$ and $C \approx 0$ yield the second-class constraint

\begin{equation}
D \equiv p - \langle p \rangle_g \sqrt{g} \approx 0
\end{equation}

which is the CMC foliation condition, as we expected from our heuristic discussion of best-matching with respect to VPCTs in the Lagrangian formulation. Moreover, $D$ generates VPCTs on the phase space $\Gamma$.

Finally, we can define the Dirac bracket as

\begin{equation}
\begin{gathered}
\{ R, Q\}_{DB} \equiv \{R, Q \} - \int d^3x d^3y \, \{ R , \pi_\phi (x)\} G(x,y) \{ \tilde{T}_{\phi} H (y) , Q \} \\
+ \int d^3x d^3y \, \{R, \tilde{T}_{\phi} H (x) \} G(y,x) \{ \pi_\phi(y) , Q\},
\end{gathered}
\end{equation}

where $G(x,y)$ is the Green's function, i.e.,

\begin{equation}
\int d^3x' \, G(x,x') \{ \tilde{T}_{\phi} H (x') , \pi_\phi (y) \} = \delta^3(x,y),
\end{equation}

and is guaranteed to exist because of the existence and uniqueness of the CMC lapse fixing, as in the toy model.

We can then impose the second-class constraints strongly and eliminate $\phi$ and $\pi_\phi$ from the theory by solving the equation $\tilde{T}_{\phi} H = 0$ for $\hat{\phi}_0$ and find it in terms of $g_{ab}$ and $p^{ab}$ and insert it into the total Hamiltonian to reduce the theory to the original phase space. As the second-class constraints are functions of both $\phi$ and $\pi_\phi$, the dynamics given by the Dirac bracket after elimination of them is identical to the one given by the Poisson brackets. It is because the additional Poisson brackets in the definition of Dirac bracket give either $\{ F(g_{ab},p^{ab}), \pi_\alpha\}$ or $\{ \pi_\alpha, T_{{\phi}_0} H_{f.c.}\}$ which are zero. The situation is analogous to what happened in the particle toy model. Now, we solve for $\hat{\phi}_0$. From (\ref{3linkingtheoryconstraints}) we have

\begin{equation}\label{3lichnerowiczyorkeq}
T_{{\phi}_0} H = \frac{e^{-6\hat{\phi}_0}}{\sqrt{g}} \left( \sigma^{ab} \sigma_{ab} - \frac{1}{6} \langle p \rangle_g^2 e^{12 \hat{\phi}_0} g - e^{8\hat{\phi}_0} \bar{R} g \right) = 0,
\end{equation}

where $\sigma^{ab} = p^{ab} - \frac{1}{3} g^{ab} p$ is the transverse-traceless momentum and

\begin{equation}
\bar{R} \equiv R - 8 e^{-\hat{\phi}_0} \bigtriangleup e^{\hat{\phi}_0},
\end{equation}

which is proportional to the transformed Ricci scalar under the conformal transformation. Equation (\ref{3lichnerowiczyorkeq}) is the \emph{Lichnerowicz-York} (LY) equation and was already derived back in 1970s for solving the initial value problem of general relativity. It has been studied extensively and is known to have unique solutions for the transverse-traceless $\sigma^{ab}$ \cite{OMurchadha.1973,OMurchadha.2005}. For our approach here, we need to only be sure that $\hat{\phi}_0 [g_{ab},p^{ab}]$ can formally be determined in terms of $g_{ab}$ and $p^{ab}$.

Finally, we arrive at the total Hamiltonian for shape dynamics:

\begin{equation}\label{3totalHamiltoniansd}
H_T = N T_{{\phi}_0} H_{f.c.} [g_{ab},p^{ab}] + \xi_a \sbullet T_{{\phi}_0} H^a [g_{ab},p^{ab}] + \rho \sbullet D[g_{ab},p^{ab}].
\end{equation}

where $N$ is a pure number and represents the reparametrization invariance of the theory. Note that $D$ is invariant under the VPCT. The Hamiltonian $H_{f.c.}$ is a non-local functional of $g_{ab}$ and $p^{ab}$, as was the first-class Hamiltonian in the particle toy-model. But as we saw, this is not a problem on its own as long as the physical equations of motion remain local and the non-locality of the Hamiltonian is merely the artifact of gauge invariance.

Note that $T_{{\phi}_0} H_a$ is first-class and generates spatial diffeomorphisms. $D$ is also first-class and generates spatial VPCTs. This is shape dynamics.

\subsection{Shape dynamics is a dual theory to GR}

As the final step in the formulation of shape dynamics, we construct the explicit dictionary between the theory we arrived at and the original ADM formulation of GR to show that shape dynamics is actually a dual theory. The ADM Hamiltonian is

\begin{equation}
H_{ADM} = N \sbullet H[g_{ab},p^{ab}] + \xi^a \sbullet H_a[g_{ab},p^{ab}].
\end{equation}

As $T_{{\phi}_0}$ is merely a canonical transformation and does not change the Hamiltonian dynamics, we impose the gauge condition\footnote{We are assuming that we can consistently impose this gauge condition. Strictly speaking, GR and SD are not inherently equivalent and the spacetime solution must be globally hyperbolic and CMC foliable.}

\begin{equation}
N \propto N_0[g_{ab},p^{ab},\hat{\phi}_0[g_{ab},p^{ab}]],
\end{equation}

and see that the theory would be equivalent to the theory (\ref{3totalHamiltoniansd}) in the gauge

\begin{equation}
\rho = 0.
\end{equation}

These gauge conditions establish the duality of GR and SD. In other words, if we fix the gauge freedom of SD corresponding to conformal transformations, and the refoliation invariance of GR, those two theories would reduce to an identical gauge theory in the phase space $\Gamma (g_{ab},p^{ab})$, which we call the dictionary. Thus,

\begin{samepage}
\begin{displayquote}
\emph{Shape dynamics is an alternative geometrical theory of gravity with the gauge group of 3D spatial diffeomorphisms and volume-preserving 3D conformal transformations, and has the same set of solutions as general relativity provided that the spacetime is globally hyperbolic and CMC foliable.}
\end{displayquote}
\end{samepage}

An important consequence of this derivation is that it shows \emph{refoliation invariance} is incompatible with \emph{spatial relationalism}. Shape dynamics possesses an absolute notion of simultaneity. The origin of this simultaneity is due to the underlying relational principles of the theory. The fact that it appeared at the heart of the dynamics of GR might not be a coincidence and a deep physical principle, still unknown, might lie beyond shape dynamics. The concept of absolute simultaneity was overthrown by special relativity and especially Minkowski's formulation of it, but it might be resurrected once again in physics as it might illuminate the problem of quantum gravity and provide some remarkable insights. Lee Smolin, as an example, is very fond of shape dynamics as an alternative to GR for precisely that reason and discusses some possible implications of this theory \cite{Unger.2015}.

Note that although shape dynamics manifestly breaks Lorentz invariance and offers an absolute notion of simultaneity, it does not necessarily violate the `empirical' content of SR. Actually, it does include a subset of the solutions of GR, and so it does predict the correct field equations compatible with SR (Klein-Gordon equations for the scalar field for example) if we take the weak-field limit and ignore the back-reaction of matter fields on the geometry (see Sec. 7.2 of \cite{Mercati.2018}). But showing that SD does successfully reconcile an absolute notion of simultaneity with the predictions of SR, such as time-dilation, requires a deeper and proper study of the \emph{clocks} and \emph{rods}, and is quite challenging. We will discuss the problem of the origin and behavior of rods and clocks in the next chapter.

\begin{figure}[h]
\centering
\scalebox{0.9}{
\begin{tikzpicture}[every text node part/.style={align=center}]
% Create an entity with ID node1, label "Fancy Node 1".
% Default for children (ie. attributes) is to be a tree "growing up"
% and having a distance of 3cm.
%
% 2 of these attributes do so, the 3rd's positioning is overridden.

% Now place a relation (ID=rel1)
\node [entity] (n1) {Linking Theory\\ $T_{\phi}H \approx 0,\quad T_{\phi}H_a \approx 0,\quad T_{\phi}\pi_\phi \approx 0$} [grow=up];
% Now the 2nd entity (ID=rel2)
\node [entity] (n2) [below right = 2 of n1]	{Dual Theory \\ (Shape Dynamics)\\ $T_{\phi}H_{f.c.} \approx 0,\quad T_{\phi}H_a \approx 0,\quad C \approx 0 $ \\ $\tilde{T}_\phi H \approx 0,\quad \pi_\phi \approx 0$};
\node [entity] (n3) [left = 9cm of n2]	{Original Theory \\ (General Relativity)\\ $H \approx 0,\quad H_a \approx 0$};
\node [entity] (n4) [below = 5.5cm of n1]	{Dictionary\\ $H_{f.c.} \approx 0,\quad H_a \approx 0$};
% Draw an edge between rel1 and node1; rel1 and node2
\path[-stealth] (n1) edge node [above right] {$\pi_\phi \approx 0$} (n2)
edge node [above left] {$\phi \approx 0$} (n3);
\path[-stealth] (n2) edge node [below right] {$\theta = 0$} (n4);

\path[-stealth] (n3) edge node [below left] {$N = N_0$} (n4);

\path[<->] (n2) edge node [above] {symmetry trading} (n3);
\end{tikzpicture}}

\captionsetup{width=0.8\linewidth}
\caption[Illustration of the symmetry trading procedure (shape dynamics)]{\small The diagram illustrating different gauge fixings of the linking theory, from which GR and SD can be obtained. The dictionary can be constructed by a final gauge fixation of both of the theories, which leads to the standard ADM formulation of GR in CMC foliation.}
\label{3linkingtheory}
\end{figure}
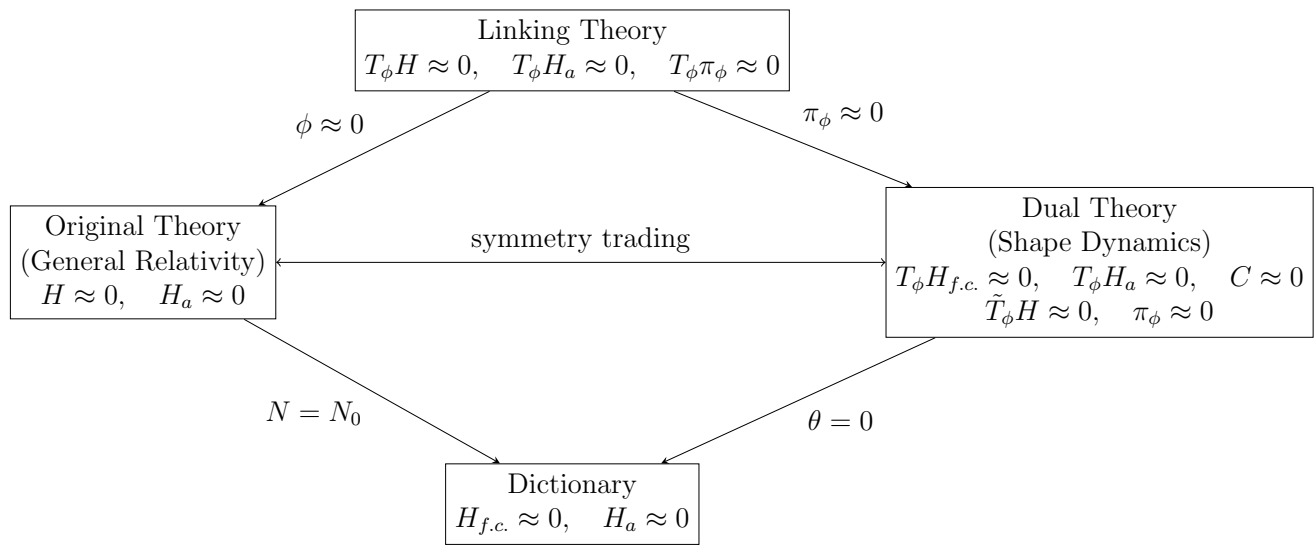

Fig. \ref{3linkingtheory} illustrates the Linking theory, GR, SD, and the relevant gauge fixings.

\subsection{Deparametrizing shape dynamics and Mach-Poincaré Principle}\label{3deparametrizingsd}

We should do the counting of the degrees of freedom of shape dynamics once again and clarify the status of the Mach-Poincaré Principle in the theory. In the canonical formulation of the theory, we constructed a theory with gauge groups corresponding to global reparametrization invariance, spatial diffeomorphisms, and spatial VPCTs. Thus, if we make the transition to the configuration space, we have a geodesic theory on conformal superspace plus volume, $CS^V \equiv CS \times \mathbb{R}^+$: an initial point and a direction in $CS^V$ determines the future evolution uniquely, and the theory satisfies the strong version of the Mach-Poincaré Principle on $CS^V$. However, the criterion of Machianity requires that we take $CS$ alone as the relational configuration space. Given the initial point and a direction in $CS$, it seems that we need two additional numbers: the initial volume and its conjugate momentum. We can see that York time, $\tau = 2/3 \langle p \rangle_g$, is actually the momentum conjugate to volume, as

\begin{equation}
\{V, \tau \} = \frac{2}{3} \{ V, \langle p \rangle_{g} \} = 1.
\end{equation}

The situation is a bit more hopeful as the initial volume V is conventional and irrelevant to the dynamics. As Barbour and Ó Murchadha noticed in \cite{Barbour.2010b}, the laws of shape dynamics are invariant under the dynamical similarity, expressed as

\begin{equation}
(g_{ab}, p^{ab}) \rightarrow (\alpha^2 g_{ab},p^{ab}),
\end{equation}

and if we transform $N$ to $N/\alpha$. The reason is that the Lichnerowicz-York equation (\ref{3lichnerowiczyorkeq}), and the diffeomorphism and conformal constraints remain invariant under this transformation and $H$ transforms to $\alpha H$.

This transformation maps solutions in $CS^V$ to other solutions, but they all describe the same curve when projected down to $CS$. Moreover, note that this transformation is not a VPCT. In this way, one can employ the dynamical similarity and set the value of the initial total size arbitrarily, and would actually need only one pure number in addition to shape degrees of freedom. Julian Barbour calls it \emph{anomaly}. One way to include this single parameter into the equations of motion is to define an external clock by deparametrizing the theory with respect to York time \cite{Mercati.2014}.

In fact, any reparametrization invariant theory can be deparametrized by using one of the variables (or a monotonic function of it) as the parameter, provided that it is monotonic. To outline the procedure, assume we have a constrained Hamiltonian system with the constraint $H(q^0,p_0,\cdots,q^n,p_n) \approx 0$ and that the evolution of $q^0$ is monotonic. We deparametrize the theory with respect to it by solving $H=0$ and find the conjugate momentum $p_0$ in terms of other variables, i.e., $p_0 = h(q^0,q^1,p_1,\cdots,q^n,p_n)$. Then, by using the implicit function theorem one can write

\begin{equation}
\frac{d q^i}{d q^0} = \{h,q^i\},\qquad \frac{d p_j}{dq^0} = \{h,p_j\}.
\end{equation}

York time is also known to be monotonic in CMC gauge \cite{Barbour.2013}, therefore, it can be used as an external measure of time and we can parameterize the solutions of SD with respect to it. In the deparametrized theory, one needs to know the `velocity' of the change of the initial shape degrees of freedom. It means that SD satisfies the `weak' Mach-Poincaré Principle in $CS$.

%% file: Chapter5.tex
\chapter{The Solutions of The Particle Shape Dynamics and The Arrow of Time}

In the last part, we cast the charming philosophical intuitions into rigorous mathematical forms and developed a relational theory of particle mechanics and geometrodynamics by carrying out the best-matching. Our job to show that relational physics is mathematically viable and has the potential to illuminate some of the conceptual difficulties of the standard classical physics is done. But we, philosophers of nature, ought to ultimately contemplate the natural order of things and phenomena, and reason why they are as they are and not otherwise.

So the next step we have to take is to probe the solutions of equations of motion of shape dynamics, find the general features of them, and clearly contrast them with those of the standard Newtonian mechanics. The equations of motion of shape dynamics in the case of geometrodynamics are a lot more complicated. Although some the solutions in some simple models have been found with striking results, such as the possibility of continuous continuation of the dynamics of the shape of the whole universe through the big bang \cite{Koslowski.2018}, we focus only on the solutions of the particle toy model here\footnote{See Chapter 13 of \cite{Mercati.2018} for a thorough analysis of the solutions of SD.}. Much further work needs to be done to make the field ontology clear. Nevertheless, particle ontology has already presented us with many profound results to ponder and develop, which is one of the main points of this chapter.

Next, we move to the issue of the arrow of time. We talked at length about the relational understanding of the passage of time, and that it is just a succession of shapes. Back in Chapter 1, we indicated that there is another aspect to the notion of time, its `direction', of which we have not yet spoken at all. What we aim for, is a `relational understanding of the arrow of time', in parallel with Mach's temporal Principle which captures the relationality of the duration of time. Next, we probe the emergence of the arrow of time in the solutions of the theory and will see that shape dynamics does not need any past hypothesis at all: The emergence of the arrow of time in the universe is typical and one does not need to posit some special initial state with high entropy to account for the asymmetry we observe in our universe. Moreover, the concept of entropy - which is drastically problematic when applied to the whole universe - can be discarded and replaced with a quantity, called \emph{complexity} which is a natural measure of the amount of structure of the universe, and this tells us that a story of the universe which is the one of more and more creation and `order', as opposed to the standard narrative loosely based on the thermodynamic concepts that the universe tends to more and more `disorder'.

These two steps are not disjointed. Actually, we will see the complexity which narrates the story of the universe is related to the very potential which controls the dynamics of the whole universe. We now move to the mathematics of the solutions of shape dynamics.

\section{The solutions of the theory}

\subsection{The generic solutions of the 3-body problem}

As we saw in Chapter 3, the particle mechanics is normally constructed in the Newtonian configuration space, $\mathbb{R}^{3N}$, coordinatized by the variables $q^{a}_I$, denoting the component $a$ of the position of the particle $I$. We identify the quotiented space $\mathbb{R}^{3N}/Sim(3)$ as the \emph{shape space}, the arena in which the relational dynamics of the whole universe unfolds.

The best-matching procedure demands the vanishing of the momentum associated with the group of transformations. We perform best-matching with respect to the Euclidean group which results in the vanishing of the translational and angular momenta,\footnote{We will shortly explain why we do not perform best-matching with respect to dilatations, as we have already promised.}

\begin{equation}
\begin{gathered}
P^a = \sum\limits_I p^I_a = 0\\
J^a = \sum\limits_I \epsilon^{abc} q^b_I p^I_c = 0
\end{gathered}
\end{equation}

as some constraints. Moreover, in the spirit of Leibniz's Principle of Sufficient Reason, we also set the energy constant $E$ which appears in the Jacobi's action (\ref{jacobii}) to zero. Hence, we can say that the relational solutions of the particle shape dynamics are equivalent to those of the Newtonian mechanics provided that $P$ and $J$ and $E$ vanish identically. In this sense, shape dynamics is equivalent to a smaller `subset' of Newtonian mechanics. The first condition does not take anything from the dynamical core of Newtonian mechanics, as the Galilean symmetry of Newton's theory already grants us to take a reference frame with respect to which $P$ vanishes. However, the other 4 constraints do indeed change the nature of the solutions of shape dynamics.

\begin{figure}[h]
\centering
\includegraphics[scale=0.2]{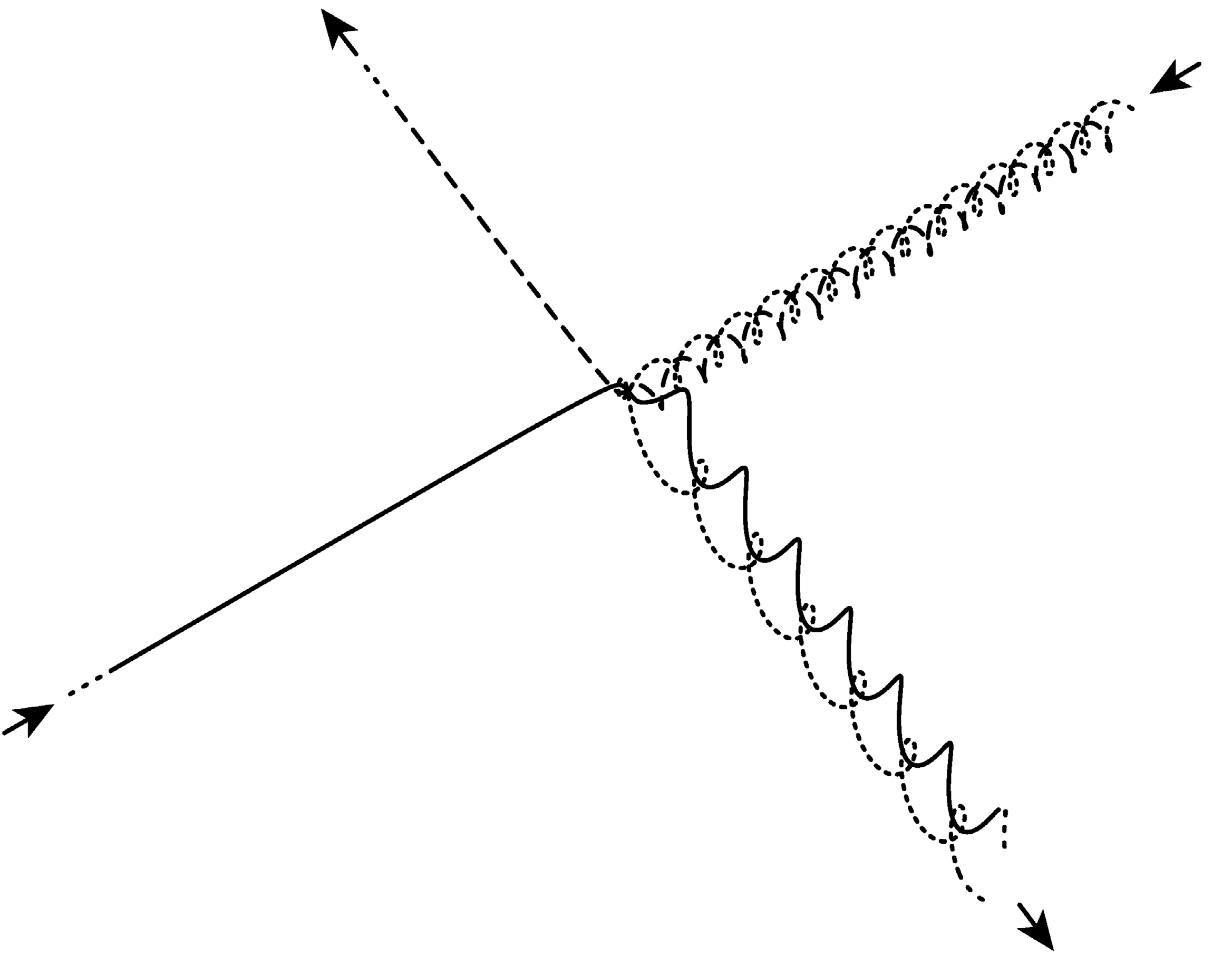}
\captionsetup{width=0.8\linewidth}
\caption[The formation of a Kepler pair and a singleton in a 3-body system in SD]{\small This plot illustrates the trajectories of the three particles with a vanishing angular momentum. In this picture, assuming an external uni-directional Newtonian time, a distant particle and a Kepler pair approach one another and engage in an irregular dance which results in a `particle swapping'. Then the other particle (or it might be the same particle) joins one of the two particles in the previous Keplerian motion, forms a new Kepler pair and a new singleton flies off. The same qualitative motion follows but in a reverse direction. Picture taken from \cite{Barbour.2020}.}
\label{4keplerpairs}
\end{figure}

To see more directly the effect of best-matching on the possible solutions of the theory, we consider the simplest case, the 3-body system. Almost all 3-body solutions with the constraints $E=0$ and $J=0$ have a period of rather irregular interaction and then develop asymptotically in both directions into a hyperbolic-elliptic escape in which a pair of particles separates from the third one and forms an elliptical Keplerian motion, while the third one escapes to infinity and asymptotically tends to an increasingly accurate inertial motion away from the pair. The pair of particles in Keplerian motion on each side is not necessarily the same. Fig. \ref{4keplerpairs} illustrates the usual Newtonian way of looking at this solution with respect to an external time.

This is the typical behavior of the 3-body system as we will see. But numerical works have established that it is actually typical of any system in particle shape dynamics! We will mention the results shortly and see why this should be the case.

The important point at this stage is that we have narrated the typical 3-body dynamics with respect to an external time. Just to sow the seeds of what will follow, I give the reader a glimpse of the explanation of the arrow of time that we will give in light of the point I mentioned: What we ultimately aim for is to narrate the 3-body particle exchange described above, in a way that the central irregular motion becomes the start of the flow of time and that time flows in both of the direction as the Kepler pair and the distant particle (called singleton) emerge. The reason is that the only objectively valid basis to posit the passage of time (or any other physical property) is the arena of shape space and the relational dynamics of the system. The dynamics we saw is vigorously presenting us with a `relational notion' of direction: The formation of the Kepler pair. Hence, we extend the principles of the doctrine of relationalism and base the emergence of the arrow of time on structure formation.

Before we do that, we quantify what we mean by structure formation, and then develop the mathematics of the three-body system and get to become more familiar with shape space in which all possible motions play out.

\subsection{Complexity and a measure of structure formation}

We define a real-valued positive-definite scale-invariant quantity on $S$ that quantifies the `clustering' of matter and formation of structure. A simple way is to use that ratio of two characteristic length of the system, one measuring the average long-range lengths, the other measuring the short-range ones. The straightforward candidate for the first one is the \emph{root-mean-square length} $l_{rms}$,

\begin{equation}
l_{rms} = \sqrt{\sum\limits_{I<J} m_I m_J r^2_{IJ}}, \qquad r_{IJ} \equiv |q_I - q_J|,
\end{equation}

where we have scaled the masses to make the sum of them identical to one. $l_{rms}$ is actually related to the center-of-mass moment of inertia of the system, $I_{CM} = l_{rms}^2$. The short-range characteristic length can be defined as the \emph{mean harmonic length} of the separations weighted by the masses,

\begin{equation}
l_{mhl} = \left( \sum\limits_{I<J} \frac{ m_I m_J}{r_{IJ}} \right)^{-1}.
\end{equation}

Then the complexity $C_S$ can be defined as

\begin{equation}
C_S = \frac{l_{rms}}{l_{mhl}}.
\end{equation}

For a given system of $N$ particles with certain mass ratios, complexity would depend on only the shape of the whole system. In other words, one can actually express the complexity in terms of the pure angles that each particle sees between any other particles. Complexity does not depend on either the global scale or the orientation of the system. For example, in the 3-body system, the particles form a triangle whose shape can be expressed through two free angles, the value of the complexity would depend on only those two angles.

Now one can see that $C_S$ is indeed a well-defined measure of the degree of non-uniformity and structure of the system. Consider the following simple observation to convince yourselves: In a specific arrangement of $N$ particles, one can move two of the particles closer to one another. Provided that the other particles are well separated and scattered, this in principle does not change $l_{rms}$ much significantly. In contrast, $l_{mhl}$ is extremely sensitive to a decrease in even one of the separations, and as the two particles approach each other, $l_{mhl}$ tends to zero, hence, making the complexity increase boundlessly. In general, $C_S$ increases with clustering and the formation of compact chunks of matter and structure.

The reader might have already observed that $l_{mhl}$ is actually related to the famous Newton potential. The significance behind this weighty fact will be clarified shortly, and we will establish that complexity is truly related to the potential term that controls the motion on shape dynamics.

Equipped with a powerful and rigorous measure of clustering, we can not state more mathematically what happens in our intuitive description of the 3-body system. As the Kepler pair forms, the increase in the relative separation between the Kepler pair and the third flying particle translates into the unbounded decrease of the ratio of the major axis of the Kepler pair to its distance from the third particle, and this is essentially the kind of process that increases the complexity, as we discussed. Hence, the typical behavior of the motion of the 3-body system in shape dynamics is that of clustering of matter in two directions away from a more or less uniform distribution of particles in the middle.

\subsection{The structure and topography of shape space}

The story of the universe plays out in the space of all possible relational degrees of freedom, the shape space. The shape space is the quotient of the Newtonian configuration space with respect to translations, rotations, and dilatations. The initial configuration space is $\mathbb{R}^{3N}$ and comes with $3N$ degrees of freedom, some of which are redundant. The similarity group is a $3+3+1 = 7$ dimensional Lie group and removes 7 degrees of freedom of the configuration space. Thus, we end up with the $3N - 7$ dimensional shape space, $S = \mathbb{R}^{3N}/Sim(3)$. For the 3-body problem, the shape space is 2 dimensional. This agrees with our expectation, as the shape of a triangle can be expressed by two angles. Hence, it can be represented by a 2-dimensional `sphere'. Shape space is always compact.

In particle shape dynamics, we use the Newton potential\footnote{$G$ has been set to one by defining a unit of time in terms of that of the length. Moreover, all the masses have been scaled so that the total mass equals one. These conventions do not change the dynamics, as will become more clear when we introduce the dynamical similarity.}

\begin{equation}
V = - \sum\limits_{I<J} \frac{m_I m_J}{r_{IJ}}
\end{equation}

and perform the best-matching with respect to translations and rotations.

Naturally, we define the scale-invariant version of the potential by multiplying the square root of the center-of-mass moment of inertia and arrive at the \emph{shape potential}

\begin{equation}
V_S \equiv \sqrt{I_{CM}} V,
\end{equation}

introduced in \cite{Mercati.2014}.

We will see more clearly how $V_S$ controls the dynamics in shape space, nevertheless, it is clear at this point that $V_S$ captures the qualitative dynamical features of the potential on shape space, as shape space is scale-free and its degrees of freedom are defined up to a redundant scale.

The profound and yet, straightforward observation is that $V_S$ is actually the negative of the complexity. This clearly means that the valleys of the shape potential are in regions with higher complexity, which means that shape dynamics typically steers almost any system towards becoming more clustered and structured: Structure formation is \emph{typical} in shape dynamics. This is already compatible with the intuition we gave about the generic 3-body solutions, but it is true for any other number of particles. Structure formation is the signature of the dynamics in shape space.

We state some commonly used concepts and definitions. First of all, the shape potential is not a monotonic function in any sense and introduces non-trivial ripples in the topographic map of shape space. The stationary points of $V_S$ are called \emph{central configurations}.

Moreover, notice that the shape potential is always negative and unbounded from below, as we discussed that complexity is unbounded from above. This means that $V_S$ must have at least one maximum for it is a continuous function defined on shape space which is compact. In the 3-body problem with equal masses, the maximum corresponds to the uniform equilateral triangle. All non-maximal stationary points are saddles, as there are no local minima apart from the infinitely deep wells of $V_S$ associated with the infinitely structured configurations.

In the 3-body problem, these non-maximal stationary points are the \emph{collinear shapes}, those with the three particles lined up. The saddles are the collinear shapes with one of the particles placed exactly between the other ones, while the minima of $V_S$ are the collinear shapes in which two of the particles coincide.

The number of saddles, deep wells, and in general the non-uniformities of the topographic features of $V_S$ increase with the number of particles. But the pivotal fact that the potential is bounded from above, but with infinitely deep wells at shapes with infinitely high complexity remains valid, and this is exactly what explains why our universe is as it is. Shape space is highly \emph{asymmetric}. What we ultimately aim for is to base the \emph{asymmetrical} behavior of our universe on this asymmetry of the shape space, but maintain the time-symmetrical (or better, timeless) laws of motion.

There are still a few more technical terms we should cover. There are a zero-measure set of solutions, called \emph{homothetic} solutions, in which a system is held at a central configuration and its shape does not change, and then `explode' and then come out again. In Newton's absolute space, such a system follows a uniform contraction until all of the particles collide in a way that does not change the shape. These solutions in which the size of the system reaches zero (in Newtonian sense) are known as \emph{central collisions}. Central collisions can only occur if the solution terminates at a central configuration.

Excluding a zero-measure subset of asymptotically homothetic solutions, all the other solutions can be extended so that they fall into the infinitely deep wells of the potential asymptotically at both ends. This is the key result we should bear in mind.

\begin{figure}[h]
\centering
\includegraphics[scale=0.5]{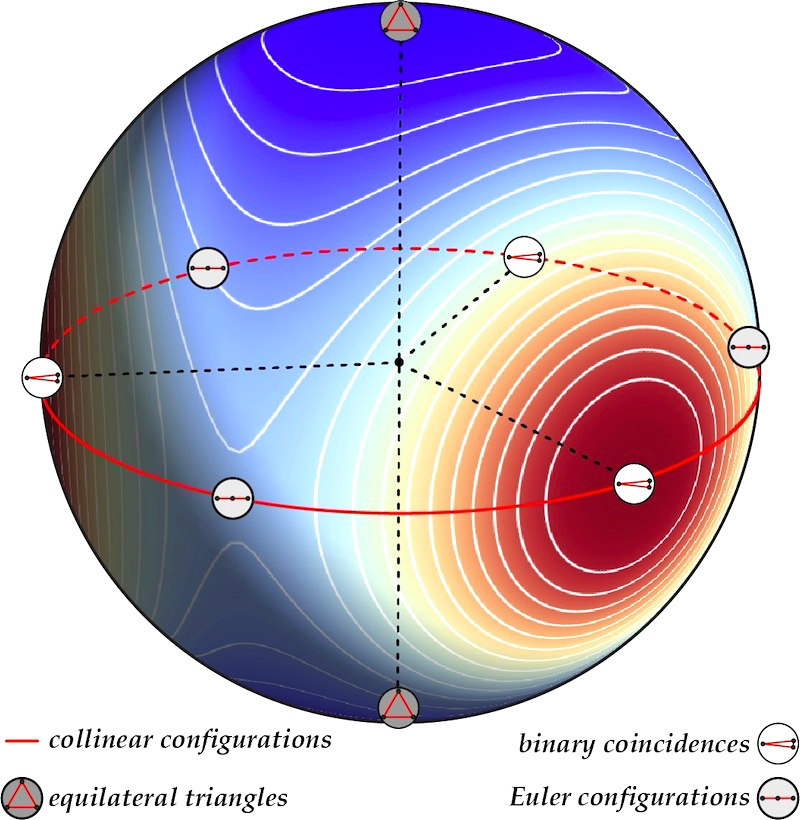}
\captionsetup{width=0.8\linewidth}
\caption[The shape sphere of the 3-body problem]{The shape space of the 3-body system with equal masses is depicted as a sphere, with a colored elevation plot of $V_S$. This representation is degenerate, as points with the same longitude and opposite latitude are the same shapes. The collinear configurations lie on the equator, including the three points corresponding to the two-body coincidences where $V_S$ is singular, and the three saddle points corresponding to the Euler configurations. The two poles represent the equilateral triangle which is the most uniform shape with the minimum value for complexity. Picture taken from \cite{Barbour.2013}.}
\label{4shapesphere}
\end{figure}

See Fig. \ref{4shapesphere} for an illustration of the shape sphere of the 3-body system with a plot of the shape potential.

\subsection{Back to the 3-body problem: the complete solution}

We now investigate the mathematical structure of the shape space of the 3-body system and find the equations of motion. Using good coordinates to satisfy the shape dynamics constraints and write down the equations of motion can be a bit tedious and difficult. We start with the Euclidean coordinates and then make 3 transformations until we finally arrive at the desired form of the equations of motion in terms of the $\theta$ and $\phi$ that coordinatize the shape sphere we introduced above.

We work in phase space, as it is a bit more straightforward. We can then make the transition to the configuration space by solving the Hamiltonian constraint and the conjugate momenta in terms of the configuration space variables. For more details see \cite{Montgomery.2002}. The results were reproduced in \cite{Barbour.2013} more compactly.

First, we start with the `Newtonian phase space' $\Gamma(q,p)$ with the Poisson bracket structure

\begin{equation}
\{q^a_I,p_b^J\} = \delta^a_b \delta^I_J.
\end{equation}

Now we make the following three canonical transformations:

\begin{enumerate}
\item The transformation\footnote{Note that $\bar{q}$ and $\bar{p}$'s here are not the same `transformed' variables we used in Chapter 3 for best-matching.}

\begin{equation}\label{4firstcanonicaltrans}
\begin{gathered}
\bar{q}^a_I \equiv \sum\limits_J M_{IJ} q^a_J \\
\bar{p}^I_a \equiv \sum\limits_J p^J_a M^{-1}_{JI},
\end{gathered}
\end{equation}

with the invertible matrix

\begin{equation}
M_{IJ} =
\begin{pmatrix}
-\sqrt{\frac{m_1 m_2}{m_1+m_2}} & \sqrt{\frac{m_1 m_2}{m_1+m_2}} & 0\\
-\frac{m_1 \sqrt{m_3(m_1+m_2)}}{m_1+m_2} & -\frac{m_2 \sqrt{m_3(m_1+m_2)}}{m_1+m_2} & \sqrt{m_3(m_1+m_2)}\\
m_1 & m_2 & m_3
\end{pmatrix},
\end{equation}

and $m_1 + m_2 + m_3 = 1$ by convention is a canonical transformation and leaves the Poisson structure unchanged. The reason that this particular transformation is useful, is that we can see that the third new momenta are the total momenta that vanish as the result of best-matching. To see that, we calculate the inverse of $M$:

\begin{equation}\label{4inversematrix}
M_{IJ}^{-1} =
\begin{pmatrix}
-\frac{\sqrt{\frac{m_1 m_2}{m_1+m_2}}}{m_1} & -\frac{m_3}{\sqrt{\left(m_1+m_2\right) m_3}} & 1 \\
\frac{\sqrt{\frac{m_1 m_2}{m_1+m_2}}}{m_2} & -\frac{m_3}{\sqrt{\left(m_1+m_2\right) m_3}} & 1 \\
0 & \frac{\sqrt{\left(m_1+m_2\right) m_3}}{m_3} & 1 \\
\end{pmatrix}
\end{equation}

and hence, we have

\begin{equation}
\bar{p}^3_a = \sum\limits_{I} p^I_a \approx 0.
\end{equation}

This helps us to apply the best-matching constraint corresponding to the translational group directly and set $\bar{p}^3_a$ weakly equal to zero. This makes sense, as $q_3$ denotes the center of mass of the system and it should not be dynamical from the relational point of view.

Another important result is that the kinetic term in terms of the new momenta is

\begin{equation}\label{4kineticterminit}
T = \sum\limits_I \frac{p^I_a p^I_a}{2m_I} = \frac{1}{2} \sum\limits_I \bar{p}^I_a \bar{p}^I_a \approx \frac{1}{2} \left( (\bar{p}^1)^2 + (\bar{p}^2)^2 \right),
\end{equation}

as can be checked directly by using (\ref{4inversematrix}).

The center-of-mass moment of inertia is also diagonal in the new coordinates:

\begin{equation}
I_{CM} = \sum\limits_{I<J} m_I r_{IJ}^2 = \bar{q}_1^2 + \bar{q}_2^2
\end{equation}

Now we have $6+6$ variables and only the angular momentum constraint to deal with,

\begin{equation}
J^a = \sum\limits_I \epsilon^{abc} q_I^b p_c^I \approx \sum\limits_{I=1,2} \epsilon^{abc} \bar{q}_I^b \bar{p}_c^I \approx 0.
\end{equation}

\item We introduce another set of conjugate variables:

\begin{equation}\label{4secondcanonicaltrans}
\begin{gathered}
w_1 = \frac{1}{2} \left( \bar{q}_1^2 - \bar{q}_2^2 \right), \qquad w_2 = \bar{q}_1 \cdot \bar{q}_2 \equiv \bar{q}_1^a \bar{q}_2^a, \qquad w_3 = |\bar{q}_1 \times \bar{q}_2|,\\
z^1 = \frac{\left( \bar{q}_1 \cdot \bar{p}^1 - \bar{q}_2 \cdot \bar{p}^2 \right)}{I_{CM}}, \qquad z^2 = \frac{\left( \bar{q}_2 \cdot \bar{p}^1 + \bar{q}_1 \cdot \bar{p}^2 \right)}{I_{CM}},\qquad z^3 = \frac{\left( |\bar{q}_1 \times \bar{p}^2| - |\bar{q}_2 \times \bar{p}^1| \right)}{I_{CM}}.
\end{gathered}
\end{equation}

These variables are rotationally invariant, and hence completely decouple the rotational degrees of freedom. This is the way we enforce the angular momentum constraint and remove $3+3$ degrees of freedom and cover the reduced phase space by the three pairs of variables in (\ref{4secondcanonicaltrans}).

A rather long calculation can show that the only non-trivial Poisson brackets are

\begin{equation}
\{w_I,z^J\} = \delta^I_J.
\end{equation}

The `magnitude' of $w$, i.e., $w_1^2 + w_2^2 + w_3^2$ equals to $\frac{I_{CM}^2}{4}$. Moreover, the kinetic term has the same simple form:

\begin{equation}\label{4kinetictermsec}
T = \frac{1}{2} \left( I_{CM} |z|^2 + \frac{|J|}{I_{CM}} \right) \approx \frac{I_{CM}}{2} |z|^2,
\end{equation}

where $|z|$ and $|J|$ are the magnitude of $z$, and $J$ respectively.

\item The final canonical transformation takes us to shape sphere, or more precisely, the phase space associated with the shape sphere parametrized by $\theta$ and $\phi$ and their conjugate momenta.

For that, we introduce the variable $R = \sqrt{I_{CM}}$ and write $w$'s in the form

\begin{equation}
w_1 = R^2 \sin \theta \cos \phi, \qquad w_2 = R^2 \sin \theta \sin \phi,\qquad w_3 = R^2 \cos \theta.
\end{equation}

The dimensionless variables $\theta$ and $\phi$ coordinatize the shape sphere Fig. \ref{4shapesphere}. Following the usual transformation to polar coordinates, we introduce the conjugate momenta $p_\theta$, $p_\phi$, and $p_r$, and arrive at\footnote{An easy way to see that is to write the kinetic term of the Lagrangian. Following (\ref{4kinetictermsec}) we deduce that $\dot{w} = R^2 z$ and therefore, $L_{kin} = \sum\limits_I \frac{1}{2R^2} \dot{w}_I^2$. Thus, in polar coordinates the kinetic term would be (\ref{4kinetictermpolar}).}

\begin{equation}\label{4kinetictermpolar}
T = \frac{1}{2} \frac{R^4 p_r^2 + p_\theta^2 + \sin^{-2} \theta p_\phi^2 }{R^2},
\end{equation}

where $p_r$ is the momentum conjugate to $r \equiv R^2$ and is

\begin{equation}
p_r = \frac{\dot{r}}{R^2} = \frac{2}{R} \dot{R}.
\end{equation}

Moreover, $\dot{R}$ is related to the dilatational momentum, as

\begin{equation}\label{4dotRasD}
\dot{R} = \frac{1}{2R} \dot{I}_{CM} = \frac{1}{2R} (2D - 2 q_{CM} \cdot P) \approx \frac{D}{R}.
\end{equation}

Thus,

\begin{equation}\label{4kinetictermfinal}
T = \frac{1}{2} \frac{p_\theta^2 + \sin^{-2} \theta p_\phi^2 + 4D^2}{R^2}.
\end{equation}

We should not neglect the potential term. The Newton potential is homogeneous of degree -1, and hence shape potential $V_S = \sqrt{I_{CM}} V$ is scale-free. Therefore, in principle, we can write $V_S$ in terms of $\theta$ and $\phi$. The details of finding the explicit form are in \cite{Montgomery.2002}. The Hamiltonian is

\begin{equation}\label{4barehamiltonian}
H = \frac{1}{2} \frac{p_\theta^2 + \sin^{-2} \theta p_\phi^2 + 4D^2}{R^2} + \frac{1}{R} V_S (\theta,\phi).
\end{equation}

\end{enumerate}

The theory is reparametrization invariant and (\ref{4barehamiltonian}) weakly vanishes. Hence, this Hamiltonian gives us a geodesic theory of the shape variables, and also the total size of the system expressed by $R$. It does not satisfy the strong-version of Mach's Principle, as we did not implement best-matching with respect to dilatations. However, in parallel with our discussion in Sec. \ref{3deparametrizingsd}, we can follow the same procedure here and use the dilatational momentum as the time variable to satisfy the weak version of Mach's Principle, as in \cite{Mercati.2014}.

It should be stressed that the $H = N H$ is the total Hamiltonian and should not be accompanied by first-class constraints. The three steps we took to enforce shape dynamics constraints already restricted the variables to the gauge-invariant ones, i.e., $w_I$ and $z^I$'s. Of course, they are not subject to any gauge transformation. The first-class Hamiltonian (\ref{4barehamiltonian}) gives us the complete equations of motion.

\subsection{Deparametrization and the equations of motion}

The dilatational momentum is monotonic, and therefore, a good candidate for parametrization of the theory. Too see its monotonicity, we derive what is known as the Euler-Lagrange equation:

\begin{equation}
\begin{gathered}
\dot{D} = \sum\limits_I \left( \dot{q}_I \cdot p^I + q_I \cdot \dot{p}^I \right) = 2T - \sum\limits_I q_I \cdot \frac{\partial V}{\partial q_I} \\
= 2T - k V,
\end{gathered}
\end{equation}

where $k$ is the degree of homogeneity of V and we have used Euler's theorem in the last line. For Newon potential $k=-1$. Assuming $T+V = E$, we have

\begin{equation}
\dot{D} = 2E - (2+k) V.
\end{equation}

In shape dynamics, we normally do not introduce any constant $E$ into Jacobi's action which means that in ephemeris time, $T+V = 0$ and $\dot{D} = -V$ for Newton potential and is always positive ($V$ is negative). Hence, $D$ is monotonic and we can deparametrize the theory with respect to it. As we discussed in Sec. \ref{3deparametrizingsd}, we have to find the variable conjugate to $D$, solve the Hamiltonian constraint, and use the result as the Hamiltonian for the reduced variables. In light of (\ref{4dotRasD}) and (\ref{4kinetictermfinal}), we should posit the fundamental Poisson bracket

\begin{equation}
\{\log R, D \} = \frac{1}{4}.
\end{equation}

which means that we should solve (\ref{4barehamiltonian}) for $4\log{R}$:

\begin{equation}\label{4deparametrizedham}
h = 4\log R = 4 \log \left( \frac{1}{2} \frac{p_\theta^2 + \sin^{-2} \theta p_\phi^2}{C_S} + \frac{2D^2}{C_S} \right).
\end{equation}

This gives us a `time'-dependent evolution of the shape degrees of freedom with respect to $D$. To make the Hamiltonian evolution more clear, we make the variables and the time parameter completely dimensionless. We introduced only a dimension of length to the initial Newtonian configuration variables. This induces dimension to all the other variables:

\begin{equation}
\begin{gathered}
[w] = length^2,\qquad [z] = length^{-3/2},\\
[p_\theta] = [p_\phi] = [D] = length^{1/2}.
\end{gathered}
\end{equation}

We can rescale the momenta $\pi = p/D_0$ where $D_0$ can be taken to be the initial dilatational momentum. We define the parameter $\zeta = D/D_0$. Although the rescaling we introduced is not a canonical transformation, the accompanying change in the parametrization leaves the equations of motion unchanged. The equations of motion in terms of the complexity are

\begin{equation}\label{4equsofmotion}
\begin{aligned}
&\frac{d \theta}{d \zeta} = \frac{ 8\pi_\theta}{\pi_\theta^2 + \sin^{-2} \theta \pi_\phi^2 + 4 \zeta^2},\qquad && \frac{d \phi}{d \zeta} = \frac{ 8 \sin^{-2} \theta \pi_\phi}{\pi_\theta^2 + \sin^{-2} \theta \pi_\phi^2 + 4 \zeta^2},\\
&\frac{d \pi_\theta}{d \zeta} = \frac{ 8 \sin^{-3} \theta \cos \theta \pi_\phi^2}{\pi_\theta^2 + \sin^{-2} \theta \pi_\phi^2 + 4 \zeta^2} + 4 \frac{\partial \log C_S}{\partial \theta}, && \frac{d \pi_\phi}{d \zeta} = 4\frac{\partial \log C_S}{\partial \phi}.
\end{aligned}
\end{equation}

We see that complexity actually controls the dynamics, as we expected. A lot can be said about these equations of motion. First, we start with the explicit `time dependence' of these equations. This is the price we had to pay to have a manifestly Machian theory. We did not implement best-matching with respect to dilatations (and we will soon explain why) and the theory is not geodesic. The amount of change of the variables with respect to $\zeta$ is dynamically relevant, and this is the only `absolute' extra ingredient we need to solve the equations of motion. (\ref{4equsofmotion}) completely determine the equations of motion if we specify the initial shape and the initial velocity. Thus, the theory satisfied the weak version of Mach's Principle.

The time-dependence of the equations has another notable consequence. It violates the Poincaré recurrence theorem. The recurrence theorem states that under any invertible Hamiltonian evolution in bounded phase space, an initial state returns arbitrarily close to itself after some long but finite time. The shape dynamics equations of motion, being explicitly time-dependent, do not satisfy the condition of invertibility and hence, the dynamics is not recurrent. This means that the evolution is `dissipative'. This dissipative aspect of the dynamics is behind the typical behavior of the trajectories of motion as they spiral and get drawn deeper and deeper into the potential wells, and lead to the formation of Kepler pair.

The dependence of the equations of motion can be eliminated by applying the non-canonical transformation

\begin{equation}
\qquad \omega_\theta = \frac{\pi_\theta}{\zeta},\qquad \omega_\phi = \frac{\pi_\phi}{\zeta},
\end{equation}

and use the parameter

\begin{equation}
\lambda = \log \zeta.
\end{equation}

This results in manifestly time-independent equations of motion, but with the extra terms $-\omega_\theta$ and $-\omega_\phi$ on the right hand side of the equations for $\frac{d \omega_\theta}{d \lambda}$ and $\frac{d \omega_\phi}{d \lambda}$. This transformation does not change the qualitative behavior of the trajectories, as the resulting equations of motion are not measure-preserving, and hence, would fortunately (!) not resurrect the recurrence theorem as one might think.

\subsection{The mysterious issue of absolute size!\footnote{A philosopher would object to my use of the word `mysterious'. If we are to be precise, mystery applies to an unknowable unknown. I do certainly hope that the issue of size is in principle knowable! Maybe I should use the word `puzzling' instead.}}

At the very moment we introduced best-matching in Chapter 3, it was obvious that one could implement this technique with respect to any Lie group. There is nothing sacred about the Euclidean group. Of course, we do not care about all the other Lie groups, but as far as the observational content of the configuration space leads us, we should have included the dilatational group. We briefly commented on this in Sec. \ref{2lagrangeequivariancebm}, and showed that the best-matching condition would be the vanishing of the dilatational momentum $D$. This condition is consistent only with a homogeneous potential of degree -2. The easiest candidate could be $V_S/R^2$. Introducing this `new' potential into the Hamiltonian results in the Hamiltonian constraint

\begin{equation}\label{4completscalefreegeo}
H = \frac{1}{2} \left( p_\theta^2 + \sin^{-2} \theta p_\phi^2 \right) - V_S.
\end{equation}

One can indeed show that the equations of motion, in this case, are equivalent to the geodesics of the conformally flat metric

\begin{equation}
g_{ab} =
\begin{pmatrix}
C_S & 0 \\
0 & \sin^2 \theta C_S \\
\end{pmatrix}
\end{equation}

defined on shape space in terms of $\theta$ and $\phi$. In other words, the completely scale-free equations of motion follow from the action

\begin{equation}
S = \int ds \, \sqrt{2g_{ab} \frac{d q^a}{ds} \frac{d q^b}{ds}}.
\end{equation}

The reason is that this action gives the Hamiltonian constraint (\ref{4completscalefreegeo}). This geodesic principle gives rise to equations of motion fundamentally different from those of (\ref{4deparametrizedham}), and to put it very succinctly, it does not conform to observations. There is no longer any dissipative evolution, and the solutions curves would not be dragged inescapably into the holes of wells of structure creation, and matter could avoid getting clumpy and clustered well enough.

Another major drawback of the model (\ref{4completscalefreegeo}) is that it does not lead to Newtonian equations of motion asymptotically. Newton potential has been tested for centuries and has been known to give valid results for weakly bounded isolated subsystems. Insisting on a $V \propto 1/r^2$ potential leads to potential experimental discrepancies for even trivial systems like our own solar system.

Absolute size is a tricky subject. We all hear cosmologists talk about the `expansion' of the universe. Julian Barbour once said `the idea of an expanding universe stinks! it's not expanding... it's changing its shape'.

The point is that expansion is not empirically verifiable at all, and does not have any role in our relational physics. The major phenomenon cosmologists point out to support our description of the expansion of the universe is the cosmological redshifts. Of course, those redshifts are real, but the point is that the standard FLRW metric explains them by presupposing some background clocks (which measure the cosmological time) and an external rod (which measures the expansion of the universe), and relates the redshift of a photon as measured by those clocks to an expansion with respect to the external rod. This story `stinks' because it takes a roundabout route and avoids the objectively real phenomenon (the redshift) and explains that in terms of some absolute structure. The alternative shape dynamic description of the redshifts involves a direct explanation of the change in the frequency of a photon as measured by certain `physical' clocks within the universe between two points.

Although this issue is still open in shape dynamics (see next subsection), it is in principle feasible to give a relational account of the redshift (and all the other evidence for an expanding universe). For that, matter must get clustered well enough and certain physical clocks must form. As the structure gets more and more compact, the time that it measures as compared to the external Newtonian time gets slower and slower, making it possible to account for the redshift of a photon as it makes its way from one of these structures (consider a galaxy) to another one. But (\ref{4completscalefreegeo}) would give us only a very limited range of redshifts, as Barbour and his collaborators already found out in the early 2000s. This is the last nail in the coffin.

This is why we cannot rely on a completely scale-free theory and had to leave the absolute size untouched. Instead, we used the dynamical similarity and deparametrization to get rid of it at the price of losing the geodesic equations of motion.

But why? Why does nature not respect a relational theory? Does it mean that nature wants us to keep a bit of the absolute structures? This may not be the case. Although the present state of this issue is hidden in deep mist, it might be eventually quantum mechanics that dispels the aura of mystery surrounding the apparent `expansion' of the universe, and the failure of shape dynamics as a classical theory to make sense of it.  

\subsection{Einstein's sin (!)}

In all successful theories in physics, whether Newtonian mechanics or relativity or even quantum field theory and string theory, we posit a spacetime and a frame of reference constructed by putting myriads of `imaginary' clocks and rods at each point to record the physical processes. But throughout this work, we have kept preaching that clocks are nothing but physical systems using the change included in a physical process as a measure of time, and clarified and defined this philosophical intuition in Chapter 2. If our physical theory is to be an inclusive and complete theory of the whole universe, and if we agree on saying that clocks and rods are themselves subject to the laws of physics, then it would be contradictory to formulate the fundamental laws of physics with respect to a reference frame outside our domain of description. As long as one clings to the `laboratory picture' of the laws of physics, this does not pose a serious challenge, but as a theory of the whole universe, this would certainly be a conceptual flaw and should be subject to meticulous scrutiny.

Einstein himself admitted that this is an issue back in 1949 in his autobiography \cite{Schilpp.19691970}:

\begin{samepage}
\begin{displayquote}
\emph{
It is striking that the theory [of special relativity] (except for four-dimensional space) introduces two kinds of physical things, i.e. (1) measuring rods and clocks, (2) all other things, e.g., the electromagnetic field, material point, etc. This, in a certain sense, is inconsistent; strictly speaking, measuring rods and clocks would have to be represented as solutions of the basic equations... not, as it were, as theoretically self-sufficient entities. The procedure justifies itself, however, because it was clear from the very beginning that the postulates of the theory are not strong enough to deduce from them equations for physical events sufficiently complete and sufficiently free from arbitrariness to base upon such a foundation a theory of measuring rods and clocks. If one did not wish to forego a physical interpretation of the coordinates in general (something that, in itself, would be possible), it was better to permit such inconsistency – with the obligation, however, of eliminating it at a later stage of the theory.}
\end{displayquote}
\end{samepage}

But Einstein never attempted to eliminate them and base them on the underlying laws of physics, and this problem was largely neglected afterward.

I think that there are two aspects to this problem:

\begin{enumerate}
\item First, a working physical model of a clock and rod must be proposed in terms of the intrinsic dynamical variables, subject to laws of physics.

\item The behavior of these physical clocks and rods must be deduced from the laws governing the whole universe and must conform to the corresponding standard principles of physics, e.g., absolute simultaneity in Newtonian mechanics and time dilation in relativity.
\end{enumerate}

Shape dynamics can potentially put this matter in a new light, as its starting point is not the spacetime, but a notion of the relational configuration of the whole universe. We will see that the particle shape dynamics we developed does provide a convincing resolution of this problem. Shape dynamics of geometry and fields, although being a more accurate physical theory of the whole universe that embodies the core of Einstein's relativity, does not solve this problem. We can address the first aspect by using the simple model in which a light ray bounces between two parallel mirrors. But the main issue is the second step, and to rigorously show that the proper distance the spacetime metric gives \emph{is the measure} of the proper time that any clock records. This is an assumption we take for granted in relativity, and its agreement with observations has enticed most physicists into accepting that as it is, without considering it critically.

But the problem can be more challenging than that. To `derive' the behavior of clocks and rods in any motion and show that they conform to the laws of relativity, requires an extra ingredient: the initial conditions. One must unambiguously specify what conditions correspond to an initial state of a clock (or a rod) in a certain motion. Normally, we do not care about this issue as we work with idealized `point-clocks' in relativity, and only the initial direction of motion is enough. But in light of the above consideration, in a more physical model of a clock, the inner structure of the clock is an essential ingredient and it requires the specification of the initial conditions for the substructure of the clock, meaning that only the initial direction of motion would not suffice.\footnote{Thanks to Mohammad Khorrami for pointing that out and a thorough discussion.}

We now proceed to the problem of rods and clocks in particle mechanics. We saw that the dynamical laws of shape dynamics for three particles lead to the formation of a Kepler pair and a singleton. These Kepler pairs grant us the pass to the first (and easy) step of the ordeal. Each Kepler pair forms a clock, and the relative motion of one of the particles against the background that the singleton puts up measures the passage of time, exactly in the same way that solar time
is defined. Moreover, the ratio of the semi-major axis of the Kepler pair and the distance from the singleton defines a measuring rod.

The formation of Kepler pairs and a singleton is typical in the solutions of shape dynamics with any number of particles. Hence, the whole universe in this picture breaks into myriads of clocks and rods, all in effective isolation from the others. The whole universe is a gigantic clock and rod factory.

More strikingly, these rods and clocks maintain their relations with one another more and more accurately as the universe evolves. The clocks `march in steps' with one another, and the rods become congruent. This is ensured in shape dynamics, following a major result in $N$-body problem \cite{Marchal.1976}. It can be shown that as the Newtonian time $t$ approaches the infinity, each subsystem (a Kepler pair and a singleton) develops the conserved quantities

\begin{equation}
\begin{gathered}
E_I (t) = E_I + O(t^{-5/3}),\\
J_I^a (t) = J_I^a + O(t^{-2/3}),\\
X_I^a (t)/t = V_I^a + O(t^{-1/2}),
\end{gathered}
\end{equation}

where here, $E_I(t), J_I^a (t), X_I^a (t)$ are respectively the energy, angular momentum, and the distance of the subsystem from the center of the mass of the whole universe. $I$ denotes the subsystems, not an individual particle.

It means that asymptotically, each substem would have constant energy and angular momentum. As these quantities encode the information of the clocks and rods, this ensures that the clocks and rods maintain their mutual relations to increasingly greater accuracy.

The complete resolution of Einstein's sin, especially in the context of geometrodynamics and field theory, requires much more sophisticated arguments and deductions, and more study and research.

\section{Time's arrow}

The arrow of time conventionally refers to the irreversible process we observe all around us: our house tends to become more disorderly, coffee spills on the ground and spreads out, we leave a bottle of ink in the open and the ink dries out, etc. Imagining the reversed processes of either one of these examples would be utterly strange: We have not experienced anything of this sort in nature. One might create these processes artificially in a laboratory, but it requires a tremendous amount of resources.

The most evident arrow of time that we observe, which is the figures prominently in all of the above examples, is the thermodynamic arrow, defined as the growth of entropy. There are two other observed arrows in the physical realm: The electromagnetic arrow, and the arrow in quantum mechanics. The first one corresponds to the propagation of the retarded waves, and not the `advanced' waves. The third one is related to the collapse of the wavefunction in the standard formulation.

Regardless of some proposed possible connections between them, these three arrows ground what we experience and identify as the arrow of time.

The existence of the arrow of time is not dubious of course, reconciling it with the known laws of physics is the problem, known as the problem of the arrow of time.

\subsection{Reflection on the problem of the arrow of time}

There are in fact two problems. The first problem is the conventional problem of the arrow of time:

\begin{enumerate}[label=\Roman*.]
\item Given that the fundamental laws of physics, apart from the weak theory that does not make a huge contribution to the evolution of the universe, are all time-symmetric, how is it that we observe an asymmetry in time? In other words, how is this symmetry of the laws of nature broken and not realized in the evolution of nature?
\end{enumerate}

This question (problem) is a logical question and points out a meaningful apparent contradiction, but there is also a presumption at the heart of the question that the laws of nature are time-symmetric.

This might be one way to dodge the inconsistency: There might be yet-to-be-found laws of nature that inherently distinguish an arrow of time.

But if we accept the laws of nature as they are, there would be one way to solve the problem: To postulate some non-symmetric initial conditions that break the time symmetry of the underlying physics \emph{at the level of the solution}. One thus has to posit an initial state with extremely low entropy and then allows the laws of nature to evolve the state and increase the entropy and lead to the emergence of an arrow of time. David Albert calls this the `Past Hypothesis'.

The general consensus among most physicists is that the Past Hypothesis is needed to solve the version of the problem of the arrow of time presented above.

But there are some concerns on both Past Hypothesis and also the generally accepted approach to this problem. First, in the known frameworks of physics, an initial condition that accounts for the qualitative behavior of our universe is extremely special and requires fine-tuning, making the Past Hypothesis very suspicious. This is the main issue, which will be addressed in relational physics.

Moreover, there are some critical remarks on the usual approach, as given in \cite{Barbour.2013,Barbour.2014,Barbour.2016,Barbour.2020}:

\begin{enumerate}
\item The concept of entropy that is widely used in the literature on the arrow of time, just like any other thermodynamic concept, is applicable to confined systems in a box which makes it possible for the system to equilibrate. However, the whole universe is \emph{not} in a box, and defining entropy for it is problematic in the first place. The walls of a box that confine a system play an essential role in pushing a system of particles to equilibrium: The particles keep bouncing off one another \emph{and} the walls.

\item Gravitational systems are anti-thermodynamic and they lead to more clustering, rather than spreading of matter. This fact is already acknowledged in shape dynamics quantitatively as complexity generally increases along the solution curves. Hence, self-gravitating systems cannot and do not equilibrate thermodynamically.

\item It is not clear how to define entropy in general relativity, and repeating the thermodynamic arguments in the context of general relativity remains highly problematic.
\end{enumerate}

This is not to overthrow thermodynamics at all. Thermodynamics remains an incredibly powerful theory \emph{within its domain of application}, as Einstein said\footnote{Einstein said of thermodynamics: ``It is the only physical theory of universal content, which I am convinced, that within the framework of applicability of its basic concepts will never be overthrown''.}, but the point is that the whole universe is definitely not in that domain. The universe is not a steam engine, or some gas in a container hovering in the vast absolute space. By thinking out of the box and going beyond the thermodynamic concepts, we can more appropriately discuss the dynamics of the whole universe in terms of its intrinsic degrees of freedom, the relational quantities we have been using in this work. Moreover, contrary to entropy, complexity is a well-defined quantity and applicable to the whole universe, on which we rely to understand the arrow of time.

A brief review of the discussions of the arrow of time can be found in \cite{Callender}.

There is another question, more metaphysical, that can be put in connection with the arrow of time:

\begin{enumerate}[label=\Roman*.]
\setcounter{enumi}{1}
\item What grounds the arrow of time? In other words, what is it from which the arrow of time emerges?
\end{enumerate}

Not all people would even accept this question as meaningful, especially in today's pragmatic era. The problem of the arrow of time arises naturally within the context of time-symmetric fundamental laws, and one can easily convince himself by invoking Past Hypothesis. But this second question is metaphysical and demands a metaphysical basis for the arrow of time. Yet, any solution to this question must also show that all of the arrows of time we observe can be `deduced' from what one suggests as the origin of time's arrow.

\subsection{The solution of the problem in shape dynamics}

We now explore the two questions we put in light of shape dynamics. As we discussed, the shape space is bestowed with a highly non-symmetric potential, and its topography induces an arrow of complexity into every region in the shape space. Moreover, the dissipative equations of motion on shape space force any generic curve to sink deeper and deeper into the attractors of infinite complexity.

As regards the first question, shape dynamics also breaks the time symmetry with a non-symmetric initial condition. But it does not give in to the Past Hypothesis. The tremendous asymmetric topography of shape space is the shield that protects us from having to posit a special initial condition to account for the emergence of the arrow of time as we observe. Apart from zero-measure homothetic initial conditions in which the shape is a central configuration and its change is set to zero, any other initial condition inevitably leads to the formation of structure, compatibly with the universe as we see.

Moreover, shape dynamics resolves the issue we raised, as it does not apply any concept of laboratory physics to the whole universe. Shape dynamics is a relational theory of the whole and describes the universe intrinsically. Our key concept in shape dynamics is complexity, which necessarily applies to a universe with gravity. However, little can be said about general relativity at the moment, as the complete set of solutions consistent with shape dynamics constraints is not yet fully studied. The famous Bianchi models for vacuum have been developed and conform to our expectations \cite{Barbour.2014,Mercati.2018}. Yamabe constant, a natural quantity defined on conformal superspace plays the role of the complexity of the geometry. But matter is an essential part of our universe, and we are hopeful that the complete relational theory of geometry and matter fields yield the same qualitative results and illuminate the third issue that we raised.

On the second metaphysical question, we can identify the complexity as the origin of the arrow of time. On Leibnizian grounds, we should base the arrow of time on the objective intrinsic properties of the whole universe. Hence, we state this principle:

\begin{displayquote}
\emph{
The universe starts with low complexity and strives to become more complex and structured.}
\end{displayquote}

Other than PSR and PII, Leibniz had another significant metaphysical principle we did not mention and is \emph{The Principle of Maximal Variety}. Leibniz said that our universe is the best one God could have created, as it contains the most possible variety, and also, every possible notion `strains towards existence' to bring more variety to the universe \cite{Leibniz.1697}. Shape dynamics, in conjunction with our above principle, encapsulate this idea beautifully.

This `master' arrow that we define, can to some extent explain the three arrows we observe. As matter gets more clustered, it makes it possible for equilibration to occur for effectively isolated and contained systems. Moreover, it hints that the electromagnetic arrow is also included in this picture of the universe. As the result of the Newtonian expansion of the universe, the system spreads and electromagnetic waves get washed away and the universe becomes clean of radiation. Typically, a process that results in the convergence of waves would be unlikely on statistical grounds. However, as we do not have a quantum theory of shape dynamics, we cannot say anything about the third arrow. We are hopeful to find out more about the notorious collapse of the wavefunction and what lies at the heart of the quantum world in the future. Of course, more rigorous works need to be done on the solutions of shape dynamics in broader contexts.

\begin{figure}[h]
\centering
\includegraphics[scale=0.4]{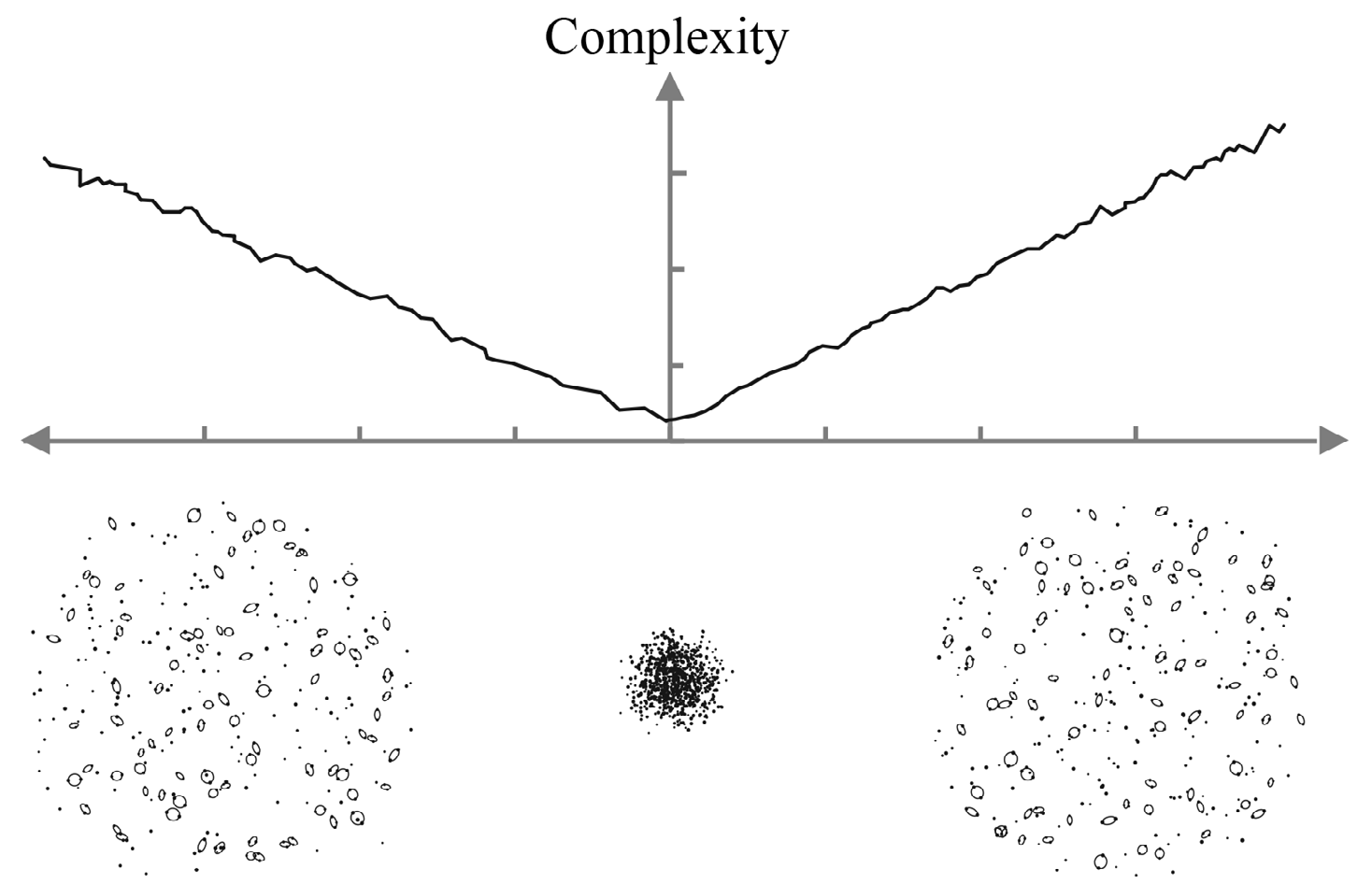}
\captionsetup{width=0.8\linewidth}
\caption[Complexity growth in both directions in a solution]{Complexity grows along both directions away from the central point with minimum complexity, as shown in this 1000-body simulation. Instead of looking at this evolution in Newtonian terms and from left to right, the intrinsic configurations \emph{define} a relational arrow that points to the formation of structures. Hence, internal observers on either sides would regard the central uniform structure as their `past'. Picture taken from \cite{Barbour.2020}.}
\label{4complexityfig}
\end{figure}

Complexity divides any inextendible generic solution curve on shape space into two curves with one end at a point with the lower complexity and another one spiraling around the attractors. Normally, excluding the asymptotically homothetic solutions, if we start with an initial condition we can extend the solution curve in the other direction. As complexity defines the arrow of time, we can interpret these complete solution curves as representing a \emph{bi-universe}, connected to one another at the point of lowest complexity, at which time points in opposite directions. See Fig. \ref{4complexityfig}.

\begin{figure}[h]
\centering
\includegraphics[scale=0.32]{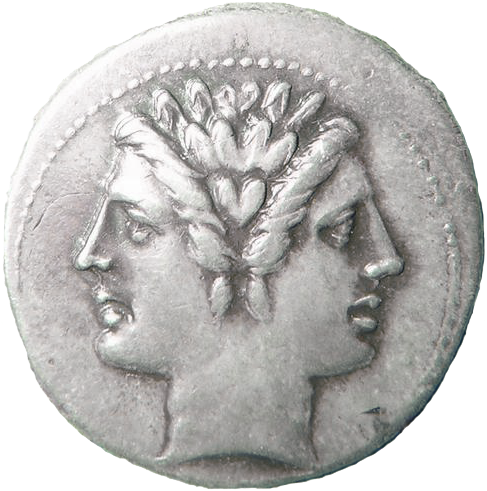}
\captionsetup{width=0.8\linewidth}
\caption[Janus depicted on a Roman coin]{\small The picture of a Roman coin with a depiction of the God `Janus'. Picture taken from Wikipedia}
\label{4januspic}
\end{figure}

This bi-universe, which starts at a central point and then evolves in opposite directions, resembles the Roman God `Janus', known as the God of gates, time, and duality. The central points with minimum complexity along the solutions curves are called \emph{Janus points}. Fig. \ref{4januspic} shows the depiction of Janus on a Roman coin, with its two faces looking in both directions.

Note that these two half-universes are not necessarily exactly the mirror reflection of one another, but qualitatively similar.

One final note on the commonly narrated story of the universe. It is usually stated that the growth of entropy leads to more disorder and the future of the universe is a drab state of more and more disorder. It is not true, as entropy is not a good candidate for describing the whole universe. On the other hand, shape dynamics tells us a story of the universe in which there would be more and more variety, structure, and order!

%% file: Conclusion.tex
\chapter{Conclusion and Outlook}

Our focus in this work has been to

\begin{enumerate}
\item Clarify the foundations and underlying principles of relationalism and delineate what exactly separates relational physics from the standard Newtonian or Einsteinian classical physics.

\item Give a rigorous and thorough formulation of shape dynamics, as the alternative relational theory in both particle mechanics and geometrodynamics.

\item Explore the implications of the particle toy model for the whole universe and reflect on the origin of time.
\end{enumerate}

Although this work presented the established results of shape dynamics, shape dynamics is a young field of study and a lot more research should be done, which might change some of these results, especially those of the third one we developed in Chapter 5.

At the moment, there are three active lines of research on shape dynamics. One of them is in the direction of doing more numerical works on the solutions of shape dynamics to reveal more elements of the potentials of shape dynamics as a theory of the whole. Some central configurations with a large number of particles have recently been simulated, with some possible interesting implications for quantum mechanics. These numerical works can suggest very valuable ideas for developing the theory.

Another research area is developing the field ontology. As I mentioned briefly, the simple Bianchi IX model of shape dynamics leads to the continuation of shape degrees of freedom of the universe through the big bang singularity \cite{Koslowski.2018}. Flavio Mercati and his research group, and David Sloan have been working on further developing this to both generalize their results and include other singularities, such as the gravitational singularity.

Lastly, let us not forget about the holy grail of modern physics, the problem of quantum gravity and quantum foundations! Very little can be said about quantum mechanics, or what I call \emph{quantum shape dynamics}. But there are some deeply insightful ideas in that direction too.

Before I mention two major guiding ideas, I recap the open questions we faced in the development of the theory in our work:

\begin{enumerate}
\item Shape dynamics is not perfectly Machian, and there is one single remnant of the absolute structures we removed: the dilatational momentum or `anomaly' as Julian Barbour calls it. How do we make sense of this non-relational variable?

\item Shape dynamics does not completely solve the problem of the rods and clocks.

\item The field ontology is not fully developed and the status of the arrow of time is still not known in that context.

\item Quantum mechanics is not taken into account.
\end{enumerate}

We are very hopeful that shape dynamics might solve these issues in the near future. We may need some new ideas for developing the theory if we are to tackle these fundamental problems.

There are two ideas, among many other ones, which are very suggestive, simple, and well-formed as compared to others.

\begin{enumerate}
\item Tim Koslowski suggests that shape dynamics should not be limited to a fixed number of particles. The number of particles might change under some law that is not yet found. This allows for particle creation and annihilation which is the key phenomenon in quantum field theory.

\item Julian Barbour suggests that complexity is \emph{literally} the time.
\end{enumerate}

The first idea is very powerful, as it opens up a possibility to reconcile quantum mechanics with shape dynamics. It makes `creation of structure' possible in shape dynamics. The second idea is also radical, especially in conjunction with the first one.

We already discussed the role of complexity as the origin of the arrow of time, and that the time all internal clocks measure are related to the global increase of complexity. But if complexity is to be the time, it means that it controls the `duration' of time as well as its direction. By looking at the mathematical expression of complexity, one sees that adding more particles generically leads to an increase in complexity. Therefore, the first idea together with the second one, if they are true, indicates that nature highly controls and constrains creation. A large number of particles cannot pop into existence right away, as it would require an infinite amount of time which is not accessible! This might solve the cosmological constant problem.

More strikingly, it might also solve the problem of gravitational collapse and singularities at the classical level! The collapse of matter increases the complexity drastically. If it is true that in a fundamental law of the whole universe complexity plays the role of time, then no gravitational singularity would form in the first place.

These are certainly worth pondering, but require more research and study to be done.

In short, now is a great time for developing shape dynamics, especially in light of the open questions we summarized, and the interesting ideas we mentioned. The current results indicate that shape dynamics might be on the right track as a theory of the whole.

%% file: Appendices.tex
\chapter{Reparametrization Invariant Lagrangian Systems and Jacobi's action}\label{jacobi}

We start with the standard Newtonian action defined on the configuration space of $N$ particles represented by the variables $q^a_I$ where $I$ denotes the particles and $a$ stands for the spatial dimensions:

\begin{equation}\label{standardactionappx}
S[q] = \int_{q^a_{Ii},t_i}^{q^a_{Ii},t_f} dt \, \left( T(\dot{q}^a_I,\cdots) - V(q^a_I,\cdots) \right) \equiv \int_{q^a_{Ii},t_i}^{q^a_{Ii},t_f} dt \, L,
\end{equation}

where $T$ is the Kinetic term

\begin{equation}
\sum\limits_{I} \frac{1}{2} m_I \dot{q}_I^2,
\end{equation}

and $V$ is an arbitrary time-independent potential.

In general, there is a procedure called \emph{Routhian reduction procedure} that expurgates the action from the cyclic variables, those which only enter the action with derivatves.\footnote{For a more detailed account of the procedure see parts 9 and 10 of Chapter VI of \cite{Lanczos.1986}, or the entry on it in Wikipedia.} One can apply this procedure to time-independent lagrangian systems and arrive at a less known, but powerful action called Jacobi's action. Before we proceed with that procedure, we note that the standard Newtonian action can be turned into an action to which this procedure would be applicable by elevating the role of the Newtonian time and treating it as a dynamical variable. Thus, along any trajectory, we can take $t$ to be a function of an arbitrary parameter $t=t(\lambda)$. Then the above action can be written as the following integral with the same boundaries:

\begin{equation}\label{repinvariant}
S = \int d\lambda \, \frac{dt}{d\lambda} \, \left( \frac{1}{2} \sum\limits_I m_I {q'}_I^2 \left(\frac{d\lambda}{dt}\right)^2 - V(q^a_I,\cdots) \right) \equiv \int d\lambda \, \bar{L}(q,q',t,t')
\end{equation}

where ${q'}_I^a$ is the derivative of the $q_I^a (t(\lambda))$ with respect to the new parameter. $\bar{L}$ is the `new' Lagrangian that generates the same trajectories in the `extended' configuration space which includes time as an independent variable, and is equal to the original Lagrangian by a factor of $t'$: $\bar{L} = t' L$.

This theory if \emph{reparametrization invariant}, meaning that we can use \emph{any} parameter to describe the system. In other words, action is invariant under the transformation $\lambda \rightarrow \lambda' = \lambda'(\lambda)$ where $\lambda'(\lambda)$ is a strictly monotonic smooth function of $\lambda$ \footnote{This smoothness condition is to assure that the derivatives of the configuration space variables with respect to the new parameter exist to any arbitrary order. The requirement of monotonicity is necessary, otherwise, two different states may be assigned to one `instant' of the new time parameter.} as can be checked simply. Actually, it is not a surprise as we found the action (\ref{repinvariant}) by introducing an `arbitrary' parameter into the original action (\ref{standardactionappx}).

If we calculate the momentum conjugate to this `new' dynamical variable, $t$, we find

\begin{equation}
p_t = \frac{\partial \bar{L}}{\partial t'} = - \left(\frac{d\lambda}{dt}\right)^2 \frac{1}{2} \sum\limits_{I=1}^{I=N} m_I {q'}_I^2 - V.
\end{equation}

What is interesting is that $p_t$ is exactly the same as the Hamiltonian function one derives from the Lagrangian in (\ref{standardactionappx}). Moreover, the conjugate momenta to the variables $q_I^a$ denoted by $\bar{p}^I_a$ remain the same:

\begin{equation}
\bar{p}^I_a = \frac{\partial L}{\partial {q'}_I^a} = \frac{1}{t'} m_I {q'}^a_I = p^I_a.
\end{equation}

It suggests that the Hamiltonian for the action (\ref{repinvariant}) vanishes:

\begin{equation}\label{Avanishhamiltcons}
\bar{H} = p_t t' + \sum\limits_{I=1}^{N} \bar{p}^I_a {q'}^a_I - \bar{L} = p_t t' + t' \left( \sum\limits_{I=1}^{N} p^I_a \dot{q}^a_I - L\right) = (p_t + H) t' = 0 .
\end{equation}

This is the price we have to pay for adding a new artificially dynamical variable to the configuration space. It means the Hamiltonian for this new theory is a constraint. This, however, does not mean that there is no evolution in phase space. If the Hamiltonian had vanished `identically' in the phase space, it would have meant that there is no evolution and the variables remain constant. But here, the Hamiltonian is a constraint and vanishes on only a `part' of the phase space that includes the true physical variables, i.e., it is a constraint. For the constrained Hamiltonian systems refer to Appx. \ref{constrained} in which we apply Dirac's approach to reparametrization invariant theories of this type and discuss them at length.

There is an important thing to note about the reparametrization invariant action (\ref{repinvariant}). The new variable, the Newtonian time, is a cyclic variable if the original Lagrangian in (\ref{standardactionappx}) is not explicitly time-dependent. For closed systems, this is always the case. Thus, for these systems the new Lagrangian $\bar{L}$ does not depend on $t$. This means that its conjugate momentum, $\bar{p}_t$ is constant. We now apply the Routhian reduction procedure to the time variable. The Routhian is

\begin{equation}
R(q^a_I,{q'}^a_I,t,p_t) = \bar{L} - p_t t' = t' L(q^a_I,{q'}^a_I) - p_t t' = t' (T-V-p_t).
\end{equation}

Because of the Legendre transformation, $t'$ is a function of other variables and we need to replace it with its explicit form. We can do that by noting that $p_t$ is constant according to the equations of motion. Let $p_t = - E$. $E$ is the total energy of the system (as we will see) and the minus sign is for later convenience. From the above calculation we saw that $p_t = - H$ where $H = T + V = \frac{1}{2} \sum\limits_{I=1}^{I=N} m_I \dot{q}_I^2 + V$. After rewriting the above expression, we have

\begin{equation}
p_t = - \frac{1}{2} \sum\limits_I m_I {q'}_I^2 \left(\frac{d\lambda}{dt}\right)^2 - V(q_I^a) = - E,
\end{equation}

thus,

\begin{equation}
t' = \sqrt{\frac{\bar{T}}{(E - V)}},
\end{equation}

where $\bar{T} = \frac{1}{2} \sum\limits_I m_I {q'}_I^2$. Hence,

\begin{equation}
R(q^a_I,q^{a'}_I,t,p_t) = 2 \sqrt{(E-V)\bar{T}}.
\end{equation}

As the result, we have the following action which in principle, is equivalent to the above action we started from

\begin{equation}\label{jacobiactionappx}
S_J = 2 \int\limits_A^B d\lambda \, \sqrt{(E-V(q^a_I))\bar{T}}
\end{equation}

This action, called Jacobi's action, is reparametrization invariant and time does not play any role in it. The $A$ and $B$ represent the fixed initial and final configurations in the configuration space. Again, time does not have any role to play in these boundaries. Minimizing this action with respect to $q^a_I$'s determines the trajectories of motion in the configuration space.

But we should need more information to find the Newtonian trajectories because the above procedure removed time completely. What happened to the dynamical role of Newtonian time now?

This one last datum is encoded in the condition $p_t = - E$. This condition determines `how fast' the system should evolve with respect to $t$. As $p_t = \frac{1}{2} \sum\limits_{I=1}^{I=N} m_I \dot{q}_I^2 + V$, we can now parametrize the trajectory we find from (\ref{jacobiactionappx}) with $t$ such that the above condition holds for a specific value for $E$.

The punchline is that Jacobi's action determines the timeless `frozen' trajectory of motion, e.g., the path of planets around the sun, and the supplementing energy condition determines how fast the planets must dance around the sun with respect to Newtonian time.

\chapter{Lie Groups and Lie Algebras}\label{lie}

Lie Groups, Lie algebras, and their actions on manifolds, are all of utmost importance to relational physics and shape dynamics. The language of relational physics is the language of gauge theories. This was clear as early as we introduced Mach-Poincaré Principle. Hence, we first review all the primary definitions and concepts in Lie group and Lie algebra theory, and then we discuss the Euclidean group in more detail as it is the cornerstone of relational particle physics.\footnote{For more details, see the Chapter 29,30 of the book \cite{Hassani.2013} and Chapter 2 of \cite{Georgi.1999}. I assume that the reader is familiar with the formal definition of group and representation. For that, see Chapter 1 of \cite{Georgi.1999}.}

\section{Lie groups and their Lie algebras}
\begin{definition}
A \emph{Lie group} $G$ is a differentiable manifold endowed with a group structure such that the group multiplication $G \times G \rightarrow G$ and the inverse map $()^{-1}: G\rightarrow G$ are smooth.
\end{definition}

If the Lie group manifold is $r$-dimensional, we say that $G$ is a \emph{$r$-parameter Lie group}.

Hence, a Lie group is basically a group of continuous elements which is also manifold in itself. It means that any concept in group theory is relevant for Lie groups, such as group multiplication, inverse, and a unique identity element denoted by $e$.

There are many examples of Lie groups. The general Linear group $GL(V)$ is a Lie group with a well-defined multiplication law (multiplication of the operators). The translation group $T(3)$, the rotation group $SO(3)$, the gauge group of electromagnetism, $U(1)$ are some other examples.

These groups are all finite-dimensional. There are also more abstract infinite-dimensional Lie groups such as the group of conformal transformations in a Riemannian manifold, the group of diffeomorphisms, etc that are of relevance to geometrodynamics.

Lie group manifolds can have various topological shapes. For instance, $U(1)$ is simply a circle, $T(3)$ is diffeomorphic to $\mathbb{R}^3$. $SO(3)$ has a more non-trivial topology and is a \emph{compact non-simply connected} manifold. It has the shape of a sphere with the two points on its surface identified with one another.

We now proceed to define Lie algebra. The concept of Lie algebra can be visualized as the \emph{tangent vector space} at a point in the Lie group manifold, usually the identity element for simplicity. Intuitively, elements of the Lie algebra (which is a vector space as we will see) are vectors in all possible directions along the Lie group manifold. You may have seen elsewhere that elements of Lie algebra are defined by considering group elements infinitesimally close to the identity. We follow a rather abstract way that ultimately leads us to a proper mathematical definition of Lie algebra. First, we have to start with the definition of \emph{left translation}:

\begin{definition}
Let $G$ be a Lie group and $g \in G$. The left translation by $g$ is the diffeomorphism $L_g : G \rightarrow G$ defined by $L_g h = gh$ for all $h \in G$ where by $gh$ the group multiplication of $g$ and $h$ is meant.
\end{definition}

A vector field $v$ defined on $G$ is called \emph{left-invariant} if for each $g\in G$, $v$ is related to itself under the pushforward induced by $L_g$. Formally, this means

\begin{equation}
L_{g\ast} v |_h = v |_{gh}
\end{equation}

for all $h\in G$.

We denote the set of all left-invariant vector fields on $G$ by $\mathfrak{g}$. In general, the space of all vector fields defined on a manifold, i.e., the vector bundle, is an infinite-dimensional vector space. We have also the Lie bracket structure defined on the manifold which is the Lie derivative of one vector field with respect to the other, $[v,w] (f) = v(w(f)) - w(v(f))$, the output of which is also another vector field. Hence, we can say that the vector bundle defined on a manifold is a \emph{Lie algebra} under the Lie bracket `multiplication'.\footnote{For a brief and formal account of Lie algebra theory see Sec. 29.2 of \cite{Hassani.2013}.}

The same thing can be constructed for Lie groups by virtue of being manifold. However, a more restricted and useful structure can be defined for left-invariant vector fields. It can be proved that the left-invariant vector fields form a Lie algebra themselves. We have the following theorem:

\begin{samepage}
\begin{theorem}
Let $G$ be a Lie group and $\mathfrak{g}$ the set of its left-invariant vector fields. Then $\mathfrak{g}$ is a real vector space of the same dimension as $G$. Moreover, $\mathfrak{g}$ is closed under Lie brackets. Hence, $\mathfrak{g}$ is a Lie algebra.
\end{theorem}
\end{samepage}

Being left-invariant, it does not make any substantial difference whether we consider the elements of the vector field at the identity or any other element. Usually, we associate the Lie algebra with the tangent vector space at $e$. Thus, we have the following definition:

\begin{definition}
The Lie algebra of the Lie group $G$ is the Lie algebra $\mathfrak{g}$ of the left-invariant vector fields on $G$. Conventionally, we think of $v\in \mathfrak{g}$ as a vector in $T_e(G)$, the tangent vector space at $e$. Then we can denote the corresponding vector field by $X_v$, i.e., $X_v |_e = v$.
\end{definition}

As we discussed, the Lie algebra comes with an algebraic structure, called the Lie bracket. In other words, the objects $[X_v,X_w]$ are also left-invariant and therefore belong to $\mathfrak{g}$. If we introduce the basis $v_i$ for the Lie algebra $\mathfrak{g}$, then we have

\begin{equation}
[X_{v_i},X_{v_j}] = \sum\limits_{k} f^k_{ij} X_{v_k},
\end{equation}

where $f^k_{ij}$ are some constant numbers independent of the group element at which the commutator has been evaluated, called the \emph{structure constants} of $G$.

Although our formal definition of Lie group and Lie algebra may seem abstract, we can study the action of the Lie groups on other more \emph{physical} manifolds, such as the configuration space, and infer many useful things from that. This is the content of the next section. We will see that it is easier to conceive the action of a group rather than the abstract group itself.

\section{Group action}\label{groupaction}
We start with the formal definition of the action of a Lie group:

\begin{definition}\label{defaction}
Let $M$ be a manifold. A local group of transformations corresponding to a Lie group $G$ acting on $M$ is a smooth map\footnote{In some mathematical literature the action of a group is not defined for all members of the group and only an open subset of it is considered. This does not quite matter at this level for the examples and physical applications we consider.} $T: G\times M \rightarrow M$ satisfying the following condition:

\begin{enumerate}
\item $g \cdot (h \cdot P) = (gh) \cdot P \quad \forall g,h \in G, P \in M$,
\item $e \cdot P = P \quad \forall P \in M$,
\item $g^{-1} \cdot (g \cdot P) = P \quad \forall g \in G, P \in M$,
\end{enumerate}
where $g \cdot P \equiv T(g,P)$.

\end{definition}

There is an important concept regarding the Mach-Poincaré Principle called the \emph{orbit} of a Lie group. The formal definition is

\begin{definition}\label{orbit}
The orbit of an element $P \in M$, denoted by $O_P$, is a set of points in $M$ obtained from $g$ by the action of $G$ on $M$\footnote{Some define the concept of orbit in a different way. However, this given definition is clear, by following which we arrive at the standard known results.}
\begin{equation}
O_P = \{ g \cdot P | g \in G \}.
\end{equation}
\end{definition}

Based on the definition of the action of a group Def. \ref{defaction} and the orbit of it Def. \ref{orbit}, it immediately follows that `belonging to the same orbit' is an equivalence relation on $M$. The reflexivity, symmetry, and transitivity conditions follow from the 2nd, 3rd, and 1st conditions in Def. \ref{defaction} respectively.

Hence, we can conceive of the quotient of $M$ with respect to the group $G$, the set of the orbits of $M$, denoted by $M/G$.

Our definition of the Mach-Poincaré Principle in Sec. \ref{contemporary} manifestly makes use of the concept of orbit. In relational physics, the configuration space $Q$ plays the role of the manifold $M$, and the spatial symmetry group is the corresponding Lie group. Then, we defined the relational configuration space as the quotiented space and got rid of the redundant degrees of freedom.

The action of a group on a manifold can also be defined infinitesimally. This, in turn, induces a homomorphism between the Lie algebra and the vector bundle of $M$. Given $v \in \mathfrak{g}$, one can consider the flow of $v$ on $G$, $F_v(t)$, parametrized by $t$. Then we consider the map $T(F_v(t),\cdot)$ as a function of $t$ on $M$. The following function defines a vector field over $M$:

\begin{equation}
X_v|_P \equiv \frac{d}{dt} \left. T(F_v(t),P) \right|_{t=0}.
\end{equation}

One can show that $[X_v,X_w] = X_{[v,w]}$ where $[\cdot,\cdot]$ is the commutation of vector fields. Hence, the function does indeed define the promised homomorphism.

In a particular coordinate system, we can assign the numbers $a^i$'s to the group elements and also coordinatize $M$ by $x^i$'s. Then the group action would have the form

\begin{equation}
x'^i = T\left(\alpha^1,\cdots,\alpha^r,x^1,\cdots,x^n\right),
\end{equation}

where $r$ denotes the dimension of the Lie group $G$ while $n$ is the dimension of $M$. Here $i$ runs from $1$ to $n$. Based on definition, if $\alpha^i = 0$ then $x'^i = x$. Therefore, we can consider the following infinitesimal changes of the coordinates in response to infinitesimal changes of the group parameters in a specific direction away from $e$:

\begin{equation}
\delta \phi^a t^i_{a j} x^j \equiv \delta x^i = T\left(\delta \phi^1,\cdots,\delta \phi^r,x^1,\cdots,x^n\right) - x^i
= \delta\phi^a \left. \frac{\partial T}{\partial \alpha^a}\right|_{\alpha^a = 0}.
\end{equation}

$t^i_{aj}$'s are called the \emph{infinitesimal generators} as they generate the transformations on the manifold corresponding to the action of $G$. Explicitly,

\begin{equation}
t^i_{aj} x^j = \left. \frac{\partial T}{\partial \alpha^a}\right|_{\alpha^a = 0}
\end{equation}

The generators can in principle be evaluated at an arbitrary $\phi$, not necessarily the identity. In that case we have

\begin{equation}\label{Bdefintarbit}
t_{a}(\phi) \equiv T^{-1} \partial_a T,
\end{equation}

where we have suppressed the indices. An important and useful relation is 

\begin{equation}\label{Bpartialtstructure}
\partial_a t_b (\phi) - \partial_b t_a (\phi) = - f^c_{ab} t_c (\phi) ,
\end{equation}

where $f^c_{ab}$ are the structure constants. This can be proved by using the definition (\ref{Bdefintarbit}):

\begin{equation}
\begin{gathered}
\partial_a t_b (\phi) - \partial_b t_a (\phi) = (\partial_a T^{-1}) \partial_b T - (\partial_b T^{-1}) \partial_a T \\
= -T^{-1} (\partial_a T) T^{-1} \partial_b T + T^{-1} (\partial_b T) T^{-1} \partial_a T\\
= - \left( t_a t_b (\phi) - t_b t_a (\phi)  \right) = -f^c_{ab} t_c (\phi).
\end{gathered}
\end{equation}

\section{Examples: $T(3)$, $SO(3)$, and $\mathbb{R}^{+}$}

So far, our discussion of Lie groups and Lie algebras has been precise but not very physically oriented. We have used abstract mathematical structures. For physical purposes, we have to work with specific groups and a certain coordinate system. We can proceed to calculate the components of the generators and the explicit form of the action of the groups on a manifold.

For particle mechanics, we consider the translation, rotation, and dilatation groups as they are the building blocks of the similarity group. We assume that the manifold is the configuration space of $N$ particles in the three-dimensional Euclidean space, parametrized by the variables $q^a_I$'s. Hence, $M = \mathbb{R}^{3N}$.

\subsection*{$T(3)$}

The group $T(3)$ is isomorphic to $\mathbb{R}^{3}$. It is a three-dimensional Lie group. We can coordinatize it by assigning the numbers $(\alpha^1,\alpha^2,\alpha^3)$ to every element of it. The product of two elements is represented by the sum of their coordinates. It then follows that the action of $T(3)$ on $M$ is given by

\begin{equation}
T\left( (\alpha^1,\alpha^2,\alpha^3),q^a_I \right) = q^a_I + \alpha^a.
\end{equation}

Thus, in this particular coordinate system, we can work out the generators on $T(3)$ by noting that

\begin{equation}
t^a_{ib} q^b_I = \delta^a_i.
\end{equation}

Thus, the generators are

\begin{equation}
t^a_{ib} = \delta^a_b \partial_i
\end{equation}

Note that we have suppressed the index $I$ in the above expressions. All the groups we study in this section act non-trivially on the spatial part of the configuration space. Therefore, the generators are dependent on the spatial indices, and not those that represent the particles. This is only a physical requirement and not a mathematical necessity. It is to only ensure that all of the particles are subject to the same transformation.

\subsection*{$SO(3)$}
The three-dimensional rotation group, characterized by the orthogonal matrices with determinant one and hence the name $SO(3)$, rotate the spatial components of $q^a_I$. The group manifold of $SO(3)$ is topologically compact and non-simply connected. We cannot cover the whole group with one coordinate system. One way is to work with Euler parameters to represent different rotations. However, as we are more curious about the generators, we just assign three numbers ($SO(3)$ is a three-parameter group) to its elements. Thus, for rotations we have

\begin{equation}
T\left( (\omega^1,\omega^2,\omega^3),q^a_I \right) = \Omega(\omega)^a_b q^b_I,
\end{equation}

where $\Omega$ is the corresponding rotation matrix. Hence,

\begin{equation}
t^a_{ib} q^b_I = \left. \frac{d}{d\omega^i} \Omega^a_b\right|_{\omega = 0} q^b_I.
\end{equation}

By a straightforward calculation, it can be shown that the derivative of $\Omega$ around the identity is a simple antisymmetric matrix with entries 1 and 0. Thus, by using the Levi-Civita symbol we have

\begin{equation}
t^a_{ib} q^b_I = \delta_{iq} \epsilon^{aqs} \delta_{sr} q^r_I.
\end{equation}

We can eliminate $q^b_I$ and arrive at

\begin{equation}
t^a_{ib}= \delta_{iq} \epsilon^{aqp} \delta_{br} q_I^r \partial_p.
\end{equation}

\subsection*{$\mathbb{R}^{+}$}

The dilatation group, as its name suggests, rescales the system, and is isomorphic to the set of positive real numbers. It is a one-dimensional group and its elements can be represented by one single number $\alpha$. Its action is given by

\begin{equation}
T\left(\alpha,q^a_I \right) = \alpha q^a_I.
\end{equation}

Thus, the generators are

\begin{equation}
t^a_{ib} q^b_I = \delta^a_b q^a_I,
\end{equation}

leading to

\begin{equation}
t^a_{ib} = \delta^a_b q^m_I \partial_m.
\end{equation}

The three groups we studied are at the heart of relational particle physics. I have summarized the generators of each one of these groups in table \ref{generators}. These are useful in our calculations for best-matching and constraints.

\begin{table}[H]
\begin{center}
\begin{tabular}{c|c|c} % <-- Alignments: 1st column left, 2nd middle and 3rd right, with vertical lines in between
\textbf{group} & \textbf{indices} & \textbf{generator} ($t^a_{\alpha b}$) \\

\hline \hline

translations, $T(3)$ & $\alpha = i = 1,2,3$ & $\delta^a_b \partial_i$ \\
rotations, $SO(3)$ & $\alpha = j+3 = 4,5,6$ & $\epsilon^{ajp} \delta_{br} q^r \partial_p$ \\
dilatations, $\mathbb{R}^{+}$ & $\alpha = 7$ & $\delta^a_b q^m \partial_m$ \\

\end{tabular}\caption[The generators of the rotation, translation, and dilatation group]{The generators of the groups $SO(3),T(3),\mathbb{R}^{+}$} \label{generators}
\end{center}
\end{table}

\chapter{Constrained Hamiltonian Systems and Dirac's Approach}\label{constrained}
In his wonderful lectures on constraints and quantization of the constrained systems \cite{Dirac.1964}, Paul Dirac wrote

\begin{displayquote}
\emph{
I feel that there will always be something missing from [alternative methods] which
we can only get by working from a Hamiltonian.}
\end{displayquote}

So do I... at least as far as shape dynamics and relational physics are concerned. The language of constrained systems is very important in relational physics. The whole point of relational physics is that we have a gauge theory in an extended configuration space or phase space. As we will prove in this appendix, whenever we have gauge theories, we have constraints. Thus, the natural language for describing the mathematical formulation of the theory is that of the constrained systems. In this appendix, we give a rather brief account of the Hamiltonian constrained systems. Lagrangian systems can also be studied deeply, but for developing relational physics - even though it is also a constrained Lagrangian system - the Hamiltonian formulation is more important.\footnote{This approach was first developed and explained very well by Dirac in \cite{Dirac.1964}. Among major references on this topic, I find \cite{Henneaux.1992} clear and thorough.}

\section{Singular Lagrangians and primary constraints}

Constrained Hamiltonian systems can be considered completely independently. However, for the sake of more clarity, we start with an action in the Lagrangian form:

\begin{equation}
S[q] = \int dt \, L(q^i,\dot{q}^i,t),
\end{equation}

where $q^i$'s belong to a $N$-dimensional configuration space and $t$ is a time parameter, not necessarily with a physical meaning at this point.

Demanding the action to be stationary along the space of all trajectories between two fixed points, gives us the Euler-Lagrange equations:

\begin{equation}\label{Ceulerlagrangeeq}
\frac{d}{dt} \left(\frac{\partial L}{\partial \dot{q}^i}\right)
- \frac{\partial L}{\partial q^i} = 0.
\end{equation}

The above equations can be written more explicitly by separating different orders as

\begin{equation}\label{Cequationsofmotion}
\ddot{q}^j \frac{\partial^2 L}{\partial \dot{q}^j \partial \dot{q}^i} + \dot{q}^j \frac{\partial^2 L}{\partial q^j \partial \dot{q}^i} - \frac{\partial L}{\partial q^i} = 0,
\end{equation}

where we have used Einstein's summation convention. The point of Eq. (\ref{Cequationsofmotion}) is that it directly shows the necessary and sufficient condition for writing the accelerations $\ddot{q}^i$ in terms of the positions and velocities, which is the invertibility of the matrix appearing next to the accelerations, i.e.,

\begin{equation}\label{Cdeterrhessian}
\det \left(\frac{\partial^2 L}{\partial \dot{q}^j \partial \dot{q}^i}\right) \neq 0.
\end{equation}

If this determinant vanishes, we say the Lagrangian is \emph{singular}. One case of such a failure is the existence of gauge freedom in the Lagrangian formulation. In this case, one can show that the Lagrangian is singular and the equations of motion do not determine the solution uniquely.\footnote{See \cite{Kiefer.2012}, Sec. 3.5 for the details.} This is actually what Noether's second theorem shows.

Singular Lagrangians, in turn, are also relevant to the Hamiltonian formulation. The starting point is the definition of the conjugate momenta

\begin{equation}\label{Cconjugatemomenta}
p_i \equiv \frac{\partial L}{\partial \dot{q}^i}.
\end{equation}

By virtue of the implicit function theorem, we know that the relations (\ref{Cconjugatemomenta}) can be used to solve for $\dot{q}^i$ in terms of the momenta if and only if (\ref{Cdeterrhessian}) holds. We first assume that the rank of $\frac{\partial^2 L}{\partial \dot{q}^j \partial \dot{q}^i}$ is constant throughout the tangent bundle of the configuration space. The failure of this matrix to be of maximal rank, i.e., (\ref{Cdeterrhessian}) fails, means that the momenta (\ref{Cconjugatemomenta}) are not all independent and must satisfy certain relations of the form

\begin{equation}\label{Cprimarycons}
\phi_a (q,p) = 0, \qquad a=1,\cdots,M.
\end{equation}

These constraints are solely the result of the definition (\ref{Cconjugatemomenta}) and reflect the structure of the phase space. Using the standard terminology of constrained systems Dirac proposed, we call (\ref{Cprimarycons}) \emph{primary constraints}. If we assume that the rank of $\frac{\partial^2 L}{\partial \dot{q}^j \partial \dot{q}^i}$ is $N-M$, then $M$ independent primary constraints would be found.

The canonical Hamiltonian function is by definition

\begin{equation}
H = p_i \dot{q}^i - L,
\end{equation}

but now that we are dealing with a constrained phase space, one wonders how to construct the Hamiltonian as a function on phase space and write down the equations of motion. For that, we state a theorem, the proof of which can be found in Chapter 1 of \cite{Henneaux.1992}.

\begin{theorem}\label{Ctheoremconstraints}
If a smooth phase space function $F$ vanishes on the surface $\phi_a = 0$, then it follows that $F = u^a \phi_a$ for some smooth functions $u^a(q,p)$. Furthermore, if we have $\lambda_i \delta q^i + \mu^i \delta p_i = 0$ for variations tangent to the constrained surface, then

\begin{equation}
\lambda_i = u^a \frac{\partial \phi_a}{\partial q^i},\qquad
\mu^i = u^a \frac{\partial \phi_a}{\partial p_i},
\end{equation}
for arbitrary smooth functions $u^a$'s.
\end{theorem}

This should be intuitively clear. The second part of the theorem is especially used for finding Hamilton's equations of motion. Following the definition of the Hamiltonian, we vary it with respect to the parameters $\dot{q}^i,q^i$ independently:

\begin{equation}
\begin{aligned}
\delta H &= \delta p_i \dot{q}^i + p_i \delta \dot{q}^i - \delta q^i \partial_{q_i} L - \delta {\dot{q}^i} \partial_{\dot{q}^i} L\\
&= \delta p_i \dot{q}^i - \delta q^i \partial_{q_i} L.
\end{aligned}
\end{equation}

Because of the existence of certain constraints, the variations appearing in the second line are tangent to the constraint surface. Thus, along the constraint surface, we have

\begin{equation}
\left( \frac{\partial H}{\partial p_i} - \dot{q}^i \right) \delta p_i +
\left( \frac{\partial H}{\partial q^i} + \partial_{q_i} L\right) \delta q^i = 0,
\end{equation}

and by using the theorem (\ref{Ctheoremconstraints}) and Euler-Lagrange equation (\ref{Ceulerlagrangeeq}) we have the following equations of motion

\begin{equation}
\begin{gathered}
\dot{q}^i = \frac{\partial H}{\partial p_i} + u^a \frac{\partial \phi_a}{\partial p_i},\\
\dot{p}_i = -\frac{\partial H}{\partial q_i} - u^a \frac{\partial \phi_a}{\partial q^i},
\end{gathered}
\end{equation}

for some phase space functions $u^a$. But we will see that certain consistency conditions may restrict them.

The above equations of motion can be derived from the phase space action

\begin{equation}
S_P[q,p,u] = \int dt \, \left(p_n \dot{q}^n - (H + u^a \phi_a) \right).
\end{equation}

Following Dirac, we introduce the concept of \emph{weak equality}, denoted by $\approx$, which means that a certain function is equal to another function \emph{on} the constrained surface. For example, for all constraints, we have $\phi_a \approx 0$. This is particularly useful for having a notion of equality restricted to the constraint surface.

\section{Consistency conditions}

The constrained system is consistent if the constraints `propagate' under the Hamiltonian evolution, meaning that $\dot{\phi}_a = 0$. Using the above equations of motion, this condition is

\begin{equation}\label{Cconsistency}
\dot{\phi}_a = \{ \phi_a , H \} + u^{a'} \{ \phi_a , \phi_{a'} \} = 0.
\end{equation}

(\ref{Cconsistency}) can result in three scenarios: either it reduces to a relation independent of the $u^{a'}$'s involving only the $q$'s and $p$'s, or it may impose a restriction on some of the $u^{a'}$'s (or all of them), or it may actually lead to an inconsistency! The last case is the least attractive one, but in the first case, the resultant relation - which is independent of the primary constraints - is called a \emph{secondary constraint}, because the equations of motion have been used for arriving at them. We can then continue the process with the secondary constraints and check the consistency conditions. Finally, if the system is consistent, this approach should come to an end and we should have a number of primary and secondary constraints

\begin{equation}\label{Ccompleteset}
\phi_{\alpha} \approx 0, \qquad \alpha = 1,\cdots,M+K,
\end{equation}

with $K$ secondary constraints that have been added. Once again, we assume that these constraints are all independent, and hence, the above set is minimal.

Now that we have found the complete set of constraints (\ref{Ccompleteset}), we can study the implication of their propagation for the $u^{a}$'s, as

\begin{equation}\label{Cinhomequation}
\dot{\phi}_{\alpha} = \{ \phi_{\alpha} , H \} + u^{a} \{ \phi_{\alpha} , \phi_{a} \} \approx 0
\end{equation}

should hold. In order to find the general form of $u^{a}$'s, we define another classification of constraints Dirac proposed.

Again, we assume the rank of $\{ \phi_{\alpha} , \phi_{\beta} \}$ is constant along the surface constraints to avoid certain mathematical problems. We call a constraint \emph{first-class} if its Poisson bracket with every other constraint vanishes weakly, and \emph{second-class} if it does not. We denote the first-class constraints by

\begin{equation}\label{Cfirstclass}
\gamma_{\alpha} \approx 0,\qquad \alpha = 1,\cdots,f
\end{equation}

and second-class constraints by

\begin{equation}\label{Csecondclass}
\chi_{\beta} \approx 0,\qquad \beta = 1,\cdots,g.
\end{equation}

It is clear that $f+g = M+K$. One can define a first-class function in general, meaning a function whose Poisson bracket with any other constraint vanishes weakly. We have defined two classifications. But we will see that the first-class versus second-class distinction is more meaningful and fundamental in the Hamiltonian formulation than the primary versus secondary one.

For solving (\ref{Cinhomequation}) in light of this analysis, we see that the coefficients of the first-class primary constraints in the Hamiltonian remain unrestricted and arbitrary, and those corresponding to the second-class primary constraints would be solved in terms of $q$'s and $p$'s. Therefore, we arrive at the total first-class Hamiltonian

\begin{equation}
H_{T} = H_{c} + u^{b}_0 \chi_{b} + v^{a} \gamma_{a},
\end{equation}

where $H_c$ is the original canonical Hamiltonian, $u^{b}_0 (q,p)$ is the solution of (\ref{Cinhomequation}), $v^a$'s are abitrary, and the indices $a$ and $b$ represent the primary constraints here, as opposed to $\alpha$ and $\beta$. Therefore, we have the final form of the equations of motion for an arbitrary phase space function:

\begin{equation}\label{Cfinaleqofmotion}
\dot{F}(q,p) \approx \{F,H_T\}.
\end{equation}

There is something interesting about the first-class primary constraints in the total Hamiltonian. Their coefficients are arbitrary, meaning that for any given initial condition on the constraint surface, we can vary $v^a$ smoothly and generate a family of solutions. This `local freedom', in turn, signals the concept of gauge transformation.

\section{First-class constraints as gauge generators}

To clarify the role of the first-class primary constraints as gauge generators, we assume that the time parameter with respect to which we wrote (\ref{Cfinaleqofmotion}) represents a physically measurable quantity, like time measured by a specific clock.

Hence, if we start with a given initial point $s$ on phase space satisfying $\phi_\alpha = 0$, evolve that for an infinitesimal amount of time from $t$ to $t+\delta t$, and change $v^a$ to $\bar{v}^a$, the change of the evolved point becomes

\begin{equation}\label{Cgaugetransform}
\delta s = \delta t (\bar{v}^a - v^a) \{s,\gamma_a\} \equiv \delta v^a \{s,\gamma_a\}.
\end{equation}

Hence, if the time parameter is physical (observable), and according to our expectation that the equations of motion should completely determine the evolution of the physical (observable) points, we deduce that the first-class primary constraints generate gauge transformations of the form (\ref{Cgaugetransform}). It can be also proved that the Poisson bracket between the first-class primary constraints and those between one first-class primary constraint and the total Hamiltonian also generate gauge transformations (see \cite{Henneaux.1992}).

Although the above argument works essentially for first-class primary constraints, it is reasonable to say that all first-class constraints generate gauge transformations. This was conjectured by Dirac. Although some counterexamples can be proposed, in our physical applications it remains completely valid and consistent to include all the other first-class constraints in the Hamiltonian. This is exactly what happens in ADM formalism and shape dynamics. Hence, we may define the extended Hamiltonian:

\begin{equation}\label{Cextendedhamil}
H_{E} = H_{c} + u^{b}_0 \chi_{b} + v^{\alpha} \gamma_{\alpha},
\end{equation}

where $\gamma_{\alpha}$ represents all the first-class constraints. Now that we have gauge freedom, we may conceive of `gauge fixation'. It is physically desirable to fix the freedom of the theory and establish a one-to-one correspondence between the physical states and the phase space variables. The gauge freedom can be fixed by imposing a number of constraints on the phase space which are second-class with the original first-class primary constraints, i.e.,

\begin{equation}
G_c(q,p) \approx 0,\qquad \{G_c,\gamma_a\} \not\approx 0.
\end{equation}

\begin{figure}[h]
\centering
\includegraphics[scale=0.6]{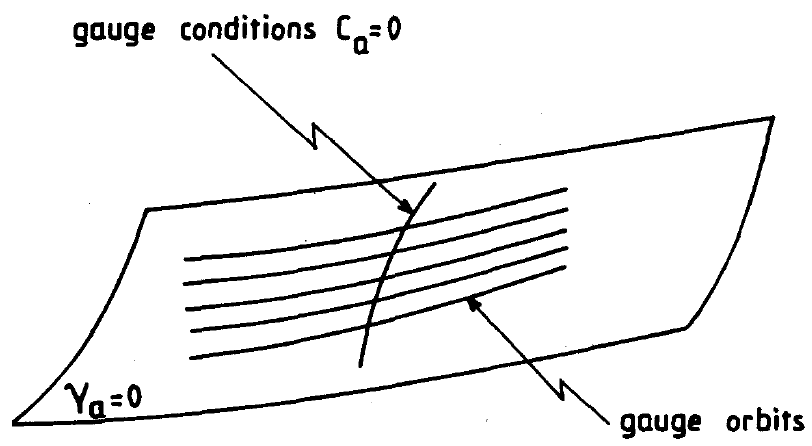}
\captionsetup{width=0.8\linewidth}
\caption[Geometrical intuition of guage fixation]{\small A complete set of gauge conditions determine a surface that intersects each one of the gauge orbits once and only once. Picture taken from \cite{Henneaux.1992}.}
\label{Cfiggaugefixation}
\end{figure}

If one wants to fix the gauge freedom completely, one in principle needs $f$ independent gauge conditions in the above form. Imposing more constraints restricts the physical part of the phase space, and is not admissible. Fig. \ref{Cfiggaugefixation} clearly illustrates what gauge conditions do. The gauge conditions cut through the gauge orbits and pick up one physical variable along each one of the orbits.

\section{Elimination of the second-class constraints}\label{Celimseconclasssec}

As we noted before, the distinction between first-class and second-class constraints is more fundamental in the Hamiltonian formulation. We dealt with the first-class ones in the previous section. Now we turn our attention to the second-class constraints. Although they do not generate gauge transformations, we will see that they can be completely removed by manipulating the Poisson brackets and imposing them strongly.

First, we begin with a simple example of two second-class constraints $q^1 \approx 0, \, p_1 \approx 0$ where $q^1,p_1$ are two canonical variables. It then becomes obvious that these constraints remove those two variables completely and we can work in the reduced phase space with a modified Poisson bracket that only includes the derivatives with respect to conjugate variables other than $(q^1,p_1)$.

Following this intuition, we can define \emph{Dirac bracket} and remove the second-class constraints in any general system. First of all, we should note that the matrix

\begin{equation}
C_{\alpha \beta} = \{\chi_\alpha,\chi_\beta\}
\end{equation}

is of maximal rank, and hence, invertible. Otherwise, in case that there exists $w^\alpha$ such that $w^\alpha C_{\alpha \beta} \approx 0$, we can identify the constraint $w^\alpha C_\alpha$ as another first-class constraint which is incompatible with our basic postulate that we classified all the constraints. From $C_{\alpha \beta}$ being invertible and asymmetric, it then follows that the number of the second-class constraints must be even.\footnote{The proof follows from the consideration that $\det C = \det C^T = - \det C$ if $C$ is of odd dimension, implying $\det C = 0$.}

Finally, we define the Dirac bracket as

\begin{equation}
\{\cdot,\cdot \cdot\}_{DB} \equiv \{\cdot,\cdot \cdot\} -
\{\cdot,\chi_\alpha\} C_{\alpha \beta}^{-1} \{\chi_\beta,\cdot \cdot\}.
\end{equation}

It is easy to show that

\begin{enumerate}
\item The Dirac bracket of any function with a first-class function, in particular, the total Hamiltonian, is weakly equal to the result of the Poisson bracket,

\begin{equation}
\{F,H_T\}_{DB} \approx \{F,H_T\}.
\end{equation}

\item The Dirac bracket of any second-class constraint with any other function vanishes weakly,

\begin{equation}
\{\chi_\alpha, \cdot\} \approx 0.
\end{equation}
\end{enumerate}

So, in light of the second property, we see that in evaluating the Dirac bracket we can set the second-class constraints equal to zero even before calculating the bracket. Furthermore, because of the first property, we can determine the evolution of the system with respect to the Dirac bracket. Hence, we deduce

\begin{equation}
\dot{F} \approx \{F,H_T\} \approx \{F,H_T\}_{DB} \approx \{\bar{F},\bar{H}_T\}_{DB},
\end{equation}

where $\bar{F}$ and $\bar{H}_T$ are $F|_{\chi_\alpha = 0}$ and $H_T|_{\chi_\alpha = 0}$ respectively.

In this way, we can eliminate all the second-class constraints by modifying the Poisson bracket. This has a far-reaching conclusion, especially for shape dynamics and non-equivariant best-matching.

We should make one final remark before ending this section. Let us count the number of degrees of freedom that are left. We started with a $2N$-dimensional phase space and found $f$ first-class constraints and $g$ second-class constraints. We noted that $g=2k$. Moreover, each first-class constraint removes 2 degrees of freedom, one of which is because of the constraint itself, the other one because of gauge fixation corresponding to that.\footnote{There is this famous saying ``the gauge always hits twice'' attributed to Teitelboim.} Hence, we are left with $2N - 2f -2k = 2(N-f-k)$ degrees of freedom, i.e., $N-f-k$ pairs of conjugate variables.

There is a caveat, however, regarding the constrained field theories. In that context, we should talk about the number of degrees of freedom `per point', as the phase space is infinite-dimensional in such theories.

\section{Reparametrization invariant theories}\label{Creparinvariant}

One particular example of a constrained system is reparametrization invariant systems. There are basically two forms of such systems. One is those systems that are initially formulated with respect to a time parameter, and then the time is taken to be a dynamical variable. This was our approach in Appx. \ref{jacobi}. However, there are also reparametrization invariant dynamical theories that are written on a well-defined configuration space that does not include any separate time parameter like the latter case. Here, we focus on general reparametrization invariant theories. Any Lagrangian theory of this sort can be subject to our analysis here.

We start with a phase space action,

\begin{equation}
S_P = \int d\lambda \, \left(p_n \dot{q}^n - H \right).
\end{equation}

Assuming that the phase space variables are scalar functions of $\lambda$, we see that the first term in the action respects the reparametrization invariance. If the action is to have this symmetry, we see that the Hamiltonian must vanish. If the Hamiltonian vanishes as a primary constraint and not identically, based on our previous analysis, the total Hamiltonian would be something like $H \equiv N H_0$, with the arbitrary function $N$. Thus, by demanding that $N$ transforms as a scalar density under reparametrization, i.e.,

\begin{equation}
N'(\lambda') = N(\lambda) \frac{d\lambda}{d\lambda'}
\end{equation}

the resultant phase space action would be reparametrization invariant.

Hence, we see that there is a strong connection between reparametrization invariance and the vanishing of the Hamiltonian as a constraint. In fact, this is the result we got to in (\ref{Avanishhamiltcons}) for a system with a dynamical time.

Therefore, we can generally write reparametrization invariant Hamiltonian systems as

\begin{equation}
H_T = N H + v^a \gamma_a,
\end{equation}

with arbitrary scalar densities $N$ and $v^a$, and the first-class constraints $H$ and $\gamma_a$.

This opens up an important conceptual issue: What is the status of the Hamiltonian as a first-class constraint in reparametrization invariant theories in light of our observation that first-class constraints generate gauge transformations? There has been quite a lot of debate over this issue in the literature. Some talk about the absence of true dynamics in such theories and that ``evolution is nothing but the unfolding of a gauge transformation''.

But there is a clearer approach to understanding the meaning of the Hamiltonian constraint proposed by Barbour and Foster \cite{Barbour.2008}. The starting point in this approach is to note that the interpretation of first-class constraints as generators of gauge transformation rested on the presumption that the time parameter itself is physical. This is obviously not the case in reparametrization invariant theories! Hence, it is reasonable to modify our original interpretation than try to apply it to a case with completely different underlying presumptions. We say that the Hamiltonian constraint generates `true' motion, but with a non-physical parameter. The gauge aspect of such theories is that we can change the $N$, resulting in a different parametrization of the solution curves.

In other words, \emph{the change in the parameter} is the gauge transformation the Hamiltonian generates, not the change in the configuration space variables. In reparametrization invariant theories, the Hamiltonian generates true dynamics with respect to an arbitrary parameter in parallel.

\chapter{ADM Formalism}\label{CADMformalism}

One of the most influential works on classical general relativity is the Hamiltonian formulation of Einstein's general relativity, known as the ADM formalism, developed in 1960s.\footnote{See the 1962 review of the work by its founders, republished in \cite{Arnowitt.2008}. Chapter 4 of \cite{Kiefer.2012} and Section 21 of \cite{Blau.} are also very thorough.} ADM formalism has the merit of describing the core dynamics underlying the theory of general relativity and melts the frozen spacetime picture of the theory, provides an initial-value formulation of the theory suitable for numerical calculations, brings many insights into quantum gravity, and builds a framework to meaningfully define energy and radiation.

It also plays a profound role in developing shape dynamics, as shape dynamics is an alternative dual theory with a different constraint algebra and its connection with the Hamiltonian formulation of the standard general relativity makes the inner working of the theory remarkably clear. We give a rather brief account of this formalism and derive the key results, starting with the standard Einstein-Hilbert action and using the Gauss-Codazzi equations.

\section{Breaking up the 4-dimensional spacetime}

We assume that spacetime $M$ equipped with the Lorentzian metric $g^{(4)}_{\mu \nu}$ is globally hyperbolic, hence, foliable to spacelike hypersurfaces with a common topology, and that there exists a time parameter $t$ that labels the `leaves' $\Sigma_t$.\footnote{See theorem 8.3.14 in \cite{Wald.1984}. The requirement of global hyperbolicity is very important and the Hamiltonian formulation works if and only if the spacetime associated with the solution of the equations is globally hyperbolic.} Thus, $M = \cup_{t} \Sigma_t \simeq \Sigma \times \mathbb{R}$, where $\Sigma$ denotes the topology of the hypersurfaces.

Now we consider a coordinate system that includes the parameter $t$ as the time coordinate. For the sake of simplicity, we assume that each one of the $\Sigma_t$'s can be covered with a single coordinate system $(x_1,x_2,x_4)$. Therefore, the $(x_1,x_2,x_4,t)$ constitutes a well-behaved coordinate system on the entire $M$.

For our analysis, we consider a general coordinate system $y^\alpha(t,x^a)$. Then, the vector field $\partial_t^\alpha \equiv \frac{\partial y^\alpha}{\partial t}$ gives us the flow of time, and the vectors $E^\alpha_a \equiv \left. \frac{\partial y^\alpha}{\partial x^a} \right|_t$ represent three tangent vectors belonging to $T(\Sigma_t)$. Moreover, The existence of foliation ensures that there is a globally well-defined unit timelike normal vector $n$ for each $\Sigma_t$. Hence, in our arbitrary coordinate system we can decompose the `time vector field' $\partial_t$ as

\begin{equation}\label{Ddecomtimevector}
\partial_t^\mu = N n^\mu + N^\mu,
\end{equation}

where $N^\mu$ is a vector field orthogonal to $n^\mu$, i.e., $g^{(4)}_{\mu \nu} n^\mu N^\nu = 0$. Note that $N^\mu$ is tangent to the leaves and can be written as $N^\mu = E^\mu_a N^a$ for the vector fields $N^a$ in the tangent bundle of the $\Sigma_t$'s. $N(x_i,t)$ is called the \emph{lapse function} and $N^a (x_i,t)$ the \emph{shift vector}. Their geometrical meaning will signify these names.

In the suitable coordinate system $(x_1,x_2,x_4,t)$ we constructed, we can use the above relations and decompose the 4D Lorentzian metric:

\begin{equation}\label{DADMdecomposition}
\begin{aligned}
ds^2 &\equiv g^{(4)}_{\mu \nu}\, dy^\mu dy^\nu \\
&= g^{(4)}_{\mu \nu} \, (N n^\mu dt + E^\mu_a N^a dt + E^\mu_a dx^a) (N n^\nu dt + E^\nu_a N^a dt + E^\nu_a dx^a)\\
&= - (N^2 - g_{ab} N^a N^b) dt^2 + 2 g_{ab} N^a dt dx^b + g_{ab} dx^a dx^b,
\end{aligned}
\end{equation}

where $g_{ab}$ denotes the induced 3D Riemannian metric on $\Sigma$. (\ref{DADMdecomposition}) is called the \emph{ADM decomposition} of the metric and is the departure point for deriving the ADM action and then pursuing the Hamiltonian formulation.

Thus, The components of the metric and its inverse in the coordinate system $(x_a,t)$ are

\begin{equation}
\begin{gathered}
g^{(4)}_{00} = -N^2 + g_{ab} N^a N^b, \qquad g^{(4)}_{0a} = g_{ab} N^b, \qquad g^{(4)}_{ab} = g_{ab},\\
g^{(4)^{00}} = -\frac{1}{N^2},\qquad g^{(4)^{a0}} = \frac{N^a}{N^2},\qquad g^{(4)^{ab}} = g^{ab} - \frac{N^a N^b}{N^2}.
\end{gathered}
\end{equation}

Another relation that becomes useful in the next section is

\begin{equation}\label{Ddeterminantmetric}
g^{(4)} \equiv \det g^{(4)}_{\mu \nu} = N g,
\end{equation}

where $g$ is the determinant of the induced Riemannian metric.

We can now clarify the geometrical meaning of $N$ and $N^a$. To see that, note that (\ref{Ddecomtimevector}) in the $(x^a,t)$ coordinate system gives

\begin{equation}
n = (N,-\frac{N^a}{N}).
\end{equation}

\begin{figure}[h]
\centering
\includegraphics[scale=0.35]{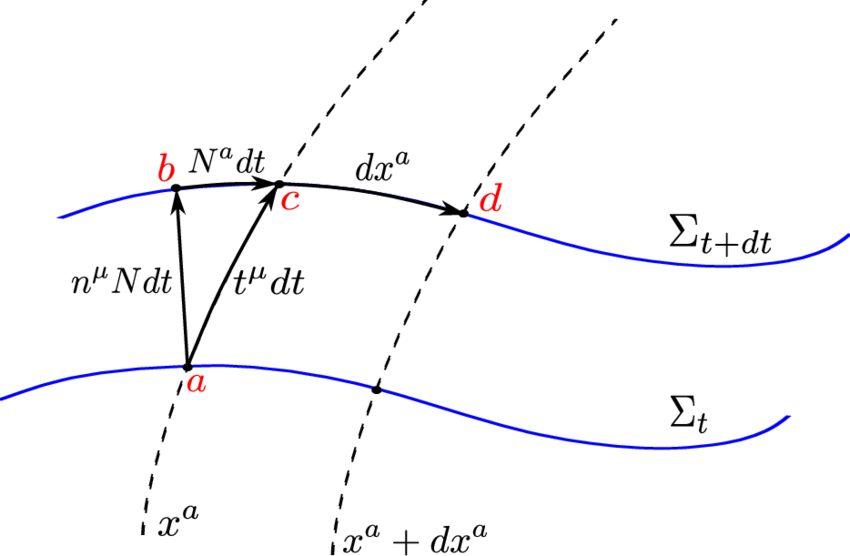}
\captionsetup{width=0.8\linewidth}
\caption[Geometrical representation of the lapse and the shift]{\small Graphical representation of the lapse function and the shift vector. Picture taken from \cite{Barrau.2014}.}
\label{Dfoliationlapseshift}
\end{figure}

Hence, $N$ expresses the proper time elapsed between two nearby points on $\Sigma_t$ and $\Sigma_{t+\delta t}$ along the direction of the normal vector, therefore, it measures the \emph{lapse} of the surfaces. $N^i$ also measures the horizontal \emph{shift} of the leaves in the direction of $n$, as illustrated in Fig. \ref{Dfoliationlapseshift}.

\section{ADM action}

We are now in the position to use the decomposition we found for a given foliation and write Einstein-Hilbert action in terms of the induced metric, and the lapse and the shift variables.

First, we define the extrinsic-curvature (the second fundamental form) as

\begin{equation}
K_{ab} \equiv E^\alpha_a E^\beta_b \nabla_\alpha n_\beta,
\end{equation}

where $\nabla_\alpha$ denotes the covariant derivative with respect to the metric $g^{(4)}_{\mu \nu}$. Using the ADM decomposition, it can be shown that $K_{ab}$ is related to the time derivative of the induced metric:

\begin{equation}\label{Dtimederivextcurv}
K_{ab} = \frac{1}{2N} \left(\dot{g}_{ab} - \mathfrak{L}_{N} g_{ab} \right).
\end{equation}

$\mathfrak{L}_N$ denotes the Lie derivative with respect to the shift vector.

Using the Gauss-Codazzi equations, we can write the 4D Ricci scalar $R^{(4)}[g^{(4)}_{\mu \nu}]$ in terms of the 3D intrinsic scalar curvature $R[g_{ab}]$ and the extrinsic curvature:

\begin{equation}\label{Dgausscodazzi}
R^{(4)} = R + K_{ab} K^{ab} - K^2 + 2 \nabla_{\alpha} \left( n^\beta \nabla_\beta n^\alpha - n^\alpha \nabla_{\beta} n^\beta \right).
\end{equation}

We ignore the last term in the action which is a total derivative.\footnote{We assume that the space is topologically closed, i.e., compact and without boundary, to avoid certain mathematical complications. This assumption is justifiable in shape dynamics. Moreover, the temporal boundary terms can also be ignored when performing fixed-end-point variation.}

Thus, in light of (\ref{Ddeterminantmetric}) and (\ref{Dgausscodazzi}), one can write the Einstein-Hilbert action

\begin{equation}
S_{EH}[g^{(4)}_{\mu \nu}] = \int d^4y \, R^{(4)}
\end{equation}

purely in terms of the spatial structures $g_{ab}$ and $K_{ab}$, referred to as the \emph{ADM action}:

\begin{equation}\label{DADMaction}
S_{ADM}[g_{ab},N,N^a] = \int dt d^3x \, \mathcal{L}_{ADM} = \int dt d^3x \, \sqrt{g} N \left(R+ K^{ab}K_{ab} - K^2 \right),
\end{equation}

where we have introduced the ADM Lagrangian density. This action includes first-order time derivatives of the metric (expressed through $K_{ab}$) and its spatial derivatives. The Ricci scalar plays the role of the `potential term' and $K^{ab}K_{ab} - K^2$ the `kinetic term'. We will see that more clearly when we introduce the BSW action at the end of this appendix. The kinetic term can be written as

\begin{equation}
K_{ab}K^{ab} - K^2 = G^{abcd} K_{ab} K_{cd},
\end{equation}

where $G^{abcd}$ is called the DeWitt supermetric and is

\begin{equation}
G^{abcd} = \frac{1}{2} \left( g^{ac} g^{bd} + g^{ad}g^{bc} \right) - g^{ab}g^{cd}.
\end{equation}

DeWitt supermetric plays the role of a metric in the space of spatial metrics. Formally, for metric deformations $\delta g_{ab}$ at a specific spatial metric, we can define the following `inner product':

\begin{equation}
\langle \delta_1 g, \delta_2 g \rangle \equiv \int d^3x \, \sqrt{g} G^{abcd} \delta_1 g_{ab} \delta_2 g_{cd}.
\end{equation}

The space of metrics is very important in shape dynamics, as the configuration space is the pivotal structure in the theory.

As that the extrinsic curvature is related to the time derivative of the metric (\ref{Dtimederivextcurv}). Hence, the ADM action is the suitable action for performing the Legendre transformation and writing general relativity in the Hamiltonian formulation.

\section{The Hamiltonian of GR}

From (\ref{DADMaction}), the momenta conjugate to $g_{ab}$ are

\begin{equation}\label{Dconjugatemomenta}
p^{ab} = \frac{\delta \mathcal{L}_{ADM}}{\delta \dot{g}_{ab}} = \sqrt{g} G^{abcd} K_{cd} = \sqrt{g} \left( K^{ab} - K g^{ab} \right).
\end{equation}

We see that $p^{ab}$ is a tensor-density of weight 1, i.e., $\frac{p^{ab}}{\sqrt{g}}$ is a tensor.

Moreover, we define the momenta conjugate to $N$ and $N^a$ which result in four primary constraints:

\begin{equation}\label{Dlapseshiftmomenta}
p_N = \frac{\delta \mathcal{L}_{ADM}}{\delta \dot{N}} \approx 0,\qquad
p_{N^a} = \frac{\delta \mathcal{L}_{ADM}}{\delta \dot{N}^a} \approx 0.
\end{equation}

It should not be a surprise. After all, the lapse function and the shift vector appear in the ADM action as some Lagrange multipliers.

We can proceed to work out the canonical Hamiltonian. The Hamiltonian density is

\begin{equation}
\mathcal{H}_{ADM} = p^{ab} \dot{g}_{ab} - \mathcal{L}_{ADM}.
\end{equation}

We can use (\ref{Dtimederivextcurv}) and (\ref{Dconjugatemomenta}) to fine $\dot{g}_{ab}$ in terms of the conjugate momenta:

\begin{equation}
\dot{g}_{ab} = 2N K_{ab} + 2 \nabla_{(a} N_{b)} = \frac{2N}{\sqrt{g}} G_{abcd} p^{cd} + 2 \nabla_{(a} N_{b)},
\end{equation}

where $G_{abcd}$ is the inverse of the DeWitt supermetric, i.e.,

\begin{equation}\label{CDeWittsupermetric}
G_{abcd} = \frac{1}{2} \left(g_{ac}g_{bd} + g_{ad} g_{bc} - g_{ab} g_{cd} \right), \qquad G^{abcd} G_{cdef} = \frac{1}{2} \left( \delta^{(a}_e \delta^{b)}_f \right).
\end{equation}

Thus, the Hamiltonian density is

\begin{equation}
\mathcal{H}_{ADM} = \frac{N}{\sqrt{g}} G_{abcd} p^{ab} p^{cd} - \sqrt{g} N R + 2p^{ab} \nabla_{a} N_{b}.
\end{equation}

The Hamiltonian density is a density of weight 1. Thus, when integrated all over the space, the Hamiltonian would be a scalar. We have

\begin{equation}
\begin{aligned}
H_{ADM}[g,p,N,N^a] \equiv \int d^3x \, \mathcal{H}_{ADM} &= \int d^3x \, \left( N(x) H[g_{ab},p^{ab},x) + N^a(x) H_{a}[g_{ab},p^{ab},x) \right) \\
&\equiv N \sbullet H + N_a \sbullet H^a,
\end{aligned}
\end{equation}

where

\begin{equation}
H = \frac{1}{{\sqrt{g}}} G_{abcd} p^{ab} p^{cd} - \sqrt{g} R
\end{equation}

is called the \emph{Hamiltonian constraint} and

\begin{equation}
H^a = -2 \nabla_{b} p^{ab}
\end{equation}
is the \emph{momentum (or diffeomorphism) constraint} (these names will be clarified shorlty) . Note that we have omitted the total derivative $2 \nabla_{a} \left( p^{ab} N_{b} \right)$ in the integral. Moreover, based on convention, $\nabla_{a} p^{ab}$ actually means $\sqrt{g} \nabla_{a} \frac{p^{ab}}{\sqrt{g}}$, as follows from a general convention for writing the covariant derivative of densities more succinctly. The bullets in the last line denote \emph{smearing}.

The Hamiltonian is a functional of the metric and its conjugate momenta, the lapse function, and the shift vector. These variables along with $p_N$ and $p_{N^a}$ constitute the phase space. We have also the following non-zero Poisson brackets:

\begin{equation}
\begin{gathered}
\{g_{ab}(x),p^{cd}(y)\} = \left( \delta^{(c}_a \delta^{d)}_b \right) \delta^3(x-y)\\
\{N(x),p_N(y)\} = \delta^3(x-y)\\
\{N^a(x),p_{N^b}(y)\} = \delta^a_b \delta^3(x-y).
\end{gathered}
\end{equation}

We have calculated the canonical Hamiltonian. But general relativity is a constrained system and Dirac's approach is applicable here. We have the primary constraints (\ref{Dlapseshiftmomenta}). Hence, we have the total Hamiltonian

\begin{equation}
H_{T} = N \sbullet H + N_a \sbullet H^a + \xi \sbullet p_N + \xi^a \sbullet p_{N^a}.
\end{equation}

The consistency conditions for the propagation of $p_N$ and $p_{N^a}$ give the secondary constraints

\begin{equation}
\{p_N(x),H_T\} = H(x) \approx 0,\qquad \{p_{N^a}(x),H_T\} = H_a(x) \approx 0.
\end{equation}

Hence, the $ADM$ Hamiltonian is a linear combination of 4 constraints. In fact, it can be shown that $H = G_{00}$ and $H_{a} = G_{0a}$ where $G$ is the Einstein tensor written in the $(x^a,t)$ coordinates. This makes sense, as we know that Einstein's equations are not all dynamical and 4 of them represent certain constraints. This is what the above constraints show for pure geometry, i.e., vacuum solutions.

Before moving to the interpretation of these cosntraints, we note that all these constraints are first-class.One can show that the Hamiltonian and the momentum constraints satisfy the following relations, known as \emph{surface deformation algebra}:

\begin{equation}
\begin{gathered}
\{N_1 \sbullet H, N_2 \sbullet H \} = \left( N_1 \nabla_a N_2 - N_2 \nabla_a N_1 \right) \sbullet H^a\\
\{ M_a \sbullet H^a, N \sbullet H \} = \left( \mathfrak{L}_{M} N \right) \sbullet H\\
\{N_a \sbullet H^a, M_b \sbullet H^b \} = \left( \mathfrak{L}_{N} M \right) \sbullet H^a.
\end{gathered}
\end{equation}

As all of these constraints are first-class, and $p_N$ and $p_{N^a}$ generate `gauge' transformations' $N \rightarrow N + \xi$, $N_a \rightarrow N^a + \xi^a$. Hence, the lapse function and the shift vector are arbitrary. Therefore, without loss of generality, we can effectively work with the ADM Hamiltonian instead and take $N$ and $N_a$ to be some arbitrary coefficients, i.e., remove $(N,p_N)$ and $(N^a,p_{N^a})$ from the phase space. Hence, we have the Hamiltonian

\begin{equation}\label{Dtotalhamiltonianadm}
H_{T}[g_{ab},p^{ab}] = N \sbullet H + N_a \sbullet H^a.
\end{equation}

and the constraints

\begin{equation}
H \approx 0, \qquad H^a \approx 0.
\end{equation}

This completes the analysis of ADM formalism.

The physical interpretation of the constraints and the surface deformation algebra is remarkable. As these constraints are first-class, they generate gauge transformations. But what do they generate?

$H_a$ generates 3D spatial diffeomorphisms. We can see that by directly calculating the transformation

\begin{equation}
\delta g_{ab}(x) \equiv \{g_{ab}(x), \delta v^a \sbullet H_a \} = 2 \nabla_{(a} \delta v_{b)}(x) = \mathfrak{L}_{\delta v} g_{ab}(x),
\end{equation}

which is indeed the transformation of the metric under a diffeomorphism corresponding to the vector field $\delta v^a$.

The interpretation of the Hamiltonian constraint is a bit more nuanced. First, it clearly shows that GR is a reparametrization invariant theory. In light of our discussion in Sec. \ref{Creparinvariant}, it generates true evolution in the phase space, but also generates gauge transformation corresponding to the arbitrary parameterization of the dynamics. Geometrically, we saw that $N$ represents the embedding of the spatial leaves along the direction of time. Changing $N$ changes the temporal alignment of the leaves `locally', that is, it changes the foliation. Time in general relativity is a local entity. It is commonly known as `many-fingered time'.

In reparametrization invariant particle mechanics, we saw that the Hamiltonian constraint with an arbitrary coefficient generates evolution with an arbitrary parametrization. Fixing the lapse function in those models amounted to defining a `duration'. But in ADM formalism, time is local and we have to fix a local lapse function. Its local nature means that we can define a `local duration': Each point in the spatial manifold evolves at its own arbitrary speed.

Hence, $N(x)$ represents the foliation of the spacetime and signifies the local duration of the evolution of 3D metrics in a spatial manifold.

Another important point is that the Hamiltonian constraint does not generate any point transformation in the space of metrics ($Riem^3$), as it mixes the metric and the momenta. But it does map certain curves in the $Riem^3$ to other curves. Each curve in the configuration space represents a spacetime. What the Hamiltonian constraint does is change the foliation of the spacetime and map that curve to another valid solution.

Finally, $H$ and $H_a$ together generate 4D diffeomorphisms. Although this is not very clear in ADM formalism, as we have broken the spacetime structure, one can observe that by writing the ADM action (\ref{DADMaction}) as the phase space action

\begin{equation}
S_P[g_{ab},p^{ab},N,N^a] = \int dt d^3x \, \left( p^{ab} \dot{g}_{ab} - \left( N(x) H[g_{ab},p^{ab},x) + N^a(x) H_{a}[g_{ab},p^{ab},x) \right) \right).
\end{equation}

By extremizing this action with respect to $p^{ab}$ and putting the solution back in the action, $S_P$ would be identical to $S_{ADM}$. Therefore, we see that the gauge group of the two actions must be equivalent, i.e., for every gauge transformation of any curve $g_{ab}(t),N(t),N^a(t)$, there exists a corresponding gauge transformation of the curve in phase space. Noting that the gauge group of $S_{ADM}$ is that of $S_{EH}$, the 4D diffeomorphisms, We then see that $H,H_a$ are indeed the generators of the 4D diffeomorphisms in phase space.

\section{BSW action}

BSW action, carrying the three names Baierlein, Sharp, and Wheeler, is Jacobi's action for GR, found in 1962 \cite{Baierlein.1962}. It manifests the local reparametrization invariance of the theory more clearly. We can simply derive it from the decomposed Einstein-Hilbert action, $S_{ADM}$. First, note that (\ref{DADMaction}) can be written as

\begin{equation}\label{DADMactionkinetic}
S_{ADM}[g_{ab},N,N^a] = \int dt d^3x \, \sqrt{g} \left( \frac{1}{4N} T + NR \right),
\end{equation}

where $T$ is the kinetic term

\begin{equation}
T =G^{abcd} \left( \dot{g}_{ab} - 2\nabla_{a} N_b \right) \left( \dot{g}_{cd} - 2\nabla_{c} N_d \right).
\end{equation}

Extremizing (\ref{DADMactionkinetic}) with respect to the lapse function $N$ gives

\begin{equation}
N = \frac{1}{2} \sqrt{\frac{T}{R}}.
\end{equation}

Substituting this expression in the action, eliminates the lapse function and gives the BSW action:

\begin{equation}\label{DBSWaction}
S_{BSW}[g_{ab},N^a] = \int dt d^3x \, \sqrt{g R T}.
\end{equation}

This formally elegant and simple action is Jacobi's action for GR, and gives the same equations of motion as $S_{EH}$ action, provided that the original spacetime is globally hyperbolic. $-R$ plays the role of the potential here.

$BSW$ action is manifestly reparametrization invariant. If we use that as the departure point for the Hamiltonian formulation, instead of the ADM action, we arrive at the canonical Hamiltonian $N_a \sbullet H^a$ and the primary constraints

\begin{equation}
H \approx 0,\qquad
p_{N^a} = \frac{\delta \mathcal{L}_{ADM}}{\delta \dot{N}^a} \approx 0.
\end{equation}

As compared to the standard ADM formalism, we see that the Hamiltonian constraint appears as a primary constraint here, but the diffeomorphism constraint is still secondary. On the whole, one can perform the Legendre transformation and arrive at the total Hamiltonian (\ref{Dtotalhamiltonianadm}). BSW action does not change that.